\documentclass[fleqn,usenatbib]{mnras}

\DeclareUnicodeCharacter{2212}{\ensuremath{-}}
\usepackage{newtxtext,newtxmath}

\usepackage[T1]{fontenc}

\DeclareRobustCommand{\VAN}[3]{#2}
\let\VANthebibliography\thebibliography
\def\thebibliography{\DeclareRobustCommand{\VAN}[3]{##3}\VANthebibliography}

\usepackage{graphicx}	
\usepackage{amsmath}	
\usepackage[dvipsnames]{xcolor} 

\title[Diverse dust vertical height in protoplanetary discs]{Diverse dust vertical height and settling strength conditions in protoplanetary discs}

\author[Antilen et al.]{Juanita Antilen,$^{1}$\thanks{E-mail: juanita.antilen.22@ucl.ac.uk}
Paola Pinilla,$^{1}$
Dafa Li,$^{2,3}$
Marion Villenave,$^{4}$
Anibal Sierra,$^{1}$
Yao Liu,$^{5}$\newauthor
Myriam Benisty,$^{6}$
and Christian Ginski$^{7}$
\\
$^{1}$Mullard Space Science Laboratory, University College London, Holmbury St Mary, Dorking, Surrey RH5 6NT, UK\\
$^{2}$Purple Mountain Observatory, Chinese Academy of Sciences, 10 Yuanhua Road, Qixia District, Nanjing 210023, China\\
$^{3}$School of Astronomy and Space Science, University of Science and Technology of China, 96 Jinzhai Road, Hefei 230026, China\\
$^{4}$Univ. Grenoble Alpes, CNRS, IPAG, F-38000 Grenoble, France\\
$^{5}$School of Physical Science and Technology, Southwest Jiaotong University, Chengdu 610031, China\\
$^{6}$Max-Planck Institute for Astronomy (MPIA), Königstuhl 17, 69117 Heidelberg, Germany\\
$^{7}$School of Natural Sciences, Center for Astronomy, University of
Galway, Galway, H91 CF50, Ireland
}

\date{Accepted 2026 May 4. Received 2026 April 24; in original form 2025 November 1}

\pubyear{2015}

\begin{document}
\label{firstpage}
\pagerange{\pageref{firstpage}--\pageref{lastpage}}
\maketitle

\begin{abstract}
The settling of dust particles plays a critical role in the growth and dynamics of dust grains. We performed a detailed modeling of the ALMA continuum substructures for six highly inclined protoplanetary discs using radiative transfer simulations, to constrain the vertical height of millimetre dust grains and the settling strength. Our modeling results are a very thin millimetre dust disc in T Cha ($\text{h}_{\text{dust}}<$ 0.1 au throughout the disc), a vertically extended dust disc in DoAr 25 ($\text{h}_{\text{dust}}$ of $\sim$ 4.7 au at 140 au) and tentatively a thin disc in MY Lup ($\text{h}_{\text{dust}}<$ 0.5 au at 70 au). From lower resolution observations we found a very thin disc for PDS 111 ($\text{h}_{\text{dust}}<$ 0.1 au throughout the disc) and a more vertically extended millimetre dust disc in V409 Tau ($\text{h}_{\text{dust}}$ of $\sim$ 1.3 au at 35 au). We could not measure the vertical height in the asymmetric disc of RY Lup. We also found that the input dust opacities are a source of degeneracy in our models. Our tentative results, assuming the Ricci dust opacities, point to a diverse settling strength in our sample and possible radial variations. We also compared the models that best fit the ALMA data with the SPHERE data to test if they can reproduce the vertical distribution of small dust grains. This comparison suggests that models that reproduce the dust density distribution in the midplane cannot reproduce the distribution of small dust grains in the upper layers, reinforcing the need for more complex models. 
\end{abstract}



\section{Introduction}
Much uncertainty still exists about the physical processes involved in planet formation and the origin of the diversity in the observed exoplanetary systems. In this context, characterising the physical conditions in protoplanetary discs can help us to understand the early phases of planet formation. One physical process that is common to all protoplanetary discs and is very relevant for planet formation is the settling of dust particles to the midplane. This process sets the vertical stratification of dust grains and accelerates the planet formation process because it increases the dust-to-gas density ratio in the midplane, favoring the growth of dust particles through collisions and the formation of planetesimals \citep{Dubrulle1995I,Youdin2007}. Elevated dust concentration in the midplane is essential for the development of streaming instability, which is a potential mechanism for the formation of planetesimals \citep{Youdin2005, Youdin2007b, Johansen2007}, and it is also essential to improve the efficiency of pebble accretion \citep{Ormel2018, Drazkowska2023}.\\ In particular, the strength of dust settling is modulated by the level of gas turbulence, which is known to play a major role in the evolution of protoplanetary discs and also affects grain growth efficiency and the formation of planetesimals \citep[e.g.][]{Pinilla2021,Lim2024}. There is still no complete understanding of the settling and turbulence conditions in protoplanetary discs.  Measurements of the strength of the settling and vertical turbulence from observations are essential to elucidate these questions, and radiative transfer simulations offer a powerful tool for inferring these quantities from observations.\\ With radiative transfer simulations, it is possible to perform a detailed modeling of the continuum substructures observed with ALMA, and simultaneously model the vertical stratification of the dust grains to reproduce the spatial thickness of the disc. This method was first presented by \cite{Pinte2016} and has been successfully applied to different systems in several works \citep[e.g.][]{Liu2022, Villenave2022, Villenave2025}, which have found vertical turbulence levels ranging from orders of magnitude $\sim$$10^{-5}$ to $\sim$$10^{-2}$. The best targets for this kind of analysis are high- and mid-inclination discs, because at different wavelengths we can directly observe the vertical distribution of the dust grains, and radial substructures can still be clearly identified, and thus a characterisation of the radial and vertical dust density distributions can be done simultaneously. Although several efforts have been made to characterise dust-settling efficiency and vertical turbulence using radiative transfer simulations, the current number of studies is limited to dozens of objects, making it difficult to set the results in the context of the large diversity of systems that can be found, for example, across different star-forming regions, evolutionary phases, and stellar types \citep[e.g.][]{Villenave2025}.\\ 
In this study, we modeled the dust distribution in six highly inclined discs to constrain the vertical height of millimetre dust grains and the strength of dust settling. For this aim, we employ radiative transfer simulations, and model the dust continuum substructures observed by ALMA. We also compare the models with SPHERE observations, to test if the best models of the dust density, found in the ALMA data modeling, can also describe the distribution of the very small dust grains in the upper layers of the discs. With this study, we expect to better characterise the distribution of the solids in this sample of protoplanetary discs and to increase the number of objects where the strength of the settling and vertical turbulence have been constrained observationally by providing a homogeneous analysis.\\
This paper is organised as follows. In Section \ref{sec:observations}, we briefly describe the data analysed in this study; in Section \ref{sec:radiativetransfer}, we explain the setup of the radiative transfer simulations;  we present the results in Sections \ref{sec:resultsalma} and \ref{sec:resultssphere}; discuss the results and compare them with previous studies in Section \ref{sec:discussion}; and finally, in Section \ref{sec:conclusions}, we summarise the main results of this work.

\begin{figure*}
\includegraphics[width=2\columnwidth]{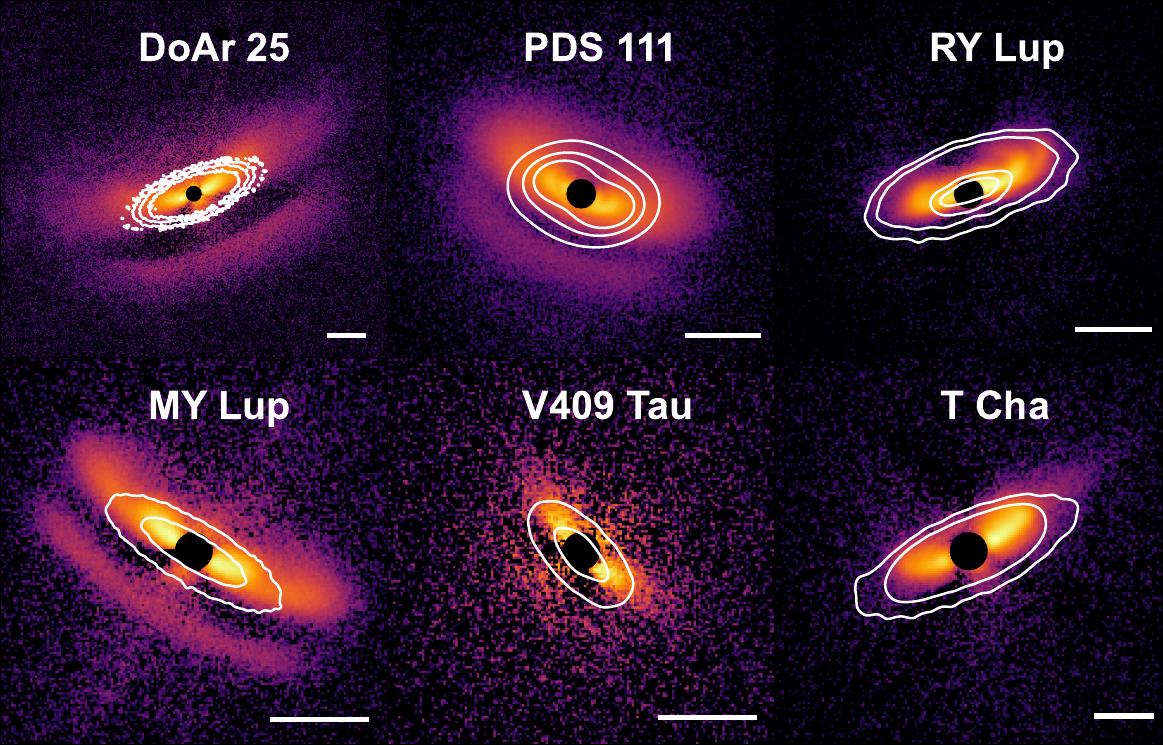}
\caption{SPHERE Q$\phi$ images in H band ($1.6\,\mu \text{m}$) are shown in colours. ALMA data is overlapped in white contours showing a fraction of the RMS noise, of 8, 40 and 100, 11 and 24, 8 and 50, 8 and 50, as well as 5 and 30, for the discs of DoAr 25, PDS 111, RY Lup, MY Lup, V409 Tau, and T Cha, respectively. Scalebars of 0.5" are shown in the lower right corner of each panel. The resolution of the ALMA observations is shown in Fig.\,\ref{fig:all_figures_alma_gallery}.}
\label{fig:gallery_sphere_data_contours}
\end{figure*}

\begin{table*}
    \centering
    Host Star Properties\\
    \begin{tabular}{cccccccccc}
        \hline \hline
        Name & Region & \textit{d}  & SpT & log $T_{\text{eff}}$  & log $L_{*}$ & log $M_{*}$ & log $t_{*}$ & log $\dot{M}_{*}$ & References  \\
         &  & (pc)  &  & (K)  & (L$_{\odot}$) & (M$_{\odot}$) & (yr) & (M$_{\odot}$\,yr$^{-1}$) &  \\
        (1) & (2) & (3)  & (4) & (5)  & (6) & (7) & (8) & (9) & (10) \\
        \hline 
        PDS 111 & foreground of Orion & 157.9 & G2V & 3.77 & 0.40 & 0.08 & 7.2 & $[-10.0,-9.3]$ & 1 \\
        DoAr 25 & Oph L1688 & 138.0  & K5 & 3.63 & −0.02  & -0.02  & 6.3 & −8.3 & 2 \\
        MY Lup & Lup IV & 156.0  & K0 & 3.71   & −0.06  & 0.09 & 7.0 & $<$−9.6 & 2 \\
        RY Lup & Lupus & 159.0 & K2 & 3.68 & 0.40 & 0.15 & $[6.3, 6.7]$ & -8.2 & 3, 4 \\
        V409 Tau & Tau L1524 & 129.7 & M0.6 & 3.56 & -0.66 & -0.40 & 6.0 & -8.0 & 5, 6 \\
        T Cha & $\epsilon$ Cha & 107.0  & G8  & 3.73 & 0.34 & 0.18 & $[6.3,7.1]$ & -8.4 & 7, 8, 9, 10 \\
        \hline
    \end{tabular}
    \caption[]{Column 1: Target name. Column 2: associated star-forming region. Column 3: distance. Column 4: spectral type. Column 5: effective temperatures. Column 6: stellar luminosities. Columns 7 and 8: stellar masses and ages. Column 9: accretion rates. \textbf{References.} (1)  \cite{Derkink2024}; (2) \cite{Andrews2018}; (3) \cite{Gravity2020}; (4) \cite{Francis2020}; (5) \cite{Long2019}; (6) \cite{Garufi2024}; (7) \cite{Huelamo2015}; (8) \cite{Schisano2009}; (9) \cite{Hendler2018}; (10) \cite{Pohl2017}.}
    \label{tab:table1}
\end{table*}

\begin{table}
    \centering
    \caption{Basic data information. The last column shows the synthesized beam FWHM and position angle.}
    \begin{tabular}{ccccc}
        \hline \hline
        Name & ALMA Band & Project ID & $\theta_{b}, $PA$_{b}$\\
         &  &  & (", \textdegree)\\
        \hline
         PDS 111 & B6  & 2021.1.01705.S & 0.357 $\times$ 0.250, -71.9  \\
         DoAr 25 & B6 & 2016.1.00484.L & 0.041 $\times$ 0.022, 70\\
         MY Lup & B6 & 2016.1.00484.L & 0.044 $\times$ 0.043, 122\\
         RY Lup & B6 & 2017.1.00449.S & 0.130 $\times$ 0.120, 95.94\\
         V409 Tau & B6 & 2016.1.01164.S & 0.130 $\times$ 0.110, -4.93\\
         T Cha & B3 & 2015.1.00979.S & 0.110 $\times$ 0.060, -18.11\\
        \hline 
    \end{tabular}
    \label{tab:table2}
\end{table}

\begin{table*}
    \centering
    Disc parameters fixed in the ALMA data modeling\\
    \begin{tabular}{ccccccccc}
        \hline \hline
        Name & i & PA & M$_{\text{disc, initial}}$& R$_{\text{out}}$ & $\gamma$ & H$_{100}$ & $\alpha$ & References  \\
         & (\textdegree) & (\textdegree)  & (M$_{\odot}$) & (au)  &  &  &  &  \\
        (1) & (2) & (3)  & (4) & (5)  & (6) & (7) & (8) & (9) \\
        \hline 
        PDS 111 & 58.2 & 66.2 & 0.013 & 98 & 1.30 & 10 & 1\,$\times\,10^{-5}$, 1\,$\times\,10^{-4}$, 1\,$\times\,10^{-3}$, 1\,$\times\,10^{-2}$  & 1 \\
        DoAr 25 & 67.0 & 111.0 & 0.220 & 166 & 1.30 & 7 & 1\,$\times\,10^{-5}$, 1\,$\times\,10^{-4}$, 1\,$\times\,10^{-3}$, 1\,$\times\,10^{-2}$  & 2 \\
        MY Lup & 73.2 & 58.8 & 0.471 & 88 & 1.30 & 7 & 1\,$\times\,10^{-5}$, 1\,$\times\,10^{-4}$, 1\,$\times\,10^{-3}$, 1\,$\times\,10^{-2}$  & 2 \\
        RY Lup & 67.0 & 109.0 & 0.022 & 136 & 1.15 & 10 & 1\,$\times\,10^{-5}$, 1\,$\times\,10^{-4}$, 1\,$\times\,10^{-3}$, 1\,$\times\,10^{-2}$  & 3, 4 \\
        V409 Tau & 69.3 & 44.8 & 0.00108 & 47 & 1.15 & 7 & 1\,$\times\,10^{-5}$, 1\,$\times\,10^{-4}$, 1\,$\times\,10^{-3}$, 1\,$\times\,10^{-2}$  & 5, 6 \\
        T Cha & 73.0 & 113.0 & 0.017 & 58 & 1.15 & 7 & 1\,$\times\,10^{-5}$, 1\,$\times\,10^{-4}$, 1\,$\times\,10^{-3}$, 1\,$\times\,10^{-2}$  & 7 \\
        \hline
    \end{tabular}
    \caption[]{Column 1: Target name. Column 2: inclination. Column 3: position angle. Column 4: total disc mass (gas and dust), set as initial guess in the modeling. Column 5: continuum radius, representing 95\% of the continuum flux, which is taken as the outer boundary of the radial grid in all the simulations. Column 6: flaring index. Column 7: scale height of the gas at 100 au in Eq. (4). Column 8: turbulence parameter. \textbf{References.} (1) \cite{Derkink2024}; (2) \cite{Huang2018}; (3) \cite{Gravity2020}; (4) \cite{Francis2020}; (5) \cite{Long2019} ; (6) \cite{Garufi2024}; (7) \cite{Huelamo2015, Hendler2018}.}
\label{tab:table3}
\end{table*}

\begin{figure*}
\includegraphics[width=1\columnwidth]{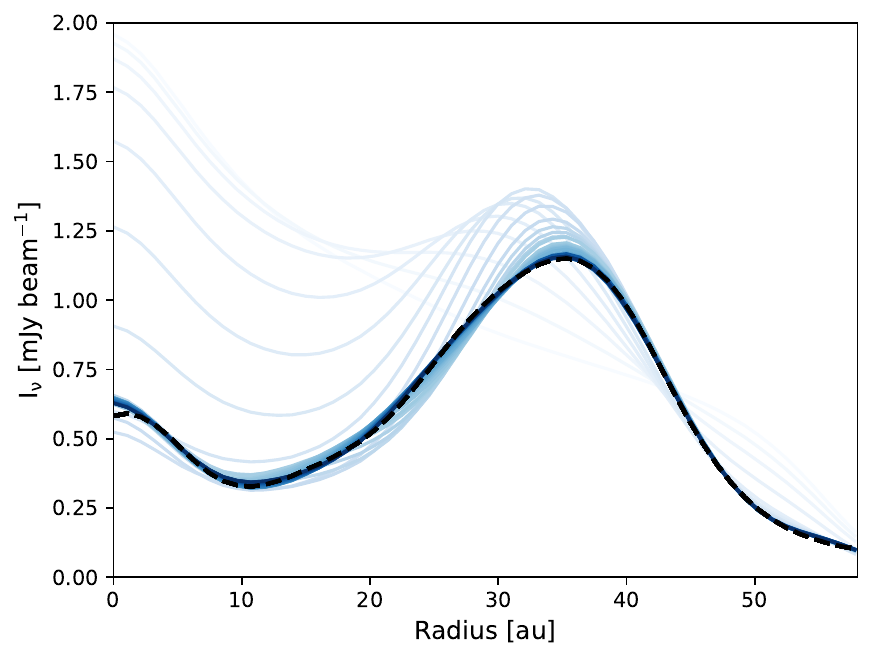}
\includegraphics[width=1\columnwidth]{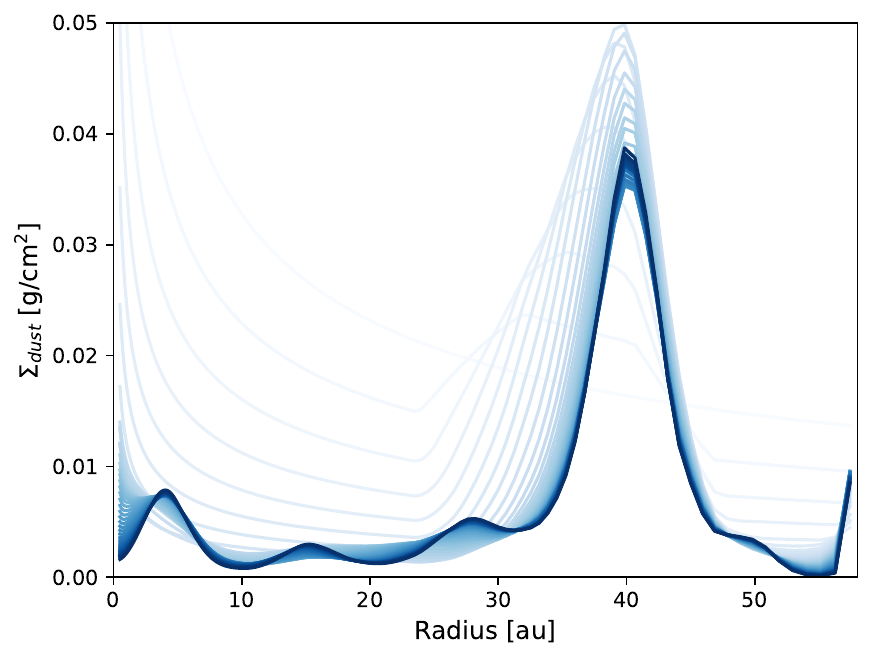}
\caption{Example of the iteration process to find the best model of T Cha for a fixed turbulence parameter of $\alpha=1\times10^{-3}$. Radial intensity profile and dust surface density distribution for the grain size $a=1257.15\,\mu \text{m}$ after each iteration, are shown on the left and right side respectively. The ALMA Band 3 radial intensity profile is represented by a dashed black line in the figure on the left. The first iterations are shown in the lightest blue colors, and the latest are shown in the darkest blue colors.}
\label{fig:iterations-tcha}
\end{figure*}

\begin{figure*}
\includegraphics[width=2\columnwidth]{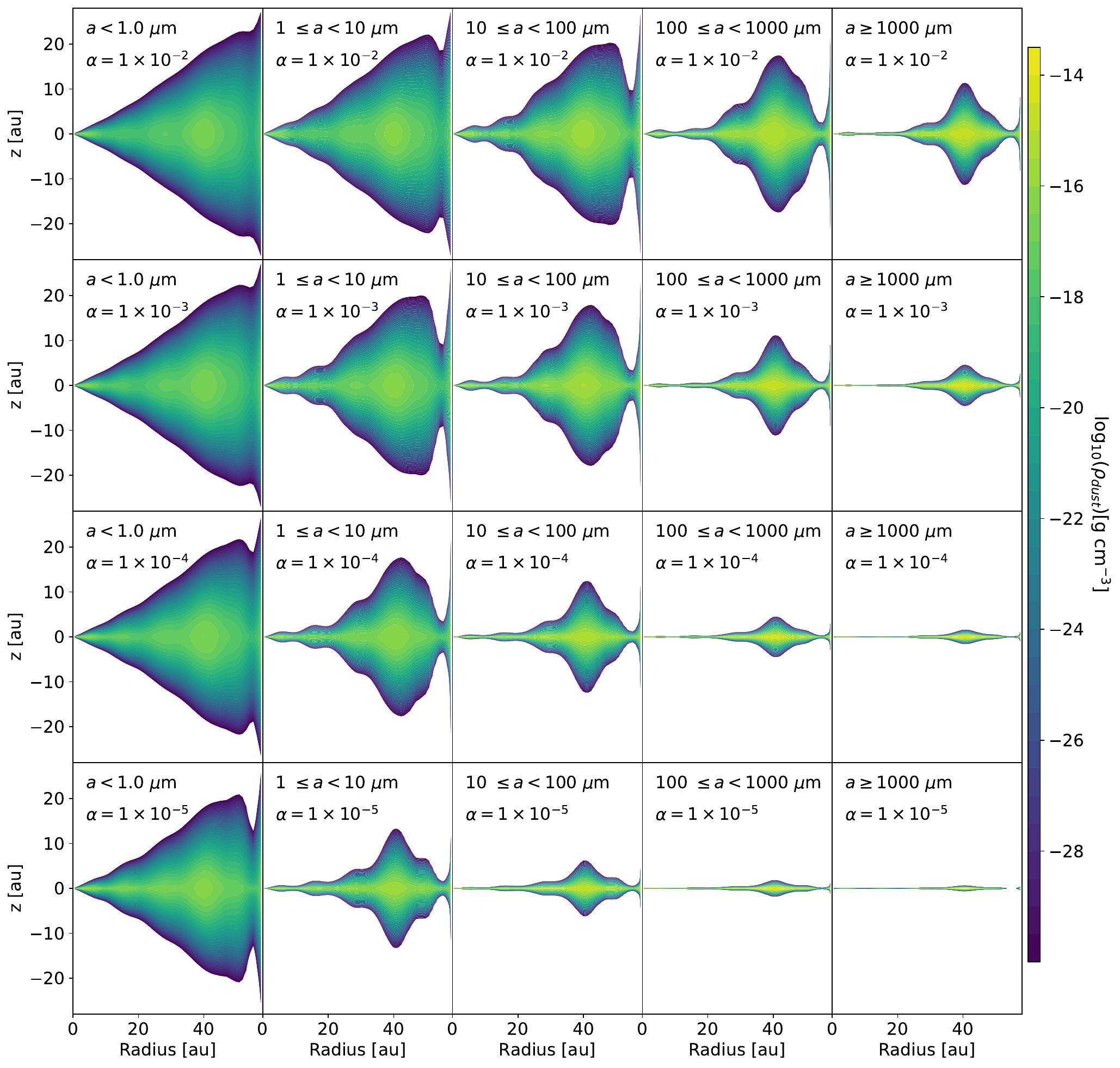}
\caption{Dust density distribution of the best models found for the disc of T Cha, for several turbulence parameters and grain size ranges. The plot does not display the full range of volumetric density to leave space for the labels.}
\label{fig:tcha_density_all-alphas}
\end{figure*}

\begin{figure*}
\includegraphics[width=0.75\columnwidth]{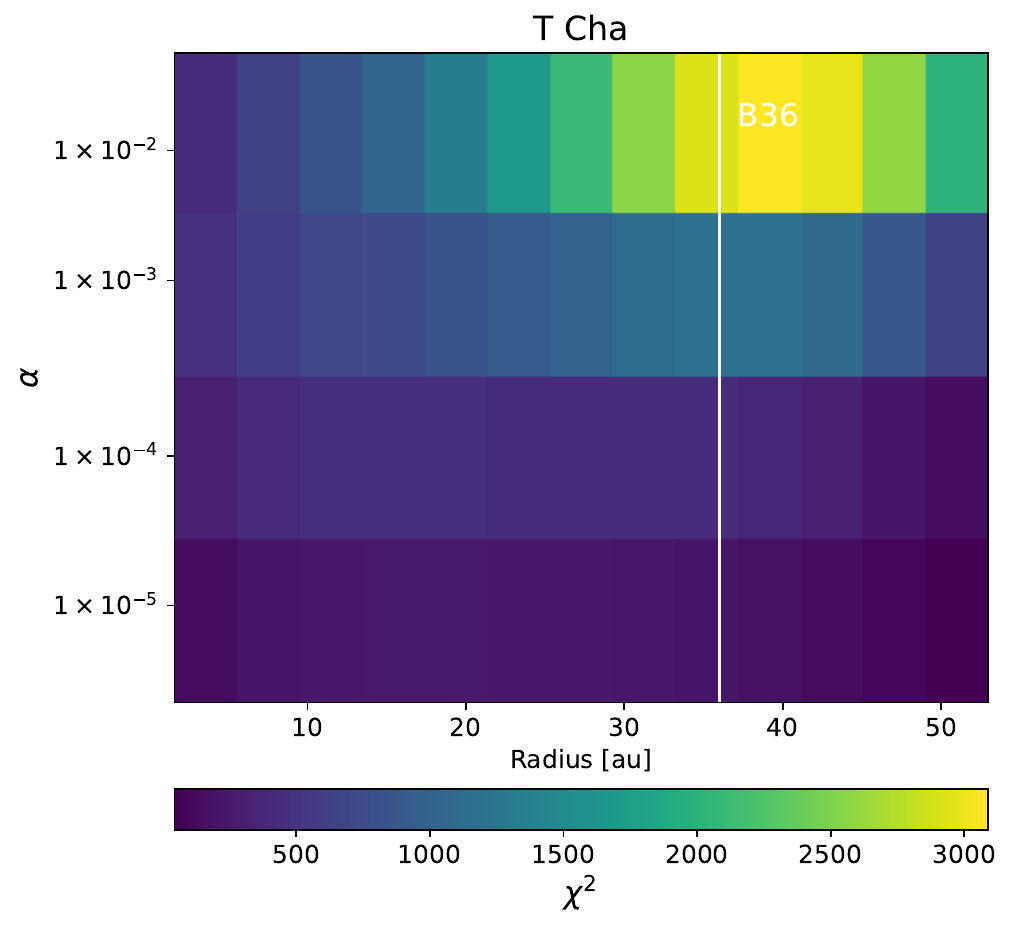}
\includegraphics[width=0.633\columnwidth]{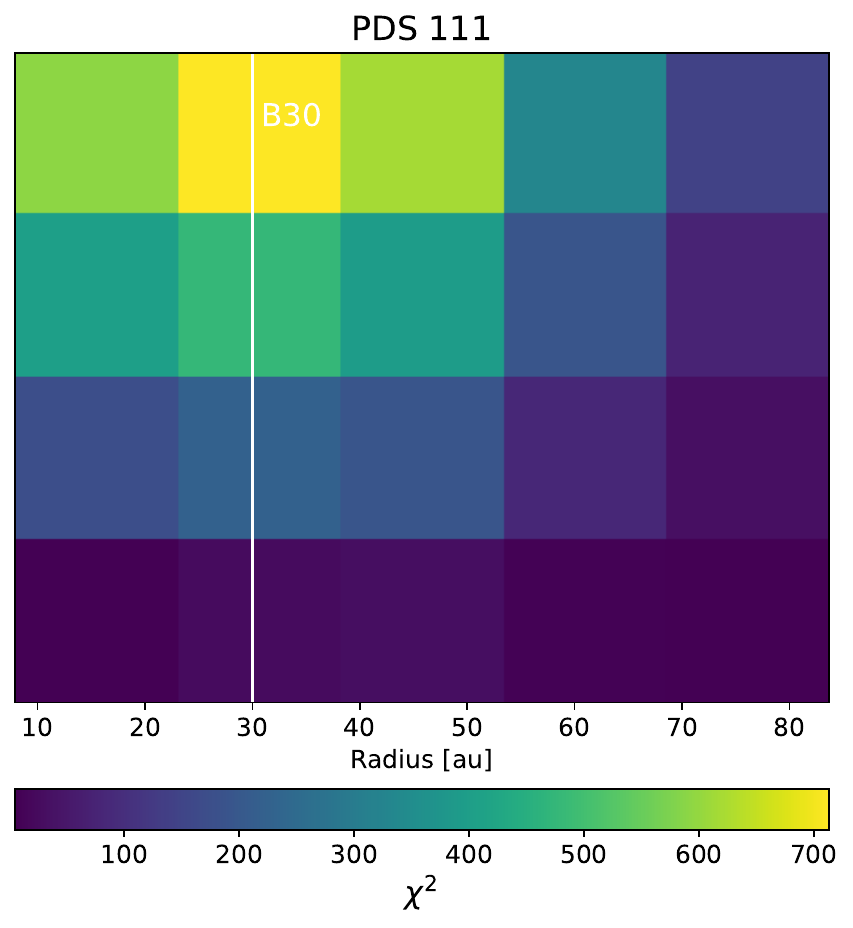}\\
\includegraphics[width=0.742\columnwidth]{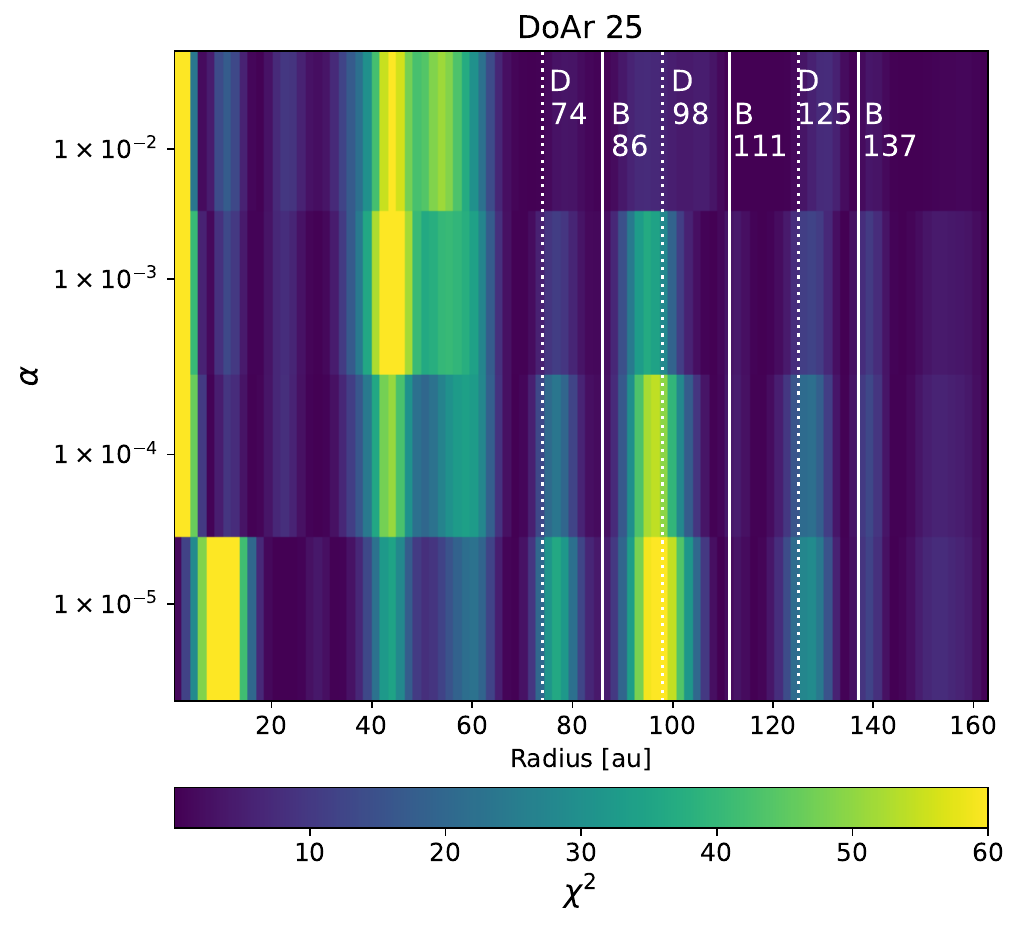}
\includegraphics[width=0.633\columnwidth]{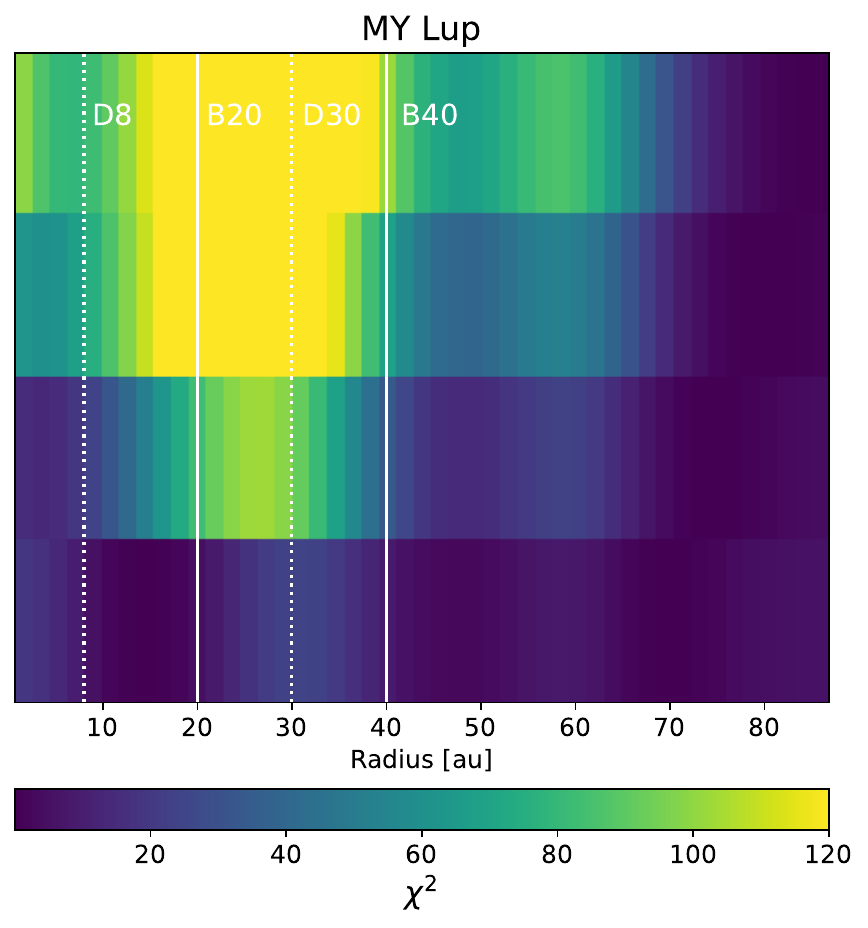}
\includegraphics[width=0.63\columnwidth]{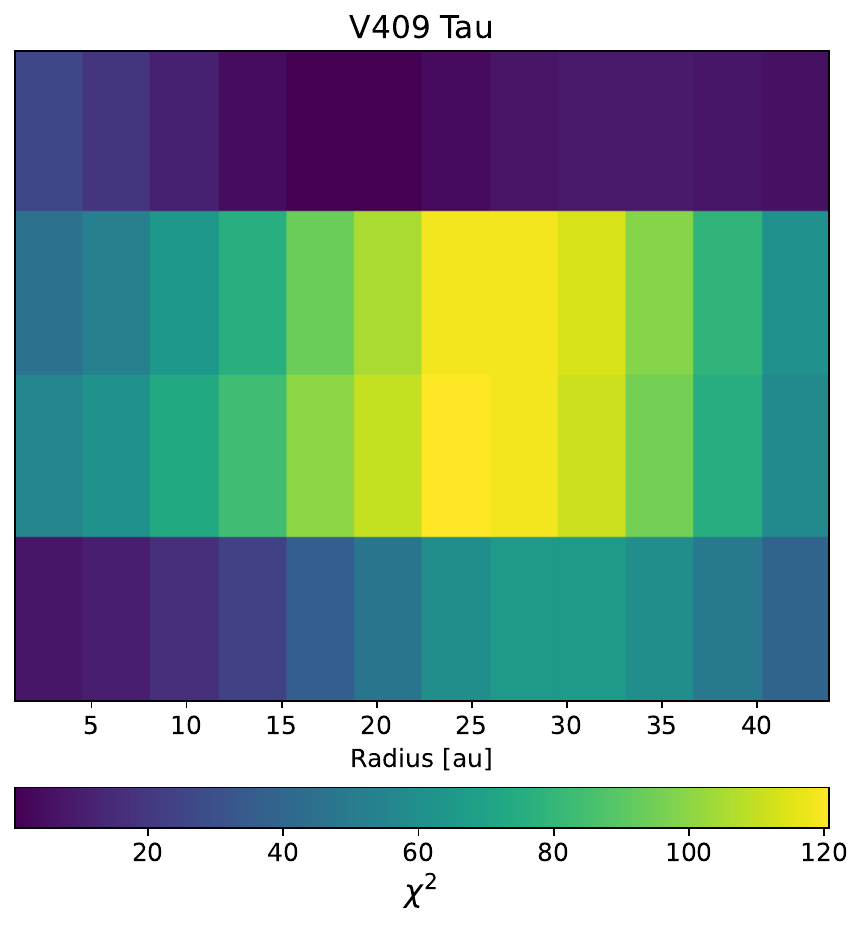}
\caption{$\chi^{2}$ between radial intensity profiles of models and observations along the semi-minor axis for DoAr 25, MY Lup, and V409 Tau respectively (left to right). Yellow tones in the figures show regions of weaker compatibility of the model to the data. In the figures of DoAr 25 and MY Lup, solid white lines mark the rings position, and dotted white lines mark the gaps position.} 
\label{fig:alphavsradius}  
\end{figure*}

\begin{figure*}
\includegraphics[width=1.8\columnwidth]{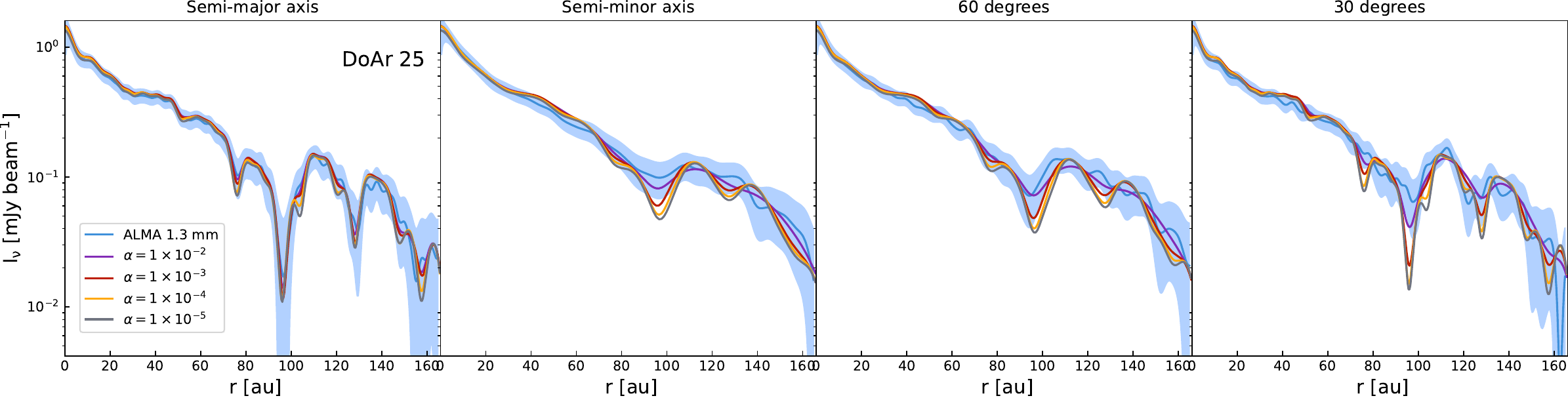}
\includegraphics[width=1.8\columnwidth]{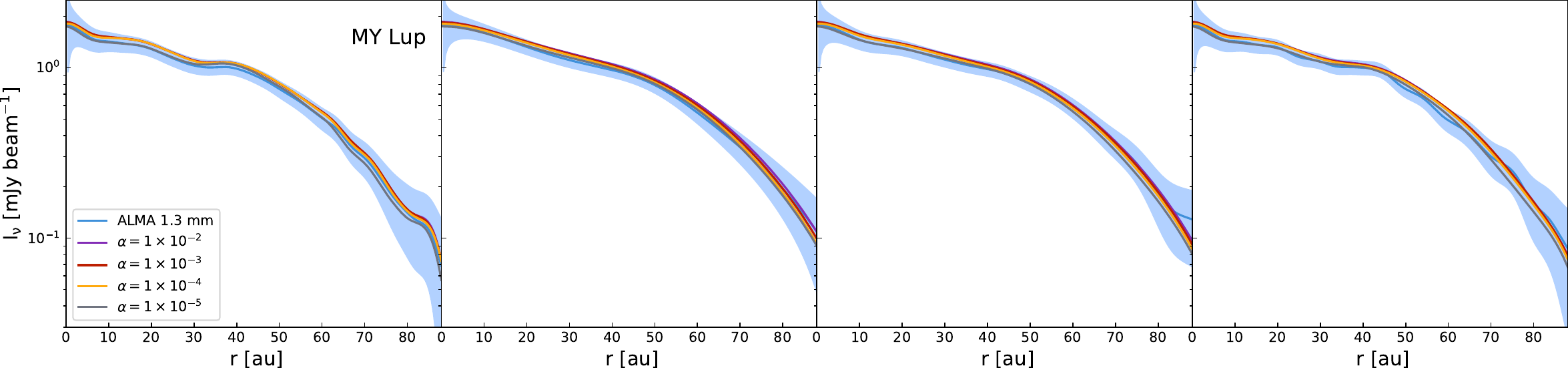}
\includegraphics[width=1.8\columnwidth]{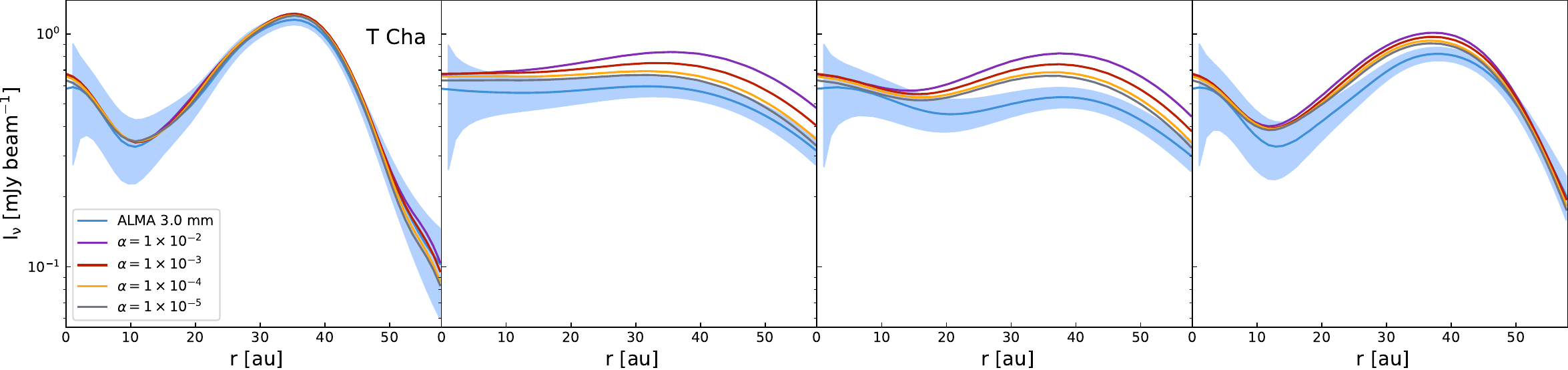} 
\includegraphics[width=1.8\columnwidth]{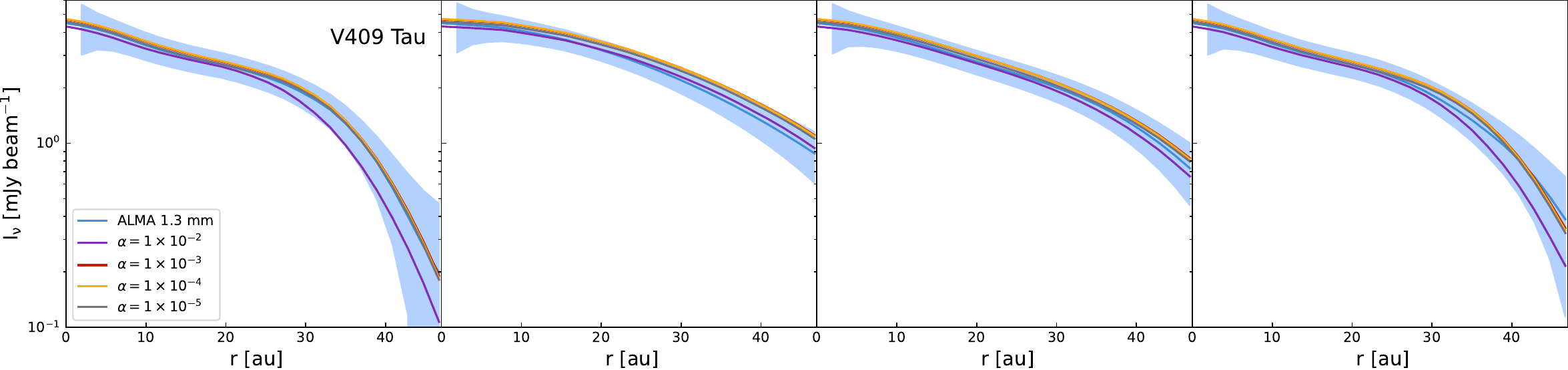}
\includegraphics[width=1.8\columnwidth]{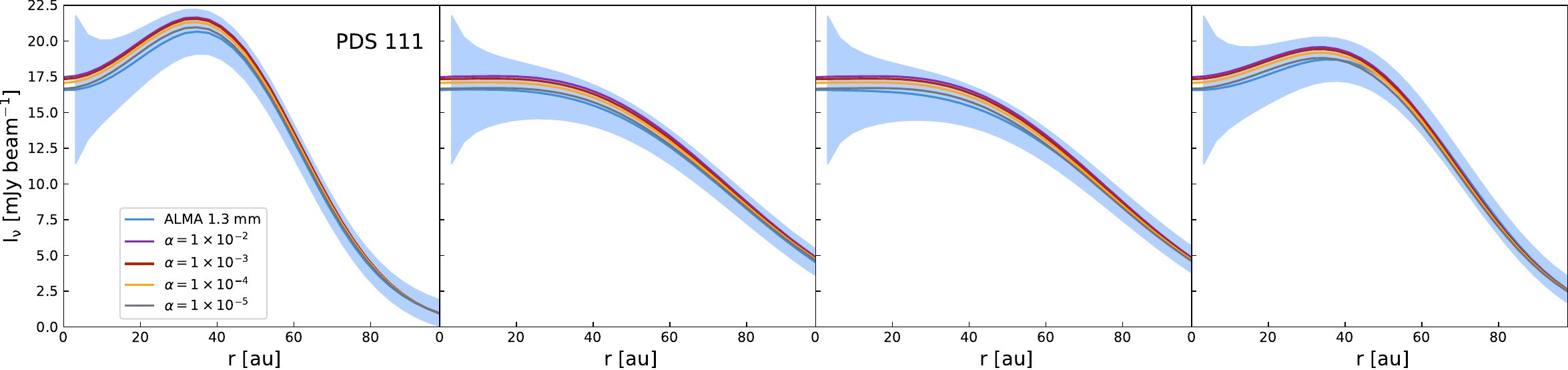}

\includegraphics[width=1.8\columnwidth]{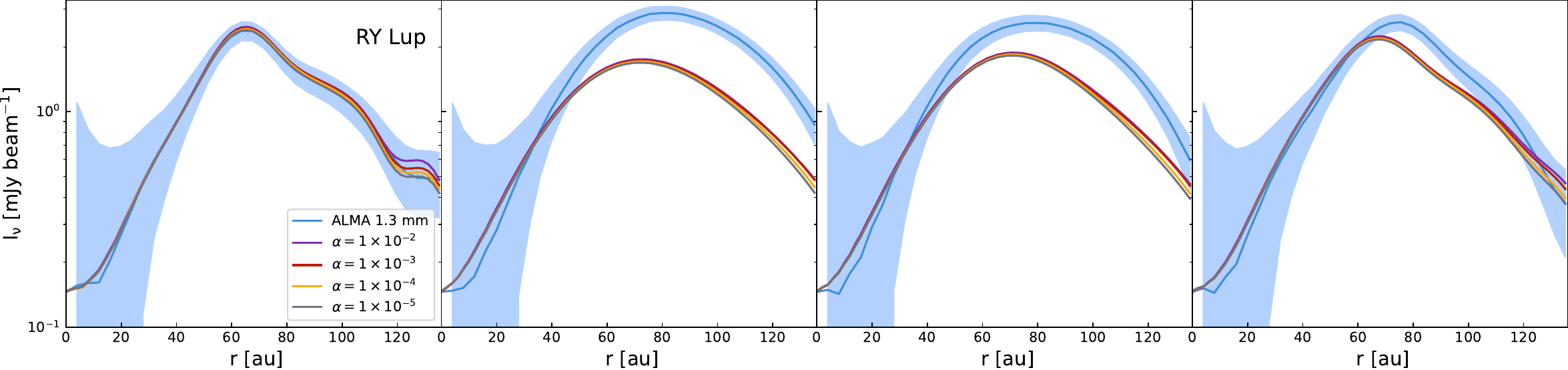}
\caption{From top to bottom, radial intensity profiles of ALMA observations and models for DoAr 25, MY Lup, T Cha, V409 Tau, PDS 111, and RY Lup respectively. These profiles are extracted along different azimuthal directions. ALMA observations and the corresponding uncertainties are shown in a blue line, and blue shaded area respectively. Best models calculated for different fixed turbulence parameters are shown in colors.}
\label{fig:all_radial_intensity_profiles}
\end{figure*}

\section{Observations}
\label{sec:observations}
\subsection{Sample selection}
Our objective is to obtain constraints on the vertical distribution of dust in protoplanetary discs and the level of vertical turbulence. As mentioned above, highly inclined discs are ideal targets for this type of study because we can have a direct view of the vertical distribution of dust. Furthermore, observations at near-infrared (NIR) and millimeter (mm) wavelengths can give us information on the dust at different regimes, NIR observations trace dust grains that are well coupled to the gas dynamics, and millimeter observations trace large dust grains decoupled from the gas. In particular, at millimeter wavelengths it is possible to resolve the dust radial substructures in the midplane of discs, and we can model these substructures to infer information not only on the vertical dust distribution but also on the radial dust distribution. For all these reasons, we select highly inclined discs (60$^{\circ}$$\lesssim\,$i$\,\lesssim\,$75$^{\circ}$), which have been observed with SPHERE / VLT and ALMA. In particular, we focus on highly inclined discs that have been observed by the disc Evolution Study Through Imaging of Nearby Stars (DESTINYS) survey \citep{Garufi2024, Ginski2024, Valegard2024} and GTO programs. The DESTINYS survey targets 85 T Tauri stars from six different star-forming regions to characterize the scattered light from the circumstellar material and constrain its evolution throughout the stages of planet formation \citep{Valegard2024}. We initially selected highly inclined discs that had been observed by these programs, and from this sample we selected all discs that had high-resolution ALMA data by November 2023. This subsample consists of seven discs with previous Band 6 and/or Band 3 ALMA observations. Finally, the modeling and the analysis are done for six of these sources because one source, PDS 415, displays a strong non-axisymmetric morphology in the continuum images, which is not possible to model with the approach of this study. Three sources in our sample show the backside in scattered light (PDS 111, DoAr 25 and MY Lup). In four cases the ALMA resolution is high enough ($\lesssim$ 0.13) to resolve the disc substructures revealing rings and gaps (DoAr 25, MY Lup, and T Cha), and in the two cases where the resolution is more coarse, the analysis is more challenging (PDS 111 and V409 Tau).

\subsection{Data}
We model ALMA continuum images of six protoplanetary discs: PDS 111, DoAr 25, MY Lup, RY Lup, V409 Tau, and T Cha. We show the H band Q$_{\phi}$ images for all the sources, with ALMA data overlapped in the contours in Fig.\,\ref{fig:gallery_sphere_data_contours}. The stellar properties are listed in Table \ref{tab:table1}, and all essential properties of the ALMA data, including the corresponding observing projects and resolution, are listed in Table \ref{tab:table2}. For the present analysis, Band 6 data were used in most cases, except for T Cha, where Band 3 data was used because these data have higher resolution than available Band 6 data. All ALMA continuum data have been previously presented and analysed in other works. We used calibrated data previously presented by the authors of the following articles: \cite{Derkink2024} (PDS 111), \cite{Andrews2018} (DoAr 25, and MY Lup), \cite{Francis2020} (RY Lup), \cite{Long2019} (V409 Tau) and \cite{Hendler2018} (T Cha), and we refer to these articles for further details on the calibration of these data. We note that all the analysis presented in this paper is done in the image plane.
\section{Radiative transfer modeling}
\label{sec:radiativetransfer}
To model the substructures and investigate the degree of dust settling, we performed radiative transfer modeling of the ALMA images. For a given disc model and star properties, dust temperature, images, and spectral energy distribution (SED) were calculated using the 3D Monte Carlo radiative transfer code RADMC-3D \citep{dullemond2012} employing the full scattering mode. In these models, we assume that the discs are only heated by stellar irradiation in all cases. We also used a two-dimensional axisymmetric disc model with a spherical coordinate system for all sources. In these models, we define radial and zenith-angle grids. Radial grid cells are logarithmically spaced (301 cells) and polar grid cells are linearly spaced (604 cells). Model images are convolved with a 2D Gaussian with the characteristics of the corresponding ALMA resolution. The number of photon packages used for the thermal and scattering Monte Carlo simulations is 10$^{7}$. 
\subsection{Dust properties}
We assumed a grain composition consisting of a mass fraction of 24$\%$ graphite, 42$\%$ amorphous carbon, and 33$\%$ water ice, consistent with the Ricci oppacity model \citep{Ricci2010}. The dust and ice (absorption and scattering) opacities were calculated using \texttt{OpTool} \citep{dominik2021} and the Mie theory, assuming compact and sphere particles. The distribution of grain sizes follows the power law $dn(a)\propto a^{−3.5}da$ with a minimum ($a_{min}$) and a maximum size ($a_{max}$), where $a$ is the grain radius, and $n(a)$ is the number of dust particles whose size is within the interval [$a, a+da$]. We consider 30 grain-size bins with a logarithmic distribution; we also fix $a_{min}$ to 0.01 $\mu$m, and $a_{max}$ to 3000 $\mu$m. All grain sizes are uniformly distributed throughout the disc.
\subsection{Disc Model}
We assume a Gaussian profile for the vertical density distribution of the dust:
\begin{equation}
\rho_d(r, z, a)=\frac{\Sigma(r, a)}{\sqrt{2 \pi} h_\text{dust}(r, a)} \exp \left[ -\frac{1}{2} \left(\frac{z}{h_\text{dust}(r, a)}\right)^2\right]
\end{equation}
where $r$ is the distance from the star along the midplane, $z$ is the altitude over the midplane, $\Sigma(r,a)$ is the dust surface density and $h_\text{dust}(r,a)$ is the dust scale height.
According to the derivation from \cite{Youdin2007}, the dust scale height is given by: 
\begin{equation}
    h_{\mathrm{dust}}(r,a)=H_{\mathrm{gas}}\left(1+\frac{\mathrm{St}}{\alpha} \frac{1+2 \mathrm{St}}{1+\mathrm{St}}\right)^{-1 / 2}
\label{eq:Hdust}
\end{equation}
where $\mathrm{St}$ is the Stokes parameter:
\begin{equation}    
\mathrm{St}=\frac{\rho_{\text{grain}}}{\Sigma_{\text{gas}}(r,a)}\frac{a\,\pi}{2}, 
\label{eq:stokes}
\end{equation}
and $\alpha$ is the disc viscosity parameter \citep{shakura1973} that sets the level of vertical turbulence. $H_{\mathrm{gas}}$ is the gas scale height, assumed to be:
\begin{equation}
H_{\mathrm{gas}}=H_{100}\left(\frac{r}{100\, \mathrm{au}}\right)^{\gamma}
\label{eq:Hgas}
\end{equation}
where $H_{100}$ is the gas scale height at radius $r = 100$\,au.\\
We iteratively build a dust surface density profile that reproduces the substructures observed with ALMA along the semi-major axis of the discs, using a strategy that has been used in previous works \citep[e.g.][]{Pinte2016,Liu2022,Villenave2025}. With this aim, we start the modeling with an initial dust surface density, given by:
\begin{equation}
    \Sigma_{i}(r, a)=\Sigma_0(a)\left(\frac{r}{100\,\mathrm{au}}\right)^{-\beta}.
\label{eq:initialsurfdens}
\end{equation}

We set stellar, observational, and disc parameters for our models (see Tables~\ref{tab:table1} and~\ref{tab:table3}), and set the following disc parameters: dust-to-gas mass ratio, $\gamma$ (flaring index), $H_{100}$, and $\beta$ (exponent for the initial dust surface density). $\gamma$ and $\beta$ are based on a combination of values that fits the SED (described in Appendix \ref{sec:discussionsed}). We also set the disc mass as an initial guess in the modeling, and $\rho_\text{{grain}}$ is fixed to 1.65 g/$\text{cm}^{3}$. To investigate the degree of dust settling, we test four orders of magnitude for the vertical turbulence parameter $\alpha$ in Eq.~\ref{eq:Hdust}; we also note that the gas surface density is calculated from the dust surface density, assuming a fixed dust-to-gas mass ratio of 0.01. To find the best-fit dust surface density profile for each $\alpha$, we follow the steps described in the following subsection.
\subsection{Fitting the ALMA image}
To model the ALMA data, we determine a dust surface density profile that reproduces the substructures observed along the semimajor axis of the discs, using an iterative procedure. We note that the fit of the data is done along the semi-major axis because the substructures are most clearly seen along this axis after the deprojection. In this procedure, we first extract the observed brightness profile $I_{v,o}(r)$ along the semi-major axis of the deprojected ALMA image, on the west side, in a wedge that covers an azimuthal range of five degrees. To calculate the first model, we start by assuming an initial dust surface density (Eq.\,\ref{eq:initialsurfdens}), calculate an initial 2D dust density distribution, calculate the dust temperature using RADMC-3D, and compute an initial model image at the corresponding ALMA wavelength. The initial model image is then convolved with the ALMA beam, which is represented by a 2D Gaussian, and we extract the brightness profile of the model on the west side $I_{v,m}(r)$ \textbf(this side provides sufficient information on the substructures) along the semi-major axis. In this way, we can calculate a correction factor $\varepsilon(r)$ for the initial surface density profile, this factor is defined as: $\varepsilon(r) =\frac{I_{v,o}(r)}{I_{v,m}(r)}$. Once we get this factor, we multiply the initial surface density profile $\Sigma_{i}(r, a)$ by $\varepsilon(r),$ and recalculate the dust density distribution to obtain an updated model image. This correction factor is applied until the residuals reach a clear minimum or until the residuals reach a threshold of 0.85; the residuals are defined as $\text{residuals} = [I_{v,o}(r) - I_{v,m}(r)]/\text{rms}_{\text{image}}$. Through this iterative procedure, we generate several models computed for different levels of dust settling. Radial intensity profiles of these models extracted from deprojected images will look almost indistinguishable along the semi-major axis, but will always display differences along the semi-minor axis; this is a projection effect that occurs because we observe the thickness of the disc through the gaps, so the gaps seem to be filled at this direction in the profiles, where the amount of filling reflects the geometric thickness of the disc. For this reason, to later differentiate between models and estimate model performance, we compare radial intensity profiles of models and observations along the semi-minor axis (Sec.\,\ref{sec:resultsalma}). We present the iteration process to model the continuum observations of T Cha in Fig.\,\ref{fig:iterations-tcha} as an example. This figure shows the radial intensity profile and the dust surface density profile for the grain size $a=1257.15\,\mu \text{m}$ after each iteration. Later, in Fig.\,\ref{fig:tcha_density_all-alphas}, we present the dust density distribution calculated for T Cha for the best models found in the iteration process. Fig.\,\ref{fig:tcha_density_all-alphas} displays the volume density for various grain size ranges. It shows that very small dust grains \textit{a}\,$<$\,$1.0\,\mu \text{m}$ are almost equally distributed vertically for different parameters $\alpha$, and, in contrast, larger dust grains, for example \textit{a}\,$>$\,$1000\,\mu \text{m}$ are much more settled for smaller $\alpha$.\\
We note that the method described in this paper is effective for a disc that is in the optically thin regime, as the intensity of the continuum emission will scale linearly with the dust surface density; we comment more on this aspect in Sec.\,\ref{sec:discussion}; and show the vertical optical depth for all the best models  in Fig\,\ref{fig:opt-depth}.

\section{Results from the ALMA data modeling}
\label{sec:resultsalma}
The results of the ALMA data modeling are presented in Fig.\,\ref{fig:alphavsradius}, Fig.\,\ref{fig:all_radial_intensity_profiles}, and Fig.\,\ref{fig:all_figures_alma_gallery}. To quantify and evaluate the performance of the models, we calculated $\chi^{2}$ vs. $r$ maps for all sources, which are shown in Fig.\,\ref{fig:alphavsradius}. $\chi^{2}$ is defined as $[I_{v,o}(r) - I_{v,m}(r)]/\text{rms}_{\text{image}}^{2}$. All maps are calculated for the brightness profiles along the semi-minor axis, where the largest contrast occurs between the models and observations (as seen in Fig.\,\ref{fig:all_radial_intensity_profiles}); the radial spacing in all maps is chosen to be one-third
of the beam size of the corresponding observations, to ensure that the signal is not under sampled. Fig.\,\ref{fig:all_radial_intensity_profiles} displays the observed brightness profiles with the brightness profiles of the best models in several azimuthal directions, and Fig.\,\ref{fig:all_figures_alma_gallery} shows the ALMA continuum data with their corresponding synthetic images calculated for the best models. The ALMA radial profiles were obtained by taking the average intensity in concentric rings and extracted from the deprojected images in an azimuthal wedge of five degrees. Flux errors are calculated following the method described in \cite{carrasco-gonzalez2019}, using the formula $\Delta I_{\nu}=\text{rms}_{\nu}/(\Omega_{\text{ring}}/\Omega_{\text{beam}})^{0.5}$, where $\Omega_{\text{ring}}$ and $\Omega_{\text{beam}}$ are the solid angles of each ring and beam, respectively. We do not consider the flux calibration uncertainty, which would give us larger error bars. The analytical equation we use to describe the settling (Eq.~\ref{eq:Hdust})  allows us to self-consistently calculate the dust scale height ($h_{\text{dust}}$) for all radii. For this calculation, we consider the grain size that dominates the emission at the corresponding ALMA wavelength, and report the dust scale height calculated for all models and all discs in Fig.\,\ref{fig:hdust1}. We also report the dust scale height (of the dust grains that emit the most efficiently at the ALMA wavelength) at the position of the rings and several radial locations, from the models that best fit the observations in Table\,\ref{tab:table4}. In this table, we also include the turbulence parameter that best fits the observations at the corresponding radial location.
\begin{table}
    \centering
    Compilation of modeling results of vertical height for millimetre dust grains at several key locations, with its corresponding $\alpha$ parameter.\\
    \begin{tabular}{cccc}
        \hline \hline
        Name & Region & $h_\text{{dust}}$ & $\alpha$ \\
        & (au) & (au)  &   \\
        (1) & (2) & (3)  & (4)\\
        \hline 
        PDS 111 & B32 & 0.03  & $1\times10^{-5}$  \\
        PDS 111 & entire disc & $\leq$ 0.09 & $1\times10^{-5}$ \\
        DoAr 25 & B86 & 2.40 & $1\times10^{-2}$ \\
        DoAr 25 & B111 & 3.68 & $1\times10^{-2}$ \\
        DoAr 25 & B137 & 4.63 & $1\times10^{-2}$ \\
        DoAr 25 & 38-62 au & $\leq$ 2.02 & $1\times10^{-5}$ \\
        MY Lup & B20 & 0.04 & $1\times10^{-5}$ \\
        MY Lup & B40 & 0.11 & $1\times10^{-5}$ \\
        MY Lup & 20-30 au & $\leq$ 0.06 & $1\times10^{-5}$ \\
        MY Lup & 60-70 au & $\leq$ 0.17 & $1\times10^{-5}$ \\
        V409 Tau & 1-8 au & $\leq$ 0.01 & $1\times10^{-5}$ \\
        V409 Tau & 8-47 au & $\leq$ 1.34 & $1\times10^{-2}$ \\
        T Cha & B36 & 0.05 & $1\times10^{-5}$ \\
        T Cha & entire disc & $\leq 0.10$ & $1\times10^{-5}$ \\
        \hline
    \end{tabular}
    \caption[]{Column 1: Target name. Column 2: radial location. Column 3: dust scale height. $h_{\text{dust}}$ is calculated for the grain size that dominates the emission at the corresponding ALMA wavelength. Grain sizes are 526.8 $\mu$m for Band 3, and 220.8 $\mu$m for Band 6. Column 4: turbulence parameter.}
\label{tab:table4}
\end{table}\\
In three cases: DoAr 25, MY Lup, and V409 Tau, the observed flux is well reproduced by the models, but there is no single model that is completely consistent with all the substructures observed along the semi-minor axis, which can be observed in Fig.\,\ref{fig:alphavsradius}. As discussed in \cite{Pinte2016}, \cite{Liu2022}, and \cite{Villenave2025}, this suggests that the degree of dust settling may vary for these sources at different radial locations. This could be the result of radial variations in the level of vertical turbulence or radial variations in the gas-to-dust mass ratio near the midplane. We note that for all the discs listed above, differences between the models and the observations lie within the error bar of the observations. We report potential radial variations in the degree of dust settling as tentative and then interpret these results carefully because the extracted brightness profile is largely affected by image noise and the substructures are dependent on the azimuthal wedge selected to extract the brightness profile. In the following, we list the results from the modeling for all the sources, one by one.

\subsection{T Cha}
ALMA continuum images of the T Cha disc in Bands 6 and 3 show an unresolved inner disc with a gap, and an outer ring \citep{Hendler2018, Francis2020}. Fig.\,\ref{fig:alphavsradius} shows that the model for the smallest $\alpha$ agrees best with the observations throughout the entire disc, and this trend is clearly seen in all azimuthal directions in Fig.\,\ref{fig:all_radial_intensity_profiles}. The results of the ALMA modeling support the idea that the T Cha disc is highly settled, and that the vertical height of mm dust grains is constrained to be smaller than 0.1 au throughout the entire disc (Fig.\,\ref{fig:hdust1}). We note that the model for $\alpha=1\times10^{-5}$ does not perfectly match the profile of the observations along the semi-minor axis and other azimuthal directions in Fig.\,\ref{fig:all_radial_intensity_profiles}, which suggests that an even more settled disc may be a better fit. Interestingly, the south side of the disc in the best model image in Fig.\,\ref{fig:all_figures_alma_gallery} displays slightly lower intensity than the ALMA image. This intensity difference could be due to a low-level asymmetry, and further research is needed to better understand this feature.

\subsection{DoAr 25}
The disc of DoAr 25 shows several narrow rings and gaps \citep{Andrews2018}. From the $\chi^{2}$ map in Fig.\,\ref{fig:alphavsradius} and the brightness profiles in Fig.\,\ref{fig:all_radial_intensity_profiles}, we observe that the largest difference between the models and the observations lies at $\sim\,$98 au, at the location of the largest gap. The model for the highest vertical turbulence $\alpha=1\times10^{-2}$ best reproduces the intensity dip in the largest gap (D98 in Fig.\,\ref{fig:alphavsradius}), implying that the disc of DoAr 25 is not so settled at that location, contrary to the case of T Cha. From the $\chi^{2}$ map in Fig.\,\ref{fig:alphavsradius}, we observe the trend that for a radius larger than 70 au, the most turbulent model appears to be the best fit throughout the outer disc. In contrast, the least turbulent model is the best fit of the inner part of the disc between $\sim$\,20 and $\sim$\,70 au. We expect that more settled models will have deeper gaps with more separate rings, and viceversa. As mentioned before, this projection effect is always best seen in the brightness profiles along the semi-minor axis and is key to discern between different models. The dust scale height for the B111 and B137 rings in DoAr 25 (Fig.\,\ref{fig:alphavsradius}) is constrained to be $\sim$\,4, and $\sim$\,4.5 au respectively (Fig.\,\ref{fig:hdust1}), which slightly differs from the upper limit of 3.0 au reported in \cite{Villenave2025} for the radial range between 86-165 au. Our results in Fig.\,\ref{fig:hdust1} are also consistent with the upper limit of 2 au for the dust scale height at 100 au from \cite{Pizzati2023}, using a different method. On the other hand, we find a dust scale height of $\sim$\,2.4 au for the radial range between 69-86 au, which is consistent with the lower limit of 0.9 au reported in \cite{Villenave2025} for this radial range, as shown in Fig.\,\ref{fig:hdust1}. There is also agreement between the best $\alpha$ found in this work for the outer disc in the radial range 69-86 au, and the lower limit of $\alpha_\text{{z,MCFOST}}$ reported for this radial range in \cite{Villenave2025}. Although both studies measure comparable dust scale heights at different radial distances, we find a difference between the best $\alpha$ in this work of $1\times10^{-2}$ and the upper limit of $\alpha_\text{{z,MCFOST}}$ reported in \cite{Villenave2025} for the outer disc in the radial range 86-165 au, of $2\times10^{-3}$. We argue that this disagreement is due to differences in the assumed optical properties of dust, as discussed in the next section. 
\subsection{MY Lup}
The MY Lup disc displays alternating shallow rings and gaps \citep{Andrews2018}, making it difficult to discriminate between models by observing the gap contrast along the semi-minor axis. However, we can still identify possible radial variations in the degree of dust settling, which are shown in Fig.\,\ref{fig:alphavsradius}. Overall, we observe the trend that up to $\sim\,$70 au, the least turbulent model fits the observations best, and beyond this radius there is a gradual increase in vertical turbulence. We found that beyond the radius of 40 au, the largest dust scale height in the outer disc, in the most turbulent model, is $\sim$ 2.2 au (Fig.\,\ref{fig:hdust1}), which is similar to the upper limit of 2 au reported by \cite{Villenave2025} in the region between 40 and 100 au. Our results in Fig.\,\ref{fig:hdust1} are also in agreement with the upper limit of 4 au for the dust scale height at 100 au reported by \cite{Pizzati2023}. In addition, our measurement of the turbulence parameter for MY Lup ($\alpha\lesssim1\times10^{-3}$) for the region between 40-80 au, is consistent with the $\alpha_\text{{z,MCFOST}}$ reported in \cite{Villenave2025} for the radial range 40-100 au. However, we  find an $\alpha$ value one order of magnitude larger ($\alpha\lesssim1\times10^{-2}$) for the outermost part of the disc (r $>$ 80 au) than in \cite{Villenave2025} ($\alpha\lesssim2\times10^{-3}$).\\We note that we use MY\,Lup and T\,Cha to test potential degeneracies of our results with the geometric parameters assumed ($\gamma$, and $\text{H}_{100}$), which are described in Appendix\,\ref{sec:geometrical-parameters}. In short, while for T\,Cha the assumed geometric parameters do not change the results, we do find differences for MY Lup; however, some of the main trends obtained in the dust scale height and settling across the MY Lup disc remain. We discuss these tests in detail in Appendix\,\ref{sec:geometrical-parameters}.

\subsection{V409 Tau}
V409 Tau shows a compact disc in the continuum, with shallow rings and gaps \citep{Long2019}. The limited angular resolution of the V409 Tau dataset (0.13"\,$\times$\,0.10") meant that it was not possible to fully resolve the substructures of the disc. However, we are able to reproduce the flux at all radii of the disc and identify possible radial variations in the degree of dust settling, which are shown in Fig.\,\ref{fig:alphavsradius}. Overall, we found that the most turbulent model best reproduces the continuum observations in the outer disc ($r$ $\gtrsim$ 8 au), where the dust scale height of the millimetre dust grains is $\lesssim$ 1.4 au, and the inner disc ($r$ $\lesssim$ 8 au) could be more settled (Fig.\,\ref{fig:alphavsradius}), where the dust scale height of the millimetre dust grains is $\lesssim$ 0.05 au.

\subsection{PDS 111}
The disc of PDS 111 shows a small disc in the continuum, and SPHERE and ALMA data provide evidence of a low contrast asymmetry in the south-west of the disc \citep{Derkink2024}. Similarly to the case of V409 Tau, with limited angular resolution (0.36"\,$\times$\,0.25"), it is not possible to fully resolve any potential fine substructures.  However, we were able to reproduce the radial brightness profile of the disc in all azimuthal directions, and our models favour a highly settled disc \textbf{($\alpha\lesssim1\times10^{-5}$)}. We also find that the vertical height of millimeter sized dust grains is smaller than 0.09 au throughout the disc (Fig.\,\ref{fig:hdust1}); as seen in Fig.\,\ref{fig:alphavsradius} and Fig.\,\ref{fig:all_radial_intensity_profiles}, the most settled model shows the best agreement with the observations for the complete extent of the disc along the direction of the semi-minor axis. Similarly to the case of V409 Tau, the results presented for this source need confirmation using higher angular resolution observations.

\subsection{RY Lup}
The ALMA continuum image of RY Lup in Band 6 displays rings with a large inner dust cavity \citep{Francis2020}. The disc in the continuum also displays a clear difference in azimuthal brightness, where the disc is brighter on the north side of the minor axis. We decided to perform the modeling and analysis for this source even though the characteristics of the disc seen in the continuum could be interpreted as an asymmetry. As seen in Fig.\,\ref{fig:all_radial_intensity_profiles}, we were able to reproduce the ALMA brightness profile along the semi-major axis of the disc, but we were not able to reproduce the flux along the semi-minor axis and other azimuthal directions, so we could not measure the dust scale height and strength of settling for this source. The reason why all model images look very similar is unclear. To check if this could be the result of high optical depth in the models, we computed the vertical optical depth (Fig.\,\ref{fig:opt-depth}), and all models are optically thin, which excludes this possibility.

\begin{figure*}
\includegraphics[width=2.0\columnwidth]{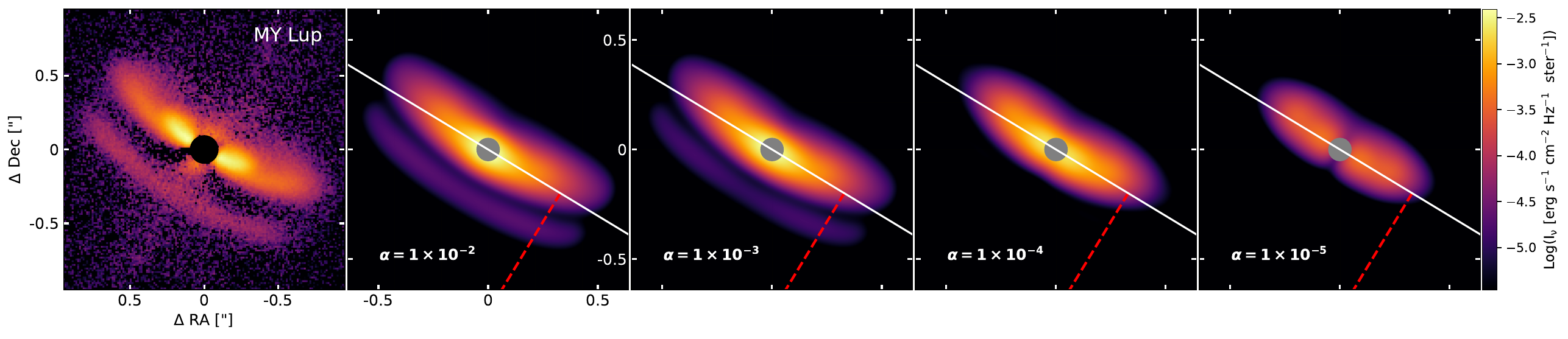}
\includegraphics[width=2.0\columnwidth]{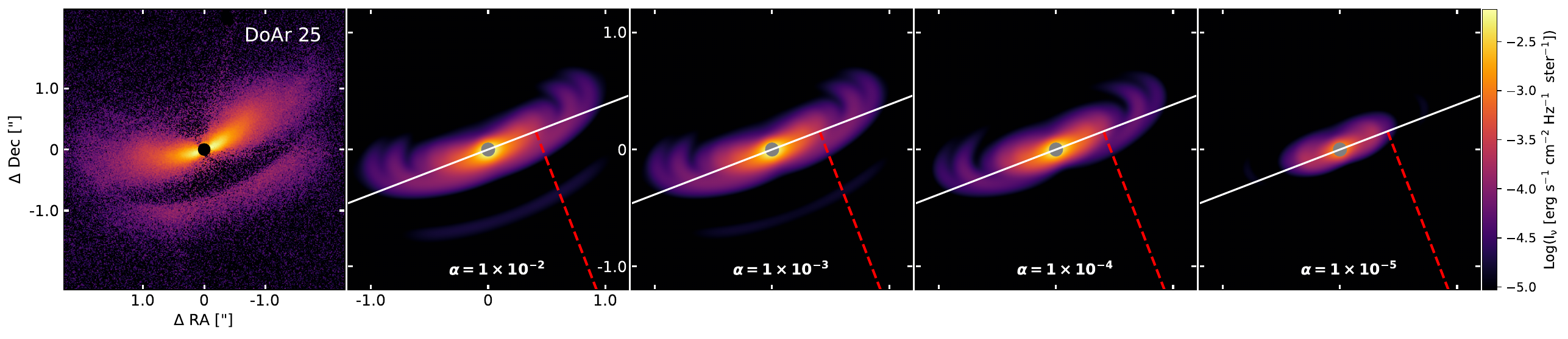}
\includegraphics[width=2.0\columnwidth]{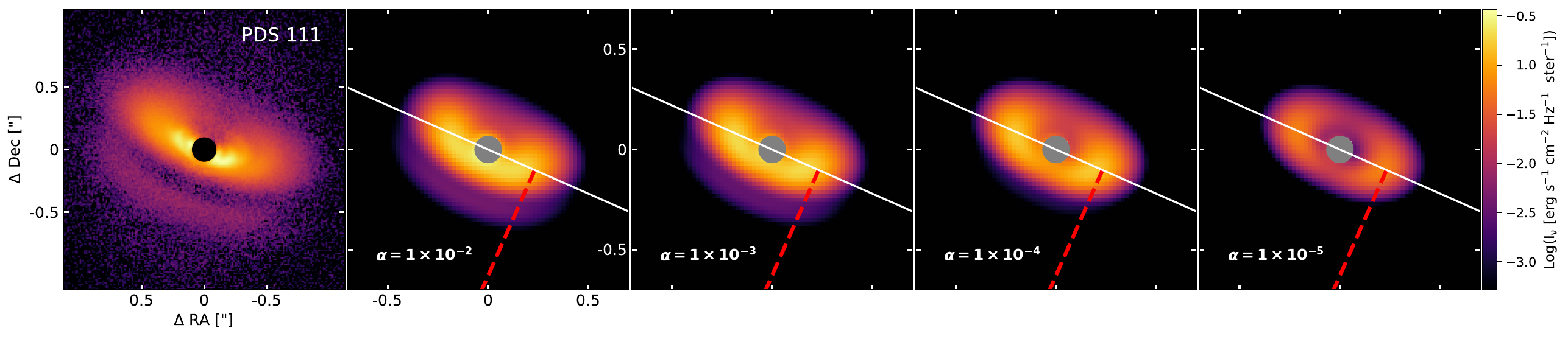}
\caption{Q$_\phi$ synthetic images at 1.6 $\mu \text{m}$ for MY Lup, DoAr 25, and PDS 111, from top to bottom, respectively. First column shows Q$_\phi$ SPHERE image in H band for each source. Each row displays a model calculated for a different degree of dust settling. All images display the same contrast of the corresponding SPHERE observations, and a synthetic coronagraph at the star's position of 92.5 mas. The white line represents the direction of the position angle of the continuum. The red line represents the direction where vertical cuts are extracted, to compare the morphology of the dark lane between models and SPHERE observations.}
\label{fig:sphere-dark-lane-gallery1}
\end{figure*}

\begin{figure*}
\includegraphics[width=0.59\columnwidth]{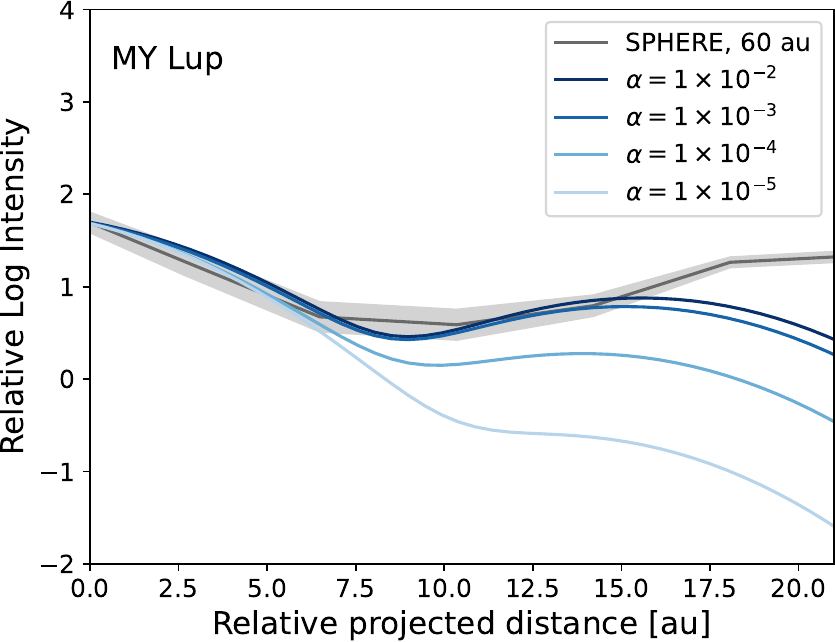}
\includegraphics[width=0.6\columnwidth]{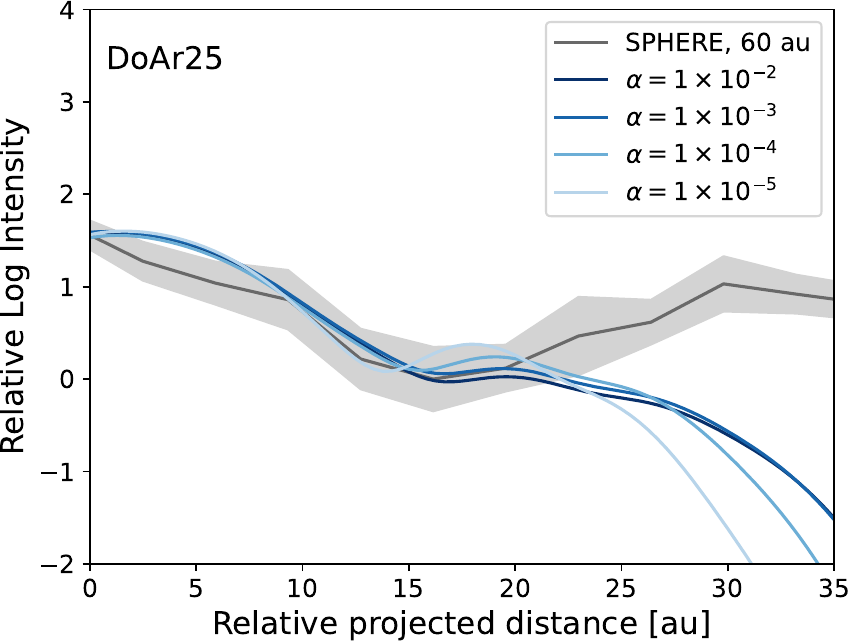}
\includegraphics[width=0.625\columnwidth]{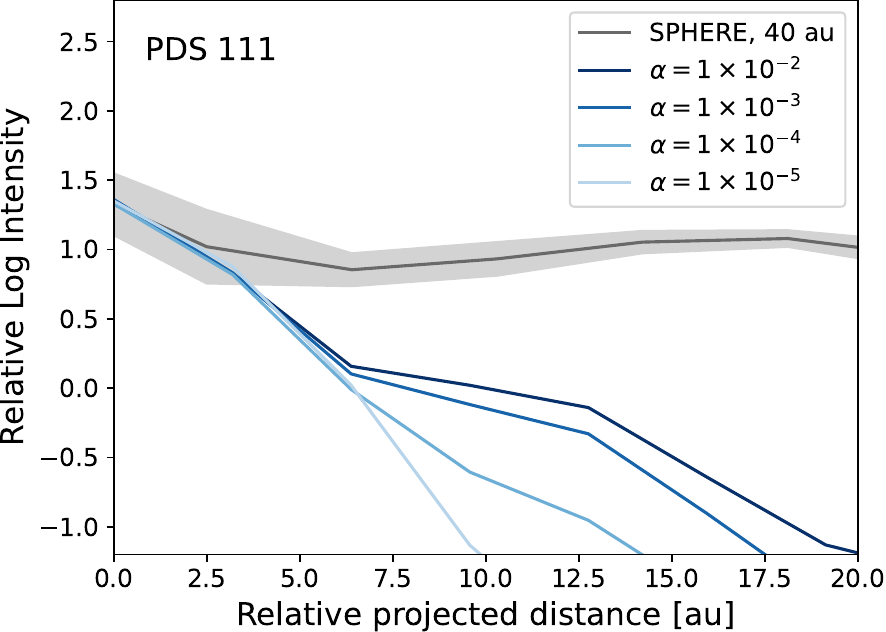}
\caption{Comparison of vertical cuts extracted along the minor-axis direction from projected SPHERE images, and SPHERE synthetic images (Fig. 6). From left to right, this comparison is presented for MY Lup, DoAr 25, and PDS 111, respectively. Vertical cuts of the synthetic images are extracted along the direction of the red line in Fig. 6. Shaded areas represent the uncertainty of the data and correspond to the standard deviation in the radial direction divided by the square root of the number of pixels. All profiles are displayed in an arbitrary intensity scale, in logarithmic scale, and compared with respect to the peak intensity of the data that is located at the top left. Profiles of models and observations are extracted at the same distance from the star, which is indicated at the top right of each panel.}
\label{fig:sphere-dark-lane-profiles1}
\end{figure*}

\begin{figure*}
\includegraphics[width=2.0\columnwidth]{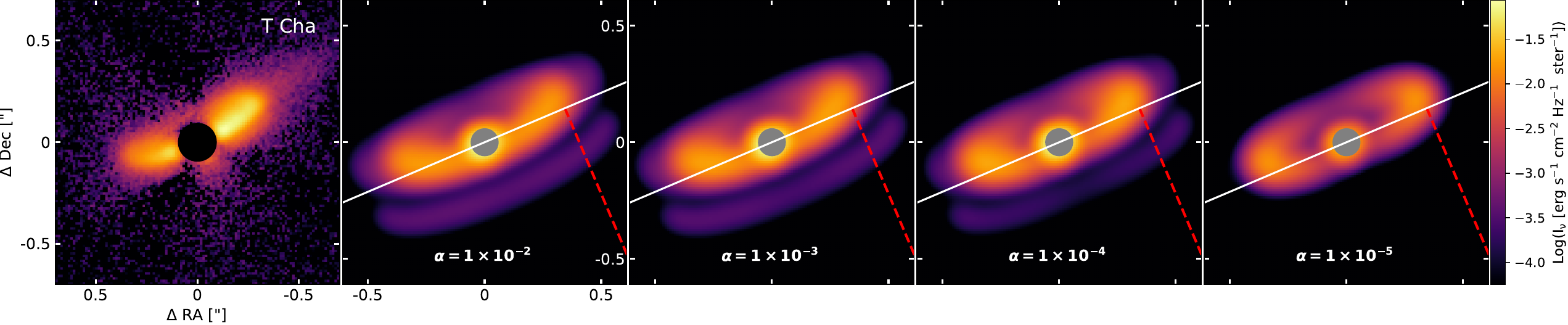}
\includegraphics[width=2.0\columnwidth]{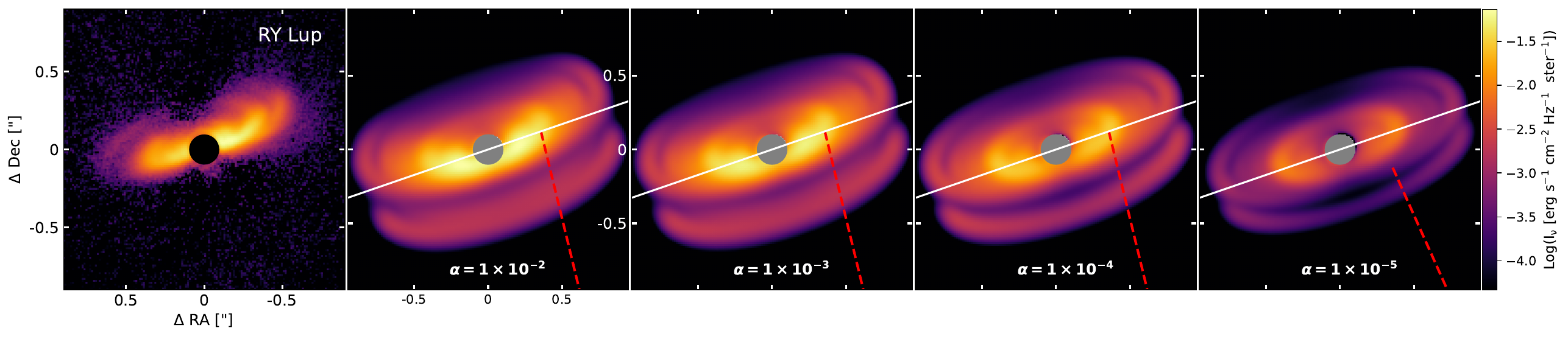}
\includegraphics[width=2.0\columnwidth]{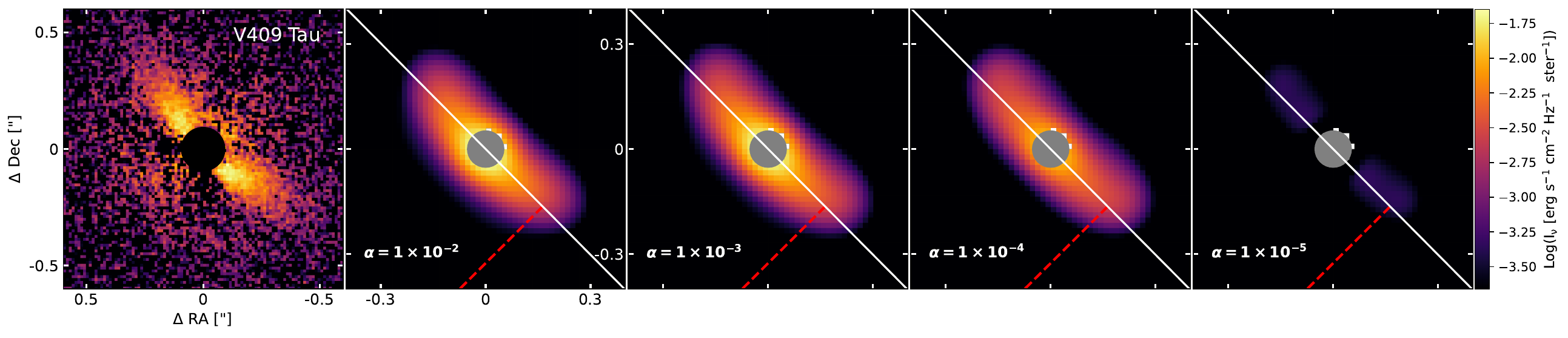}
\caption{From top to bottom, Q$_\phi$ synthetic images at 1.6 $\mu m$ for T Cha, RY Lup, and V409 Tau, respectively. Annotations follow from Fig. 6. }
\label{fig:sphere-dark-lane-gallery2}
\end{figure*}

\begin{figure*}
\includegraphics[width=0.617\columnwidth]{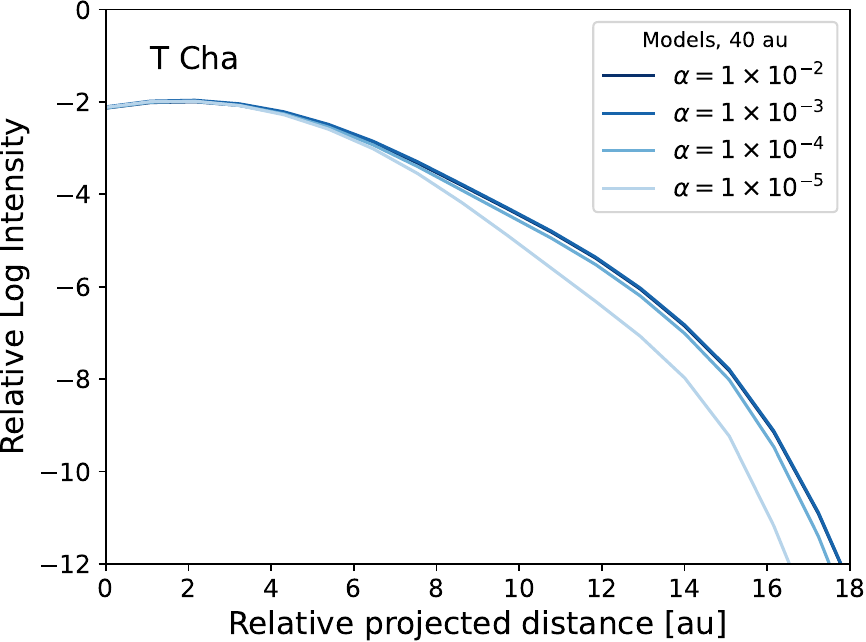}
\includegraphics[width=0.619\columnwidth]{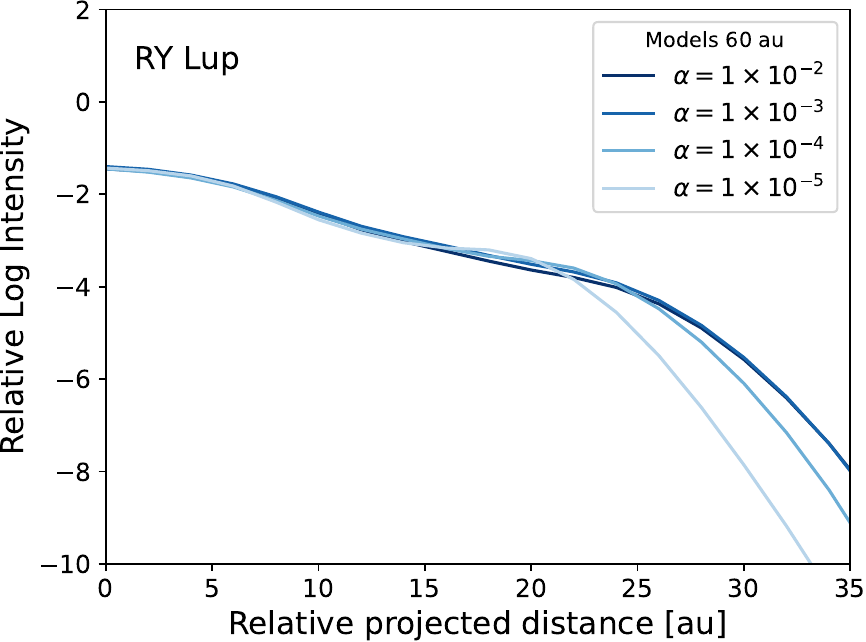}
\includegraphics[width=0.6\columnwidth]{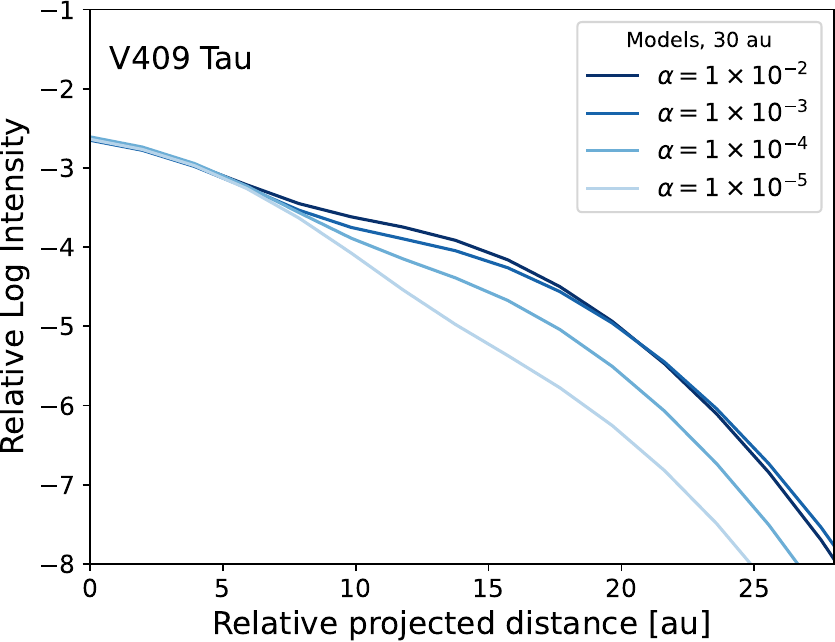}
\caption{Comparison of vertical cuts extracted along the minor-axis direction from projected SPHERE synthetic images (Fig. 8). From left to right, this comparison is presented for T Cha, RY Lup, and V409 Tau, respectively. Vertical cuts of the synthetic images are extracted along the direction of the red line in Fig. 8. All profiles are displayed in an arbitrary intensity scale, in logarithmic scale. Profiles of models are extracted at the distance from the star indicated at the top right of each panel.}
\label{fig:sphere-dark-lane-profiles2}
\end{figure*}

\begin{figure*}
\includegraphics[width=1\columnwidth]{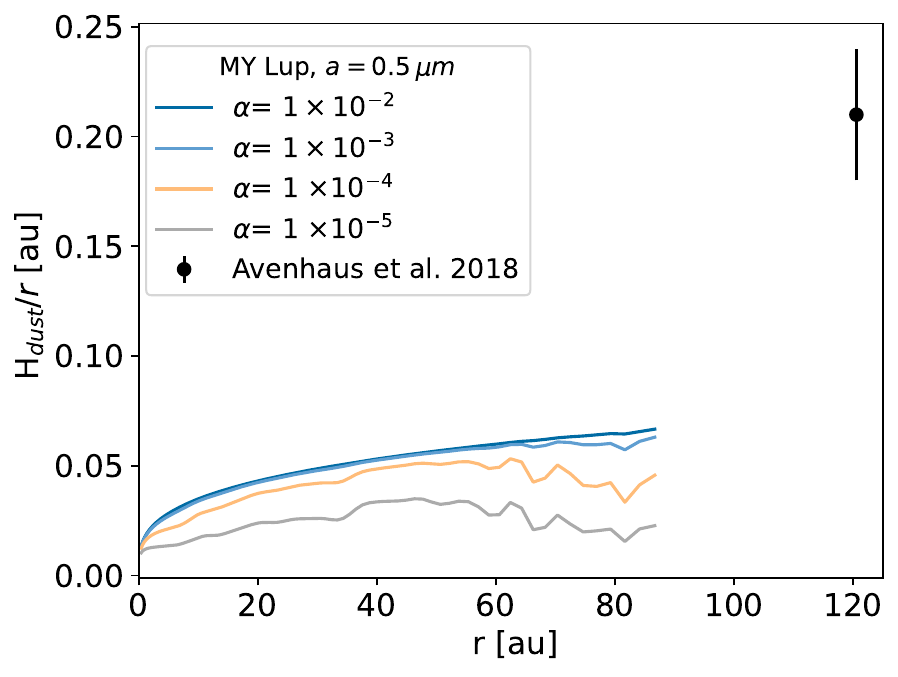}
\includegraphics[width=1\columnwidth]{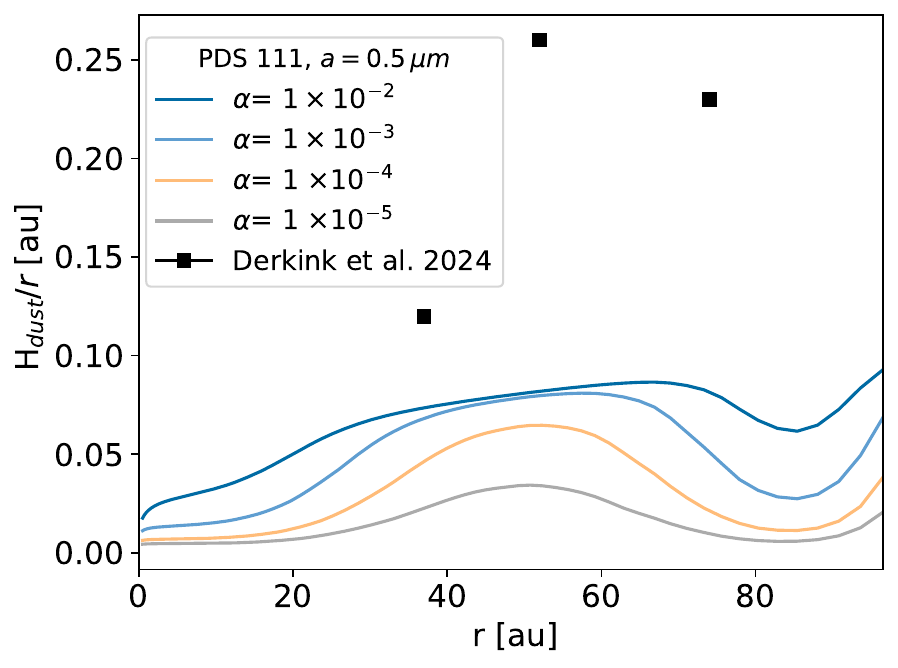}
\caption[]{Comparison of the aspect ratio of the grain size that dominates the scattering at 1.6 $\mu \text{m}$, and the aspect ratio of the surface measured from SPHERE data $Q_{\phi}$ in \textit{H} band, for MY Lup, and PDS 111, from left to right, respectively. We consider the aspect ratio of the last scattering surface measured in \cite{Avenhaus2018}, and \cite{Derkink2024}.}
\label{fig:sphere-alma}
\end{figure*}

\section{Results from the comparison of the models with SPHERE data}
\label{sec:resultssphere}
All of the sources selected for this study have been observed by the DESTINYS large program in the H band of SPHERE/VLT. In this section, we investigate whether the vertical distribution of the dust grains of the models that best fit the ALMA data can also reproduce these observations. For this purpose, we calculate synthetic observations in the SPHERE H-band (1.6 $\mu \text{m}$) for all sources. We compare the morphology of the dark lane observed in the midplane in models and observations for the three sources where the front and back disc surfaces are detected: PDS 111, MY Lup, and DoAr 25 (Sec.\,\ref{subsec:darklane}). We also compare the vertical height of very small dust grains in the models with the height of the surface from SPHERE observations in two cases: MY Lup, and PDS 111 (Sec.\,\ref{subsec:backside}). In addition, in all cases we investigate the possible detection of the backside of the discs. We used RADMC-3D in full scattering configuration to calculate dust temperature and NIR images for the best models found in the previous section, using the same input grain properties as in the ALMA data modeling (described in Sec.\,\ref{sec:radiativetransfer}). To compare the models with the SPHERE observations, we calculate $Q_{\phi}$ by combining the Q and U components of the Stokes vector, according to \cite{Avenhaus_2017}. Finally, all synthetic images are convolved with a 2D Gaussian with a FWHM of 0.04", which mimics the resolution from SPHERE. These synthetic images are later presented with a mask at the star's position of radius 92.5 mas, which represents the N\_ALC\_YJH coronagraph from SPHERE. We note that the radial grid of the synthetic images is limited to the radial extent of the continuum ($R_{out}$ in Table \ref{tab:table3}); therefore, they display a smaller radius than the SPHERE images. Due to this limitation, we can only compare the distribution of small dust grains in the models and observations in the vertical direction, within the radial range of the ALMA observations. 

\subsection{Dark lane detection at 1.6 {$\mu \text{m}$}}
\label{subsec:darklane} 

Scattered-light observations of highly inclined discs usually display two surfaces separated by a dark lane, which traces the absorption of starlight in very dense regions in the midplane \citep[e.g.][]{watson2007,Duchene2024}. This intensity drop in the midplane could provide information on the properties of dust grains, the vertical density structure of discs, and, in particular, could inform us about the degree of settling \citep[e.g.][]{Duchene2024,Villenave2024b,George2025}.
We compare the characteristics of the dark lane feature between the models and SPHERE observations by qualitatively comparing the SPHERE observations with model images at the same contrast as observations in all cases. We also investigate possible variations in its thickness for different degrees of dust settling. In particular, we compare cuts along the minor axis of the models and SPHERE images at a given distance from the star. 

\subsubsection{PDS 111, DoAr 25, and MY Lup}
We compare cuts along the minor axis of the models and SPHERE $Q_{\phi}$ images, at a distance of 40 au (PDS 111) and 60 au (MY Lup, and DoAr 25) from the star; we show this comparison in the Figures\,\ref{fig:sphere-dark-lane-gallery1}, and \ref{fig:sphere-dark-lane-profiles1}. We show a gallery of the $Q_{\phi}$ models in Fig.\,\ref{fig:sphere-dark-lane-gallery1}, where all the figures show the same contrast of the corresponding SPHERE observations. Cuts along the minor axis are extracted along the direction of the red dashed line of the panels. The position of the cut is selected so that it is not close to the star's position or close to the disc edge. We note that both the models and the SPHERE images are not deprojected. The comparison of the vertical cuts in Fig.\,\ref{fig:sphere-dark-lane-profiles1} for these three sources shows that none of the models can reproduce the characteristics of the dark lane from the SPHERE observations. For MY lup, DoAr 25 and PDS 111, we observe that for higher turbulence the backside is more conspicuous and the width of the dark lane becomes distinguishable. This trend was also observed in the recent work of \cite{George2025}, who investigates the visibility of the backside for different dust disc configurations. Another finding is that the models with higher turbulence progressively approach the data (Fig.\,\ref{fig:sphere-dark-lane-profiles1}); an important implication of this finding is the possibility that the dust distribution models that reproduce the continuum emission from the midplane cannot describe the vertical distribution of the very small grains in the upper layers of the disc. This result also raises the possibility that the level of vertical turbulence in the upper layers of the disc could be larger than that in the midplane.\\
These results should be interpreted with caution because the thickness of the dark lane could change for a disc that is more extended radially. However, we note that the detectability of the dark lane does not change for a larger disc based on the results by \cite{George2025}.\\
In Fig.\,\ref{fig:sphere-dark-lane-gallery1}, we note that the backside of the MY Lup, DoAr 25, and PDS 111 discs is only visible for the two models with the highest turbulence, which suggests that the turbulence in the upper layers of these sources could be high. These results differ from our previous findings from the ALMA data modeling, where we found a very settled disc for PDS 111 and MY Lup (tentatively and with radial variations), and also support the idea that the level of turbulence in the disc atmosphere could be higher than in the midplane; the detection of the backside depends on several factors, such as inclination, dust distribution in the outermost part of the disc, flaring of the disc, among others \citep{George2025}. In the case of DoAr 25, these findings, together with the results of the ALMA data modeling, where we found that high turbulence is the best fit of the ALMA data, might indicate that high turbulence is affecting the midplane and even higher turbulence is affecting the surface; more research is needed to examine this case and confirm this hypothesis. We will discuss these results in more detail in the next section.
\subsubsection{The cases of T Cha, RY Lup, and V409 Tau}
For these three sources, the backside is not visible in SPHERE observations, so we cannot compare the characteristics of the dark lane feature between models and observations. However,  we can evaluate whether the backside is visible or not in the models. Fig.\,\ref{fig:sphere-dark-lane-gallery2} shows the models in the same contrast as in the SPHERE observations, and Fig.\,\ref{fig:sphere-dark-lane-profiles2} shows a comparison of vertical cuts between the models. Fig.\,\ref{fig:sphere-dark-lane-gallery2} shows that in the case of T Cha, the backside is not visible only for the model with the highest settling and the lowest turbulence $\alpha=1\times10^{-5}$; this is in agreement with the SPHERE observations, as the backside is not visible, and is also in agreement with the results found in the ALMA data modeling, because the most settled disc best fits the ALMA observations. It may therefore be the case that the turbulence could be low across the vertical extent of the T Cha disc.\\
In the case of RY Lup, the backside is visible for all models in Fig.\,\ref{fig:sphere-dark-lane-gallery2}, which is in disagreement with the SPHERE observations, as the backside is not visible. This indicates that none of the models is capable of reproducing most of the vertical structure of the disc, and also indicates that small dust grains could be less extended vertically than in the models. Finally, in the case of V409 Tau, the backside is not visible for any of the models, which is consistent with the SPHERE observations (Fig.\,\ref{fig:sphere-dark-lane-gallery2}). We propose that the backside is not visible even for the less settled disc models of V409 Tau, because the dust density in the uppermost layers is too sparse to efficiently scatter the starlight in all models; this effect is extensively discussed in recent work by \cite{George2025}. 
\subsection{Vertical height of the very small dust grains}
\label{subsec:backside}
We also investigate whether the best models found in Sec.\,\ref{sec:resultsalma}, can reproduce the vertical height of the small dust grains traced by the SPHERE observations. For this purpose, we calculate the height of the dust grains that dominate the scattering at 1.6$\,\mu$m, which have a size of 0.5$\,\mu$m, using Eq.\,\ref{eq:Hdust}, and then compare the corresponding aspect ratio of the models with the aspect ratio of the last scattering surface from the SPHERE data. We make this comparison for the MY Lup and PDS 111 discs, because the height of the surface was reported by previous studies. We consider the aspect ratio of the last scattering surface measured in \cite{Avenhaus2018} and \cite{Derkink2024}, where concentric ellipses are fitted to the disc substructures in the IR, and refer to these works for further details. Fig.\,\ref{fig:sphere-alma} shows the results of this comparison. We found that in both cases, MY Lup, and PDS 111, the aspect ratio of all the models is located under the aspect ratio of the surface from SPHERE. This result suggests that none of the models is able to reproduce the vertical height of the very small dust grains in these two sources. In the previous section, we found that low vertical turbulence would dominate at the midplane location for PDS 111 and tentatively for MY Lup; therefore, these results might indicate a rise in level of turbulence from the midplane to the surface, where high turbulence could be stirring up the dust in the upper layers of these discs. However, we note that these findings are limited by the fact that the surface scale height of the discs is not expected to be identical to the pressure scale height and is expected to be located above the pressure scale height \citep[e.g.][]{Dullemond2010}. We also note that the results for MY Lup should be interpreted with special caution because the radial extent of the MY Lup models is smaller than the radius at which the aspect ratio was measured from the SPHERE data. We will discuss these results in the next section in more detail.

\section{Discussion}
\label{sec:discussion}
\subsection{Challenges to infer the level of vertical turbulence from the gap contrast method}
Using the method described in this paper, we can obtain the vertical height of millimeter dust grains $\text{h}_{\text{dust}}$ traced by ALMA, from the geometry of the disc; and find a combination of optical depth, density and dust opacity that fits the ALMA observations, but other combinations are also possible, as discussed by  \cite{Villenave2025}. Regarding the constraints on $\alpha$, several parameters in the settling prescription we adopt (Eq.\,\ref{eq:Hdust}) are uncertain, making it difficult to infer the level of vertical turbulence $\alpha$ directly from $\text{h}_{\text{dust}}$. These parameters are mainly the scale height of the gas H$_{\text{gas}}$, which we parameterise (Eq.\,\ref{eq:Hgas}), and the Stokes parameter St (Eq.\,\ref{eq:stokes}), which depends on the grain size and the gas surface density $\Sigma_\text{{gas}}$. These parameters are also difficult to measure from observations, and the effect of all these parameters in the derivation of vertical turbulence has been previously discussed in other works \citep[e.g.][]{Liu2022, Pizzati2023, Villenave2025}. In the present work, we found that the assumed dust opacities also influence the measurement of vertical turbulence $\alpha$, and we discuss this effect in more detail in the next subsection. Additionally, to test the potential influence of the geometric parameters $\gamma$ and H$_{100}$ in the $\alpha$ constraints, we performed extra simulations for MY Lup and T Cha, for several combinations of geometric parameters. Our results show that when the rings and gaps are well resolved from observations, such as in the T Cha disc, the vertical turbulence constraints  across the disc remain robust and independent of the assumed H$_{\text{gas}}$ or $\gamma$, whereas when the ring and gaps are not well resolved, there are some trends that remain independent of the assumed H$_{\text{gas}}$ or $\gamma$, but differences can exist in the inferred values of $\alpha$. We show and discuss these tests in Appendix\,\ref{sec:geometrical-parameters}.  

\subsection{Degree of dust settling and level of vertical turbulence}
\label{subsec:settling}
We modeled the dust distribution of six highly inclined discs and could estimate the degree of dust settling in five sources in our sample: T Cha, DoAr 25, MY Lup, V409 Tau, and PDS 111. Overall, we find that the degree of dust settling that reproduces the discs' observations is diverse, which reflects the diversity in the structure of protoplanetary discs. For T Cha, V409 Tau and PDS 111, there are no previous constraints in the literature using the same method; but, DoAr 25, MY Lup were included in the recent study by \cite{Villenave2025} using a similar method. Both works use different prescriptions to describe the dust scale height; \cite{Villenave2025} implements the prescription from \cite{Fromang2009}, and we use the prescription from \cite{Youdin2007}. After making several tests, we confirmed that both lead to negligible differences in dust scale height. The main difference between \cite{Villenave2025} and our study is that both works assume a different dust grain composition and therefore different dust opacities. Our choice of dust grain composition from \cite{Ricci2010} is motivated by recent results from \cite{Stadler2022}, which show that the prescription \cite{Ricci2010} leads to a higher flux in radiative transfer models than the composition of DSHARP \citep{Birnstiel2018}. In addition, the results from \cite{Delussu2024} show that in population synthesis models the Ricci opacity model for compact grains best reproduces the spectral index distribution and the flux distribution at 1.0 mm from observations for substructured discs. We note that the iterative method described in the present paper does not converge to a good fit when using the DSHARP prescription, contrary to the good fits when using the Ricci opacities. This is because the models computed for the Ricci prescription systematically display lower mass and optical depth.\\
The dust grains in \cite{Villenave2025}'s study are composed of 62.5\% astronomical silicates and 37.5\% graphite; and their opacities peak at a smaller grain size at the corresponding wavelength of the ALMA observations; we show these differences in Fig.\,\ref{fig:opacities}. Interestingly, for these two systems: DoAr 25, and MY Lup, both works measure comparable dust scale heights for the corresponding grain size that dominates the emission at 1.25 mm, which are 77.3\,$\mu \text{m}$ in \cite{Villenave2025}, and 220.8\,$\mu \text{m}$ in this work (as discussed in Sec.\,\ref{sec:resultsalma}). However, there are differences between the turbulence parameters found for DoAr 25 in both works. The best $\alpha$ measured for DoAr 25 from the gap contrast seen at the outer disc ($1\times10^{-2}$) differs from the $\alpha_\text{{z,MCFOST}}$ upper limit of \cite{Villenave2025} in the region between 86 and 165 au, which is at least one order of magnitude lower ($\leq2\times10^{-3}$). We attribute these differences to the different opacities and the dependence of the dust scale height on the grain size, since larger dust grains tend to have a smaller scale height. Thus, because the grain size that dominates the emission at 1.25 mm is larger in our work (Fig.\,\ref{fig:opacities}), larger turbulence is necessary in the models to lift the dust grains to a vertical height that reproduces the geometry of the disc observed along the semi-minor axis. To better illustrate this idea, we have computed additional ALMA continuum models for MY Lup for a different opacity prescription (shown in Fig.\,\ref{fig:opacities}, and hereafter referred to as "astrosilicates prescription"), and included the results in Appendix \ref{sec:severalopacities}. Fig.\,\ref{fig:hdust1} for MY lup, clearly shows that different opacity prescriptions  retrieve different vertical heights for the grain size in which the emission is most efficient and for the same $\alpha$. Interestingly, and aside from this study, other works have shown that dust opacities can also affect the inferred maximum grain size distribution derived from the multiwavelength observations analysis, and affect the estimates of dust mass from radiative transfer simulations \citep[e.g.][]{Guidi2022, Li2023, Liu2024, Sierra2025}.\\
The choice of the input dust opacities would therefore influence the measured vertical turbulence in radiative transfer simulations, and these effects on the modeling from the chosen dust opacities have not been shown in previous studies to date. This result highlights the importance of the dust opacity in determining the vertical turbulence from the contrast of rings and gaps in highly inclined discs, and it motivates further studies to determine the dust grain composition in protoplanetary discs with more accuracy.\\ 
Regarding the measurement of vertical turbulence by the method discussed in this paper, the prescription from \cite{Youdin2007} does not consider particle feedback. The prescription determined by \cite{Lim2024} considers the feedback of the dust to the gas and is expected to provide vertical turbulence values higher than the prescriptions used in previous studies, which do not consider the feedback of particles.\\
Using a different approach to this study, \cite{Doi2021} and \cite{Jiang2025} constrained the dust scale height in the HD 163296 disc and several protoplanetary discs, respectively, by modeling azimuthal intensity variations observed in ALMA continuum images. They found radial variations in the dust scale height and argued that these variations can be explained by radial variations in vertical turbulence or dust grain sizes. Regarding the effect of grain sizes in our study, we do not expect that a variation in the maximum grain size would affect the results and conclusions. We fixed $a_{\text{max}}$ to 0.3 cm, and from Fig.\,\ref{fig:opacities} we observe that dust grains larger than 0.1 cm do not contribute significantly to the dust emission observed at 1.25 mm.\\
Because ALMA continuum observations probe the thermal emission from large dust grains at the midplane, the constraints we find for the degree of dust settling from the ALMA data modeling are representative of the spatial scales at the midplane in the disc. Regarding the physical origin of the turbulence in this region, Magneto-rotational instability (MRI) is predicted to be active in the innermost part of the disc (r$\,<\,$1\,au) and the outer disc, where both MRI active regions are radially separated by a dead zone \citep[e.g.][]{Dzyurkevich2010, Flock2015, Delage2022}. MRI studies predict orders of magnitude for $\alpha$ from $10^{-4}$ to $10^{-2}$ \citep{Delage2022}. Other instabilities can be present within the dead zone, such as Vertical Shear Instability (VSI); and Rossby Wave Instability (RWI) can possibly be triggered at the edges of the dead zone \citep{Lesur2023}. 

\subsection{Comparison of the models with SPHERE scattered light images} 
The level of turbulence of the gas is expected to fluctuate in different regions of the disc, where the strength may vary in the radial and vertical directions \citep{Lesur2023}. Vertically, MRI active regions in the atmosphere of the disc are expected to display high turbulence \citep{Flock2011, Sorathia2012, Simon2012}, while dead zones in the midplane would display low turbulence \citep{Bai2013}. On the other hand, MHD simulations of VSI have shown that the level of gas turbulence can increase vertically in the disc, from the midplane \citep[e.g.][]{Flock2020}; recent stellar-irradiated hydrodynamic models of VSI display large turbulence ($\alpha$\,$\propto$\,$10^{-2}$) in the atmosphere of the disc \citep{Zhang2024}; and the highly turbulent gas in the upper layers can lift small dust grains to high altitudes, preventing them from settling in the midplane \citep[e.g.][]{Fukuhara2025}. In addition, planet disc interactions can drive gas vertical flows, also affecting the vertical distribution of dust grains \citep[e.g.][]{Bae_2016, Szulagyi_2022, Petrovic2024}. Our results from the comparison between the models that fit the ALMA data and SPHERE observations suggest that a higher level of turbulence than in the midplane is required in the disc upper layers of MY Lup, DoAr 25 and PDS 111, which is consistent with the predictions from MRI, and VSI models previously mentioned. Although two of our models cannot reproduce the vertical height of the small dust grains in the observations (MY Lup, and PDS 111) and the detection of the dark lane feature in three cases, the most turbulent models tend to reproduce better the SPHERE observations for the three discs. Additional research is needed to develop a more complex analytic method that can reproduce the vertical height of large and small dust grains simultaneously, potentially implementing a variable vertical turbulence; and also to model the different radial extent observed by ALMA and SPHERE.

\subsection{Vertical height of millimetre dust grains and general trends}
As pointed out earlier, several efforts have been made to constrain the dust scale height of millimetre dust grains and the strength of dust settling from ALMA observations. Previous studies that have used the same method \citep[e.g.][]{Pinte2016, Liu2022, Villenave2025}, have reported that the dust scale height of millimetre dust is typically a few au at 100 au distance from the star, and our results are consistent with those results (Table\,\ref{tab:table4}, and Fig.\,\ref{fig:hdust1}). Additionally, in this study, we observe that the vertical height of millimetre dust can be even smaller than 0.1 au for the most settled discs in our sample, which is also in agreement with previous constraints for other highly settled discs such as HL Tau, and Oph 163131 \citep{Pinte2016, Villenave2022}. It is also interesting to note that, despite the difficulty in measuring a completely meaningful vertical turbulence parameter from the present method, according to current observational evidence, it can span several orders of magnitude, since each study can be considered an independent ruler to measure the settling strength, and it can also vary with radius.

\section{Conclusions}
\label{sec:conclusions}
The purpose of the present study was to measure the vertical height of millimetre dust grains and the strength of dust settling for six highly inclined protoplanetary discs. With this objective, we performed a detailed modeling of the ALMA continuum substructures using radiative transfer simulations, where all radiative transfer
models presented are computed for the Ricci dust opacity prescription. The modeling results of this investigation show that the strength of the settling and vertical turbulence is diverse in all discs in our sample. In addition, we calculated synthetic SPHERE images (1.6 $\mu \text{m}$) for all the best models of the dust density distribution, and compared the models with the SPHERE data. In particular, we compared the vertical height of the very small dust grains, the morphology of the dark lane at the midplane, and evaluated the detectability of the backside. Our main results can be summarised as follows.
\begin{enumerate}
    \item From the ALMA data modeling, we could provide measurements of the dust scale height for five sources, for the grain size in which the emission dominates at the corresponding ALMA wavelength; and found that the vertical height is smaller than 5 au in all cases, which is similar to values reported in the literature; in three cases these are the first measurements reported.
    \item From the ALMA data modeling, we could also measure the degree of dust settling and infer the dust distribution on five of the six discs in our sample: DoAr 25, MY Lup, V409 Tau, T Cha, and PDS 111. For the RY Lup disc we could not measure the strength of the settling due to its asymmetric nature. For PDS 111 we found a very settled disc ($1\times10^{-5}$). For DoAr 25, MY Lup and V409 Tau, we found possible radial variations in the degree of dust settling, where the models favour large vertical turbulence ($1\times10^{-2}$) for the outer disc of DoAr 25 and V409 Tau. Finally, for T Cha our models clearly favour low vertical turbulence ($1\times10^{-5}$) and a very settled disc.
    \item Our analysis shows that the input dust opacities can have a large impact on the measurements of the level of vertical turbulence derived from comparing observations with radiative transfer simulations. If dust opacities peak at a larger grain size, the vertical turbulence derived from this method is larger.
    \item From the comparison of the SPHERE synthetic images and the SPHERE data for MY Lup, DoAr 25, and PDS 111, we observe that none of the models can reproduce the dark lane shape from the SPHERE data and the vertical height of the dust grains traced by SPHERE (for MY Lup and PDS 111). The shape of the dark lane may differ for a disc that is more radially extended; however, we note that the detectability of the dark lane is independent of the disc size \citep{George2025}. In addition, we observe that models with the highest vertical turbulence tend to get closer to the observations in most cases. This is because for models with high turbulence, the dark lane feature is more conspicuous and the backside is brighter. 
    \item The results of the ALMA data modeling for MY Lup, DoAr 25, and PDS 111, and the comparison of the best models of the vertical dust density distribution with SPHERE data for these sources, suggest that the level of vertical turbulence in the upper layers of these discs may be higher than in the midplane, which is in agreement with predictions from MRI and VSI models.
    \item The qualitative comparison of synthetic images of scattered light and SPHERE data, together with the results of the ALMA data modeling, suggests that the turbulence could be high ($\geq1\times10^{-2}$) across the vertical extent for DoAr 25, and low ($1\times10^{-5}$) across the vertical extent for T Cha. 

\end{enumerate}
Further modeling work will have to be conducted to determine whether a rise in vertical turbulence from the midplane would be able to reproduce the distribution of large dust grains in the midplane and small dust grains in the upper layers in some sources of this sample of protoplanetary discs.

\section*{Acknowledgements}
We thank the referee for their constructive comments and suggestions, which
improved the manuscript.\\
J.A. acknowledges the support from the Science and Technology Facilities Council (STFC), part of UK Research and Innovation (UKRI), through a PhD studentship.\\ 
P.P. and A.S. acknowledge funding from the UK Research and Innovation (UKRI) under the UK government’s Horizon Europe funding guarantee from ERC (under grant agreement No 101076489).\\
Y.L. acknowledges the financial support by the Natural Science Foundation of Sichuan Province of China (grant No 2025ZNSFSC0060) and the Fundamental Research Funds for the Central Universities (grant No 2682025CX028).\\
We are thankful to Nienke van der Marel for sharing calibrated ALMA data of T Cha and RY Lup.\\
This paper makes use of the following ALMA data: ADS/JAO.ALMA\#2015.1.00979.S, ADS/JAO.ALMA\#2016.1.0048\\4.L, ADS/JAO.ALMA\#2016.1.01164.S, ADS/JAO.ALMA\#2017.1.\\00449.S, and ADS/JAO.ALMA\#2021.1.01705.S. ALMA is a partnership of ESO (representing its member states), NSF (USA) and NINS (Japan), together with NRC (Canada), NSTC and ASIAA (Taiwan), and KASI (Republic of Korea), in cooperation with the Republic of Chile. The Joint ALMA Observatory is operated by ESO, AUI/NRAO and NAOJ.\\
This research has made use of the VizieR catalogue access tool, CDS, Strasbourg, France (DOI : 10.26093/cds/vizier). The original description 
of the VizieR service was published in in 2000, A\&AS 143, 23.
\section*{Data Availability}
The data from observations and models underlying this article will
be shared on request to the corresponding author. The uncalibrated
ALMA data is publicly available at \url{https://almascience.nrao.edu/aq/} using the project codes 2015.1.00979.S, 2016.1.01164.S, 2016.1.00484.L, 2017.1.00449.S, and 2021.1.01705.S. 



\bibliographystyle{mnras}
\bibliography{bibliography} 




\appendix
\section{Complementary figures}
\label{sec:complementary-figures}
\begin{figure}
\includegraphics[width=1\columnwidth]{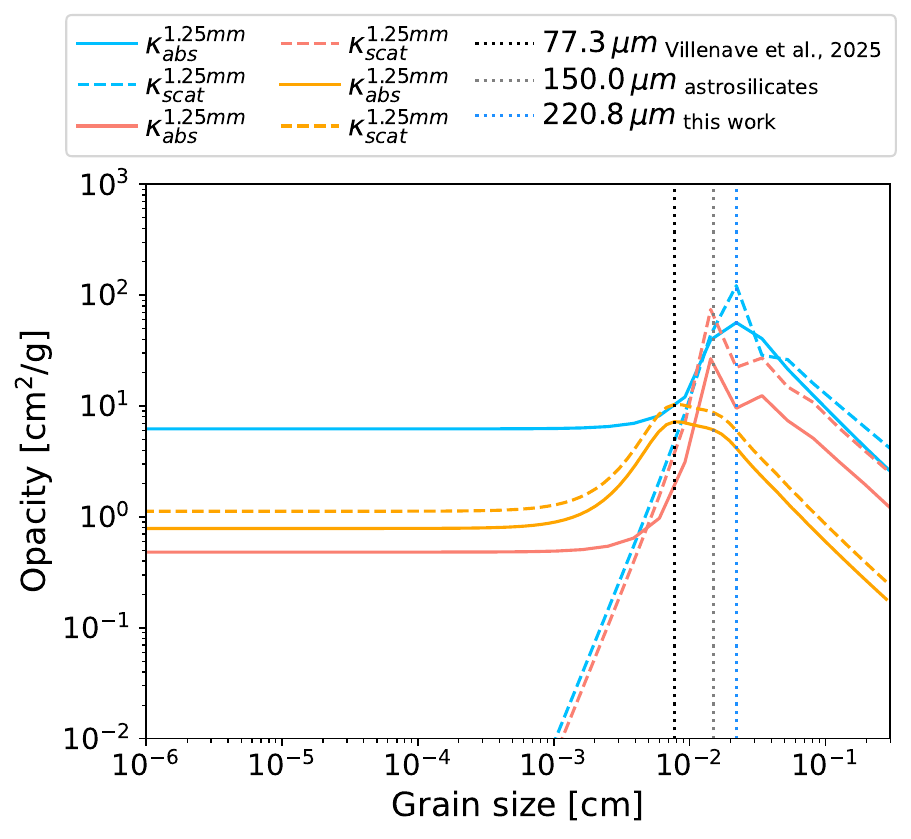}
\caption{Scattering and absorption coefficients at $\lambda=1.25$ mm as a function of grain size \textit{a}. The light blue lines show the Ricci model dust grain properties used in this work, the salmon lines show the astrosilicates prescription, and the dust model used in Villenave et al. 2025's paper is indicated with the orange lines.}
\label{fig:opacities}
\end{figure}

\begin{figure*}
\includegraphics[width=2\columnwidth]{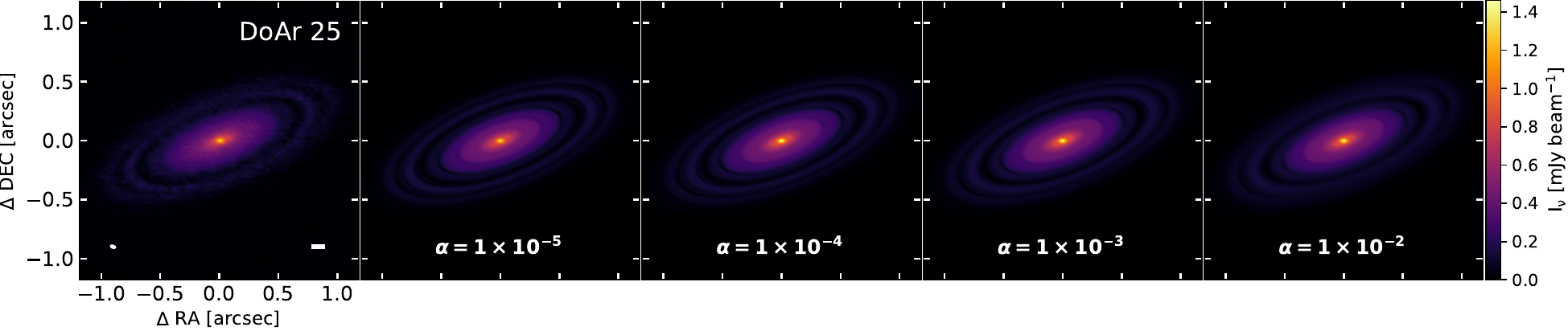}
\includegraphics[width=2\columnwidth]{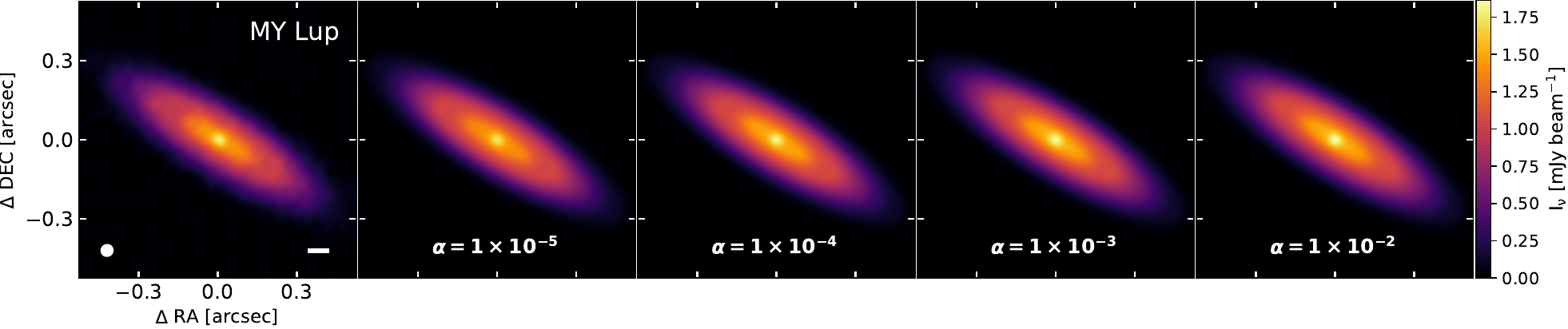}
\includegraphics[width=2\columnwidth]{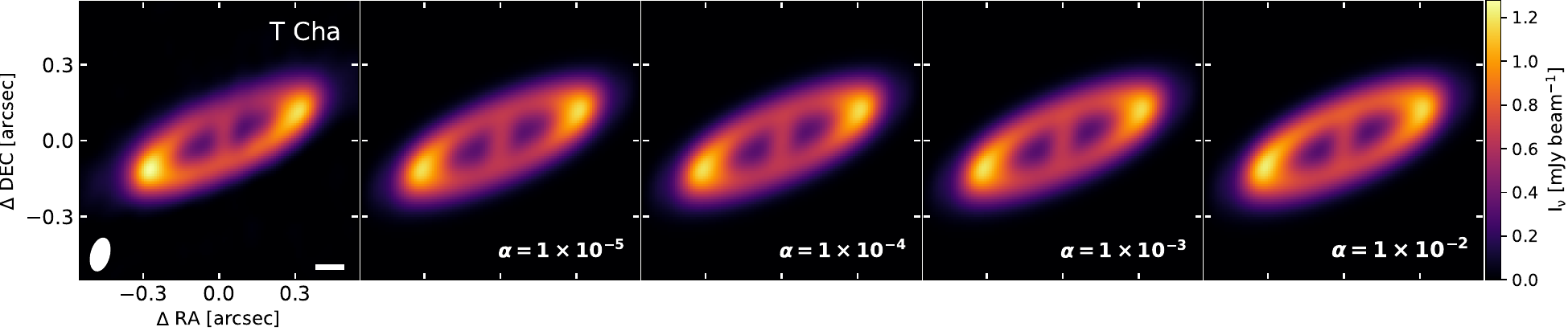}
\includegraphics[width=2\columnwidth]{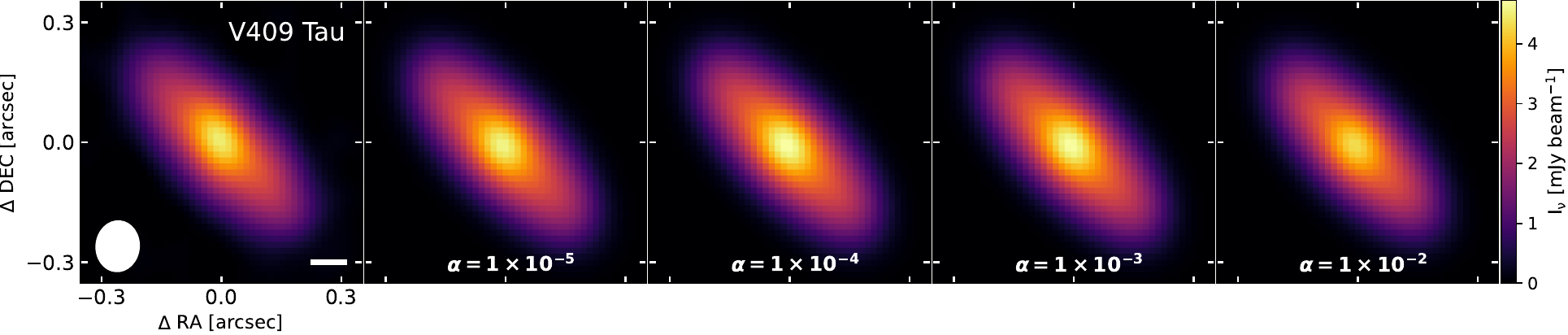}
\includegraphics[width=2\columnwidth]{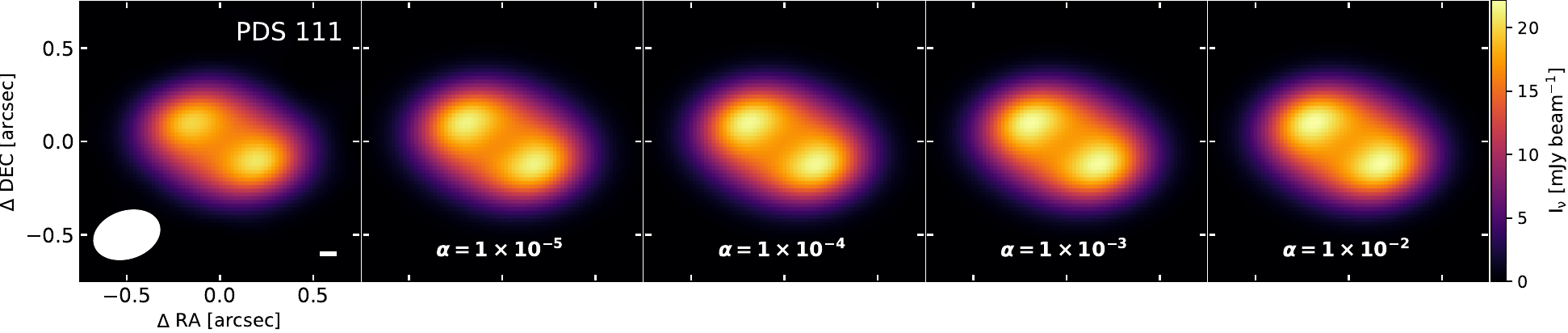}
\includegraphics[width=2\columnwidth]{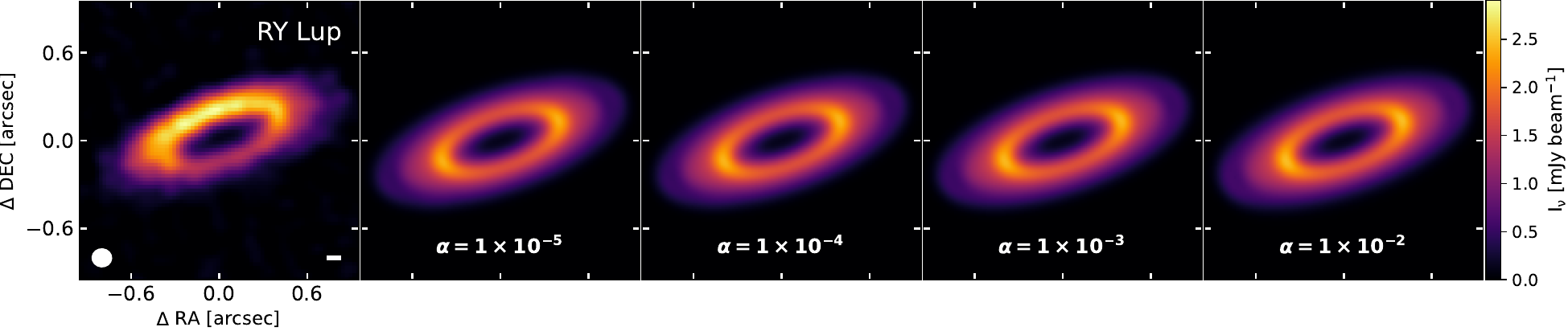}
\caption{Gallery of best models found for DoAr 25, MY Lup, T Cha, V409 Tau, PDS 111, and RY Lup respectively, from top to bottom. All models are calculated for a different degree of dust settling or $\alpha$. The first column of each row corresponds to the ALMA image for each source. Beam sizes and 10 au scalebars are shown in the lower left and right corners of each ALMA image, respectively. }
\label{fig:all_figures_alma_gallery}
\end{figure*}

\begin{figure*}
\includegraphics[width=0.95\columnwidth]{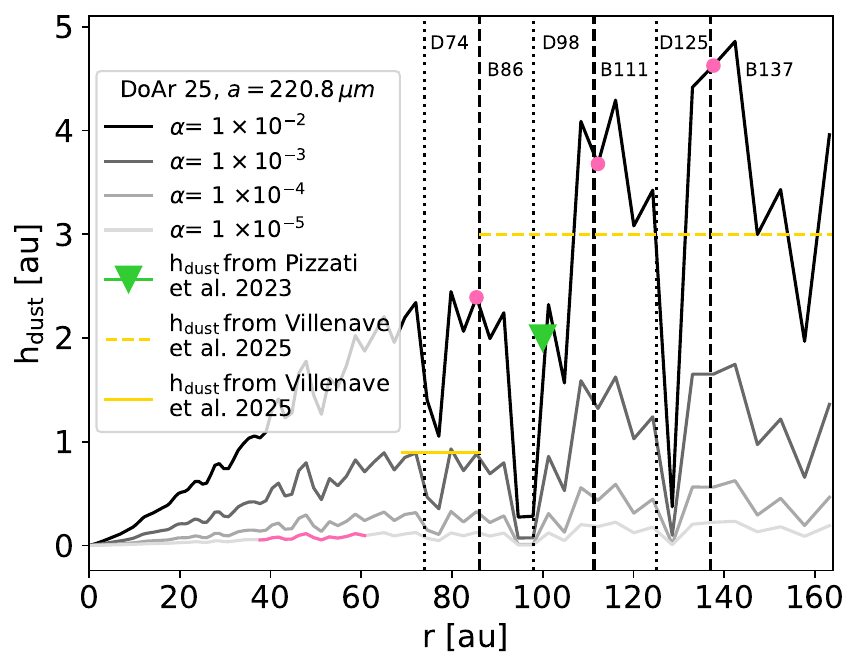}
\includegraphics[width=0.97\columnwidth]{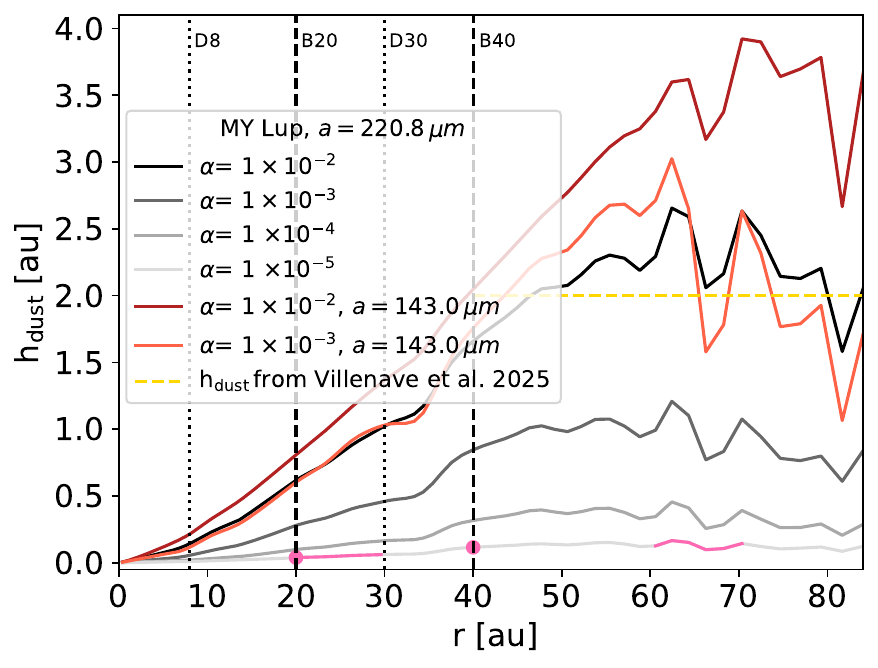}
\includegraphics[width=0.96\columnwidth]{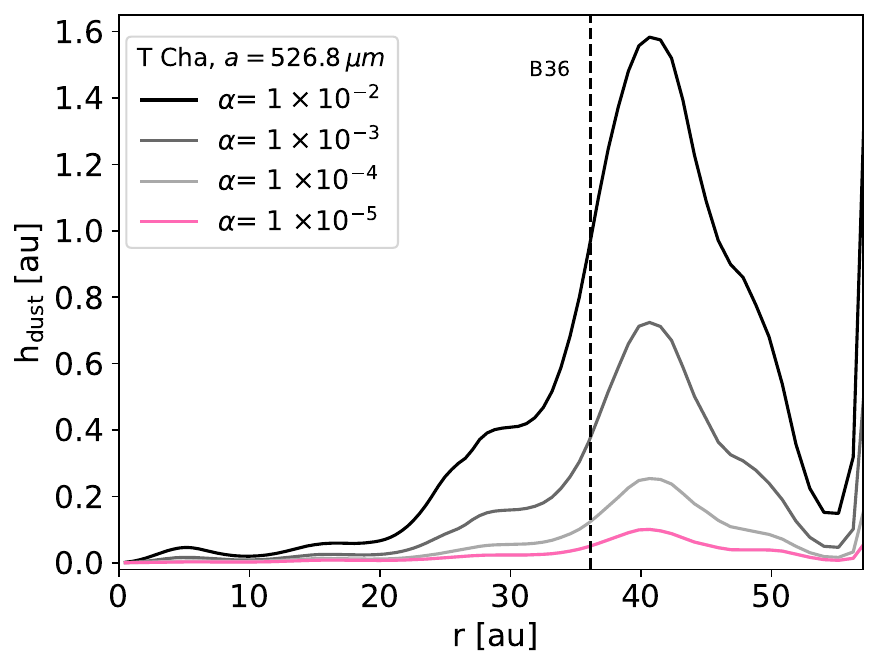}
\includegraphics[width=0.98\columnwidth]{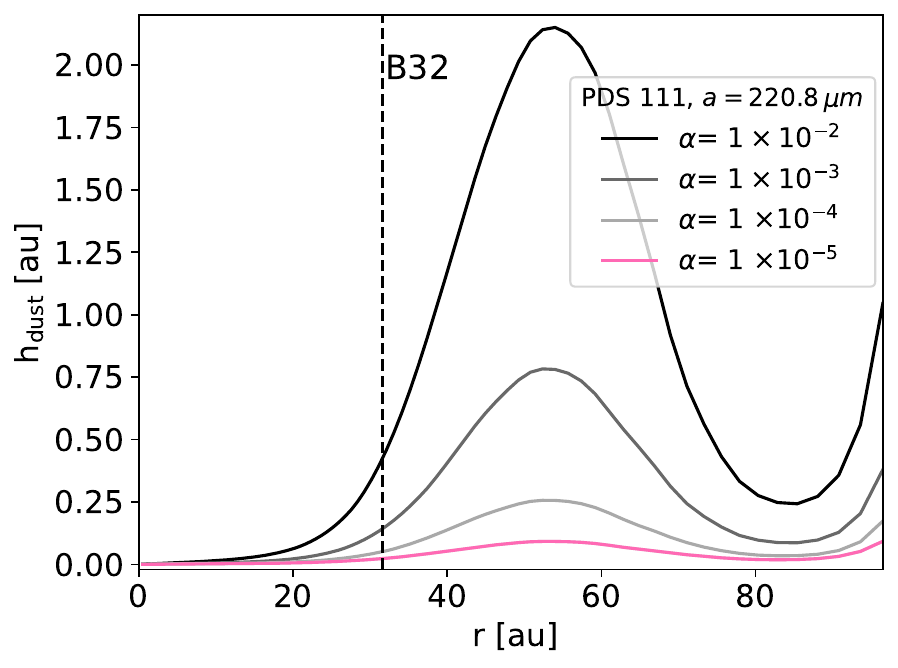}
\includegraphics[width=0.98\columnwidth]{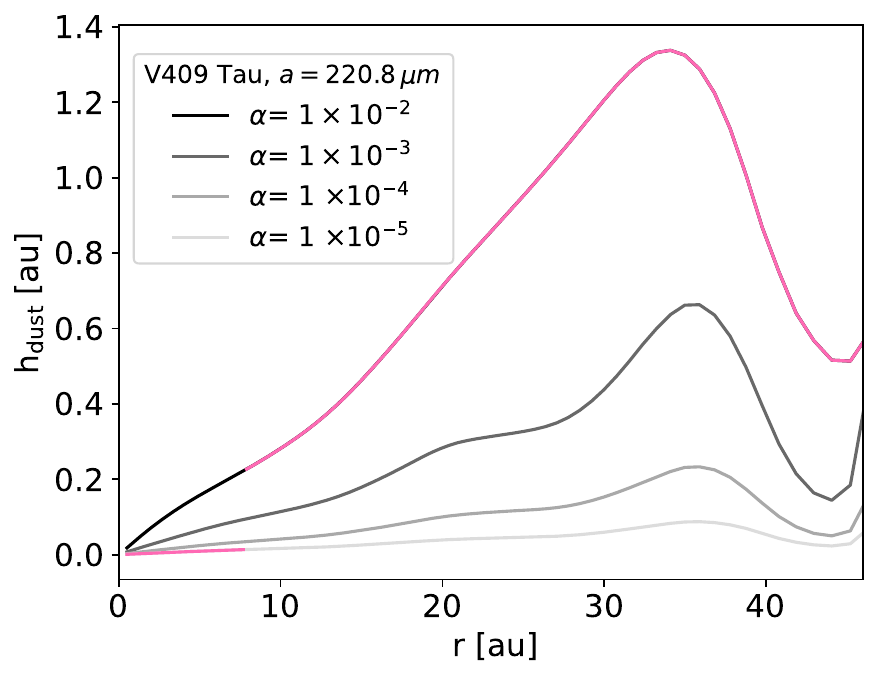}
\includegraphics[width=0.98\columnwidth]{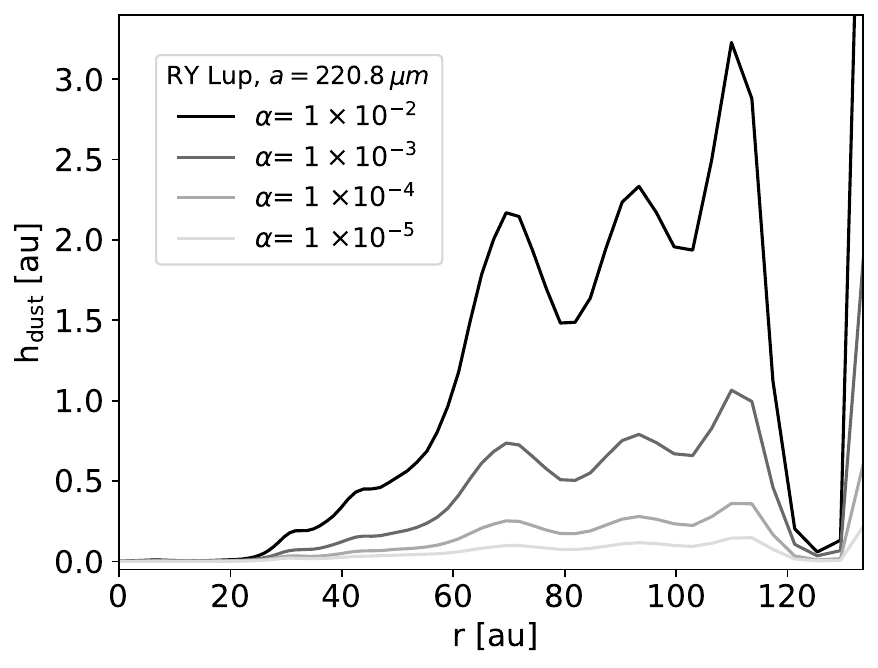}
\caption[]{Scale height of the dust ($\text{h}_{\text{dust}}$) for the grain size that dominates the emission ($a=220.8\mu \text{m}$ at 1.25 mm, and $a=526.8\mu \text{m}$ at 3.00 mm), calculated for models with a different fixed turbulence parameter $\alpha$. The name of the source is displayed at the top left part of each panel. Dashed black lines mark the location of rings, and dotted black lines mark the gaps position. Pink solid lines and dots represent $\text{h}_{\text{dust}}$ of the models with the best fit. Yellow dashed lines in the figures of DoAr 25 and MY Lup represent upper limits from \cite{Villenave2025}, and the yellow solid line represents a lower limit from the same article. The green upside down triangle in the figure from DoAr 25 is the upper limit at 100 au from \cite{Pizzati2023}. The red and orange lines in the MY Lup panel, represent the dust scale height from the models computed for the astrosilicates prescription discussed in Sec.\,\ref{subsec:settling} and Appendix \ref{sec:severalopacities}.}
\label{fig:hdust1}
\end{figure*}

\begin{figure*}
\includegraphics[width=0.73\columnwidth]{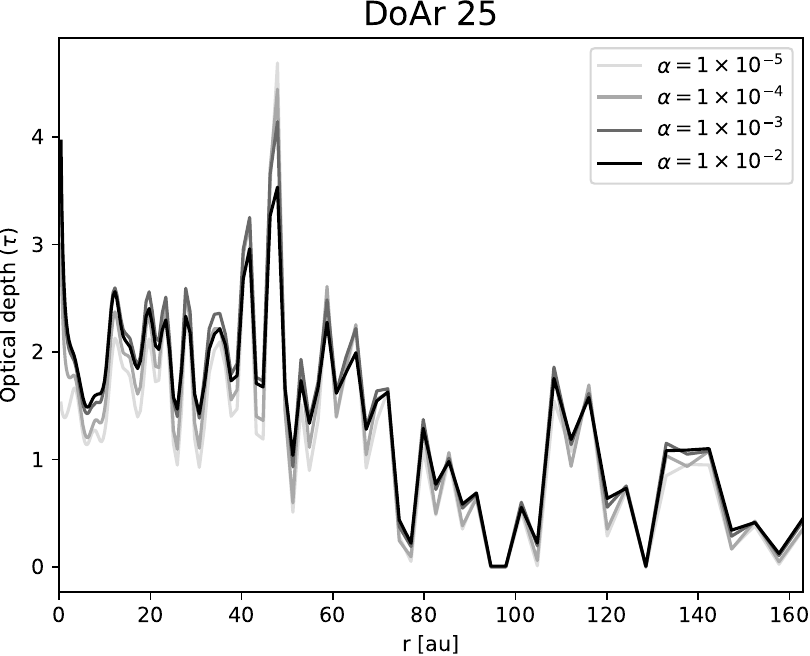}
\includegraphics[width=0.725\columnwidth]{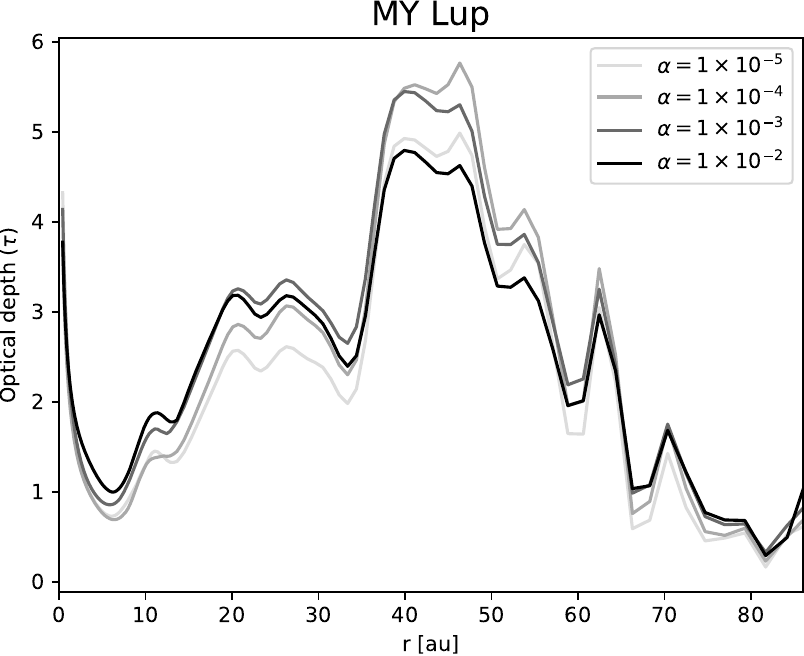}
\includegraphics[width=0.73\columnwidth]{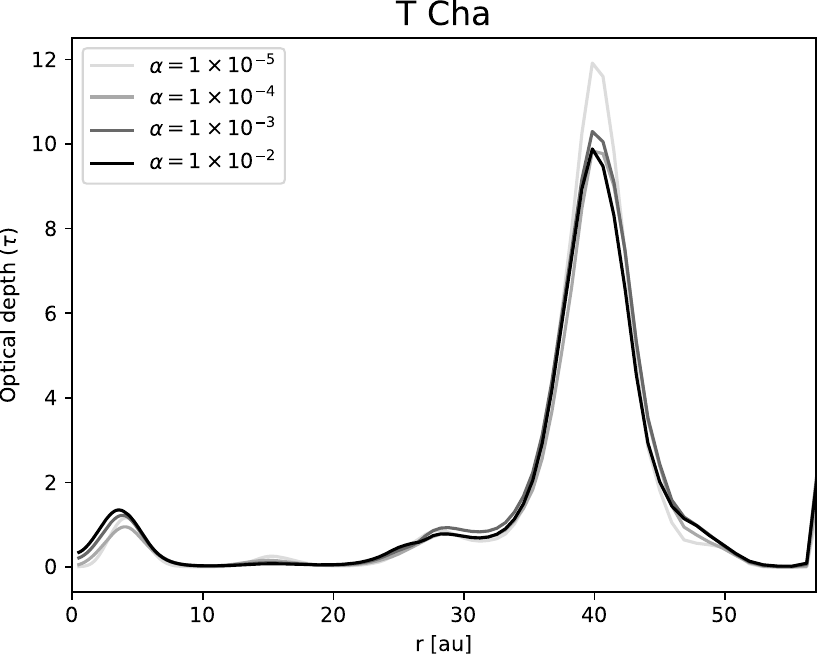}
\includegraphics[width=0.735\columnwidth]{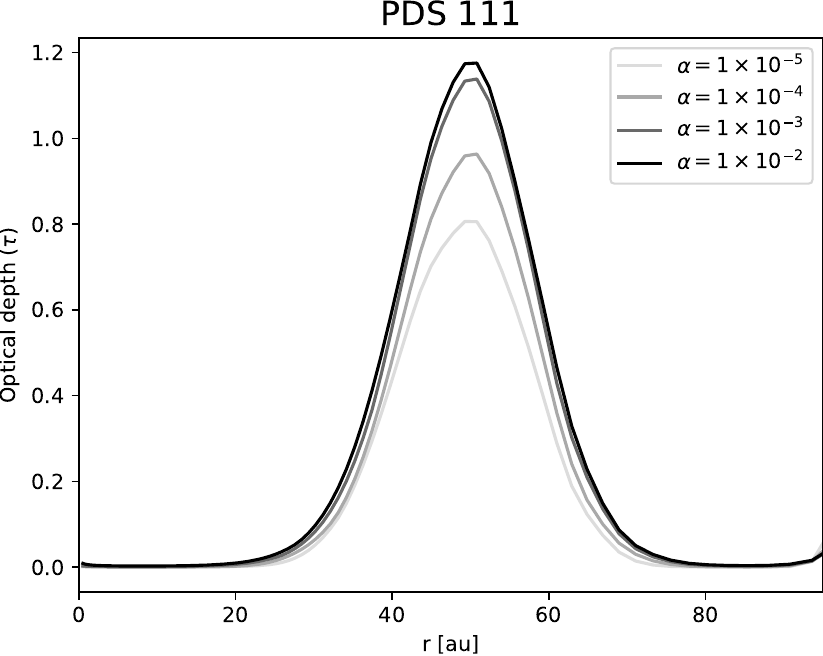}
\includegraphics[width=0.73\columnwidth]{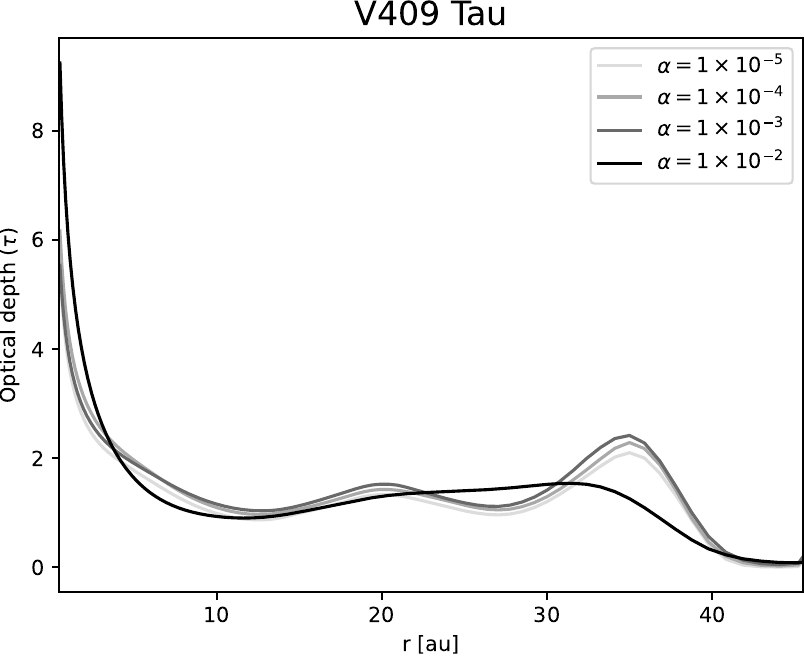}
\includegraphics[width=0.744\columnwidth]{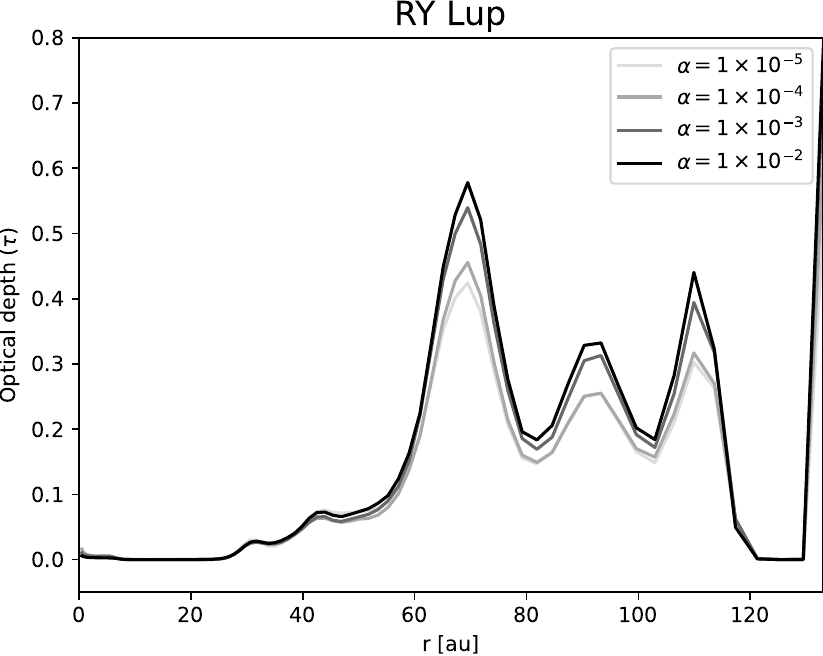}
\caption{Vertical optical depth ($\tau$) as a function of radius calculated for the best models and all the sources included in the analysis.}
\label{fig:opt-depth}
\end{figure*}

Fig.\,\ref{fig:opacities} shows the optical properties of the dust grains that we consider for the analysis. It also shows the optical properties considered in the recent study from \cite{Villenave2025}.
\\
Fig.\,\ref{fig:all_figures_alma_gallery} shows the best model images calculated for each source and different degree of dust settling.
\\
In Fig.\,\ref{fig:hdust1} we report the dust scale height (H$_{\text{dust}}$) for the grain size that dominates the emission at 1.25 mm, or 3.0 mm (ALMA Band 6 or Band 3), for all sources of our sample. We calculate H$_{\text{dust}}$ for all models with a different degree of dust settling using Eq.\,\ref{eq:Hdust}.
\\
We have computed the vertical optical depth of the best models using the formula $\tau (r) = 2\int_{0}^{\infty}[\int\kappa_{v}(a)\rho(a,r,z)da]\,\text{cos ($i$)}^{-1}dz$ and display the results in Fig.\,\ref{fig:opt-depth}.
\\
In Table\,\ref{tab:table5} we summarise the estimates of $\alpha$ for DoAr 25 and MY Lup from the literature, using the same method.
\begin{table}
    \centering
    \begin{tabular}{cccc}
        \hline \hline
        Name & Region & $\alpha_{\text{z,MCFOST}}$ & $\alpha$ \\
        & (au) &  &   \\
        (1) & (2) & (3)  & (4)\\
        \hline 
        DoAr 25 & 68-86 & $\geq2\times10^{-4}$   & $1\times10^{-2}$  \\
        DoAr 25 & 86-165 & $\leq2\times10^{-3}$ & $1\times10^{-2}$ \\
        MY Lup & 40-100 & $\leq2\times10^{-3}$ & $\leq1\times10^{-2}$ \\
        \hline
    \end{tabular}
    \caption[]{Column 1: Target name. Column 2: radial location. Column 3: $\alpha_{\text{z,MCFOST}}$ derived in Villenave et al. 2025. Column 4: $\alpha$ derived in this work.}
\label{tab:table5}
\end{table}\\

\section{Testing a different opacity prescription}
\label{sec:severalopacities}
\begin{figure*}
\includegraphics[width=2.0\columnwidth]{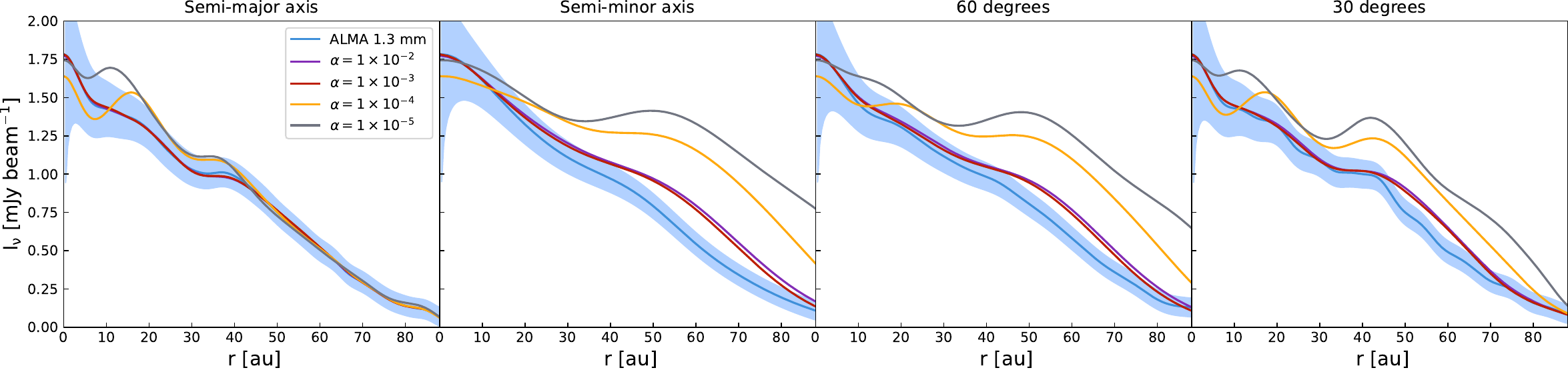}
\caption{Radial intensity profiles for the best models found for MY Lup, computed for the astrosilicates prescription.}
\label{fig:profilesastrosil}
\end{figure*}

\begin{figure*}
\includegraphics[width=0.85\columnwidth]{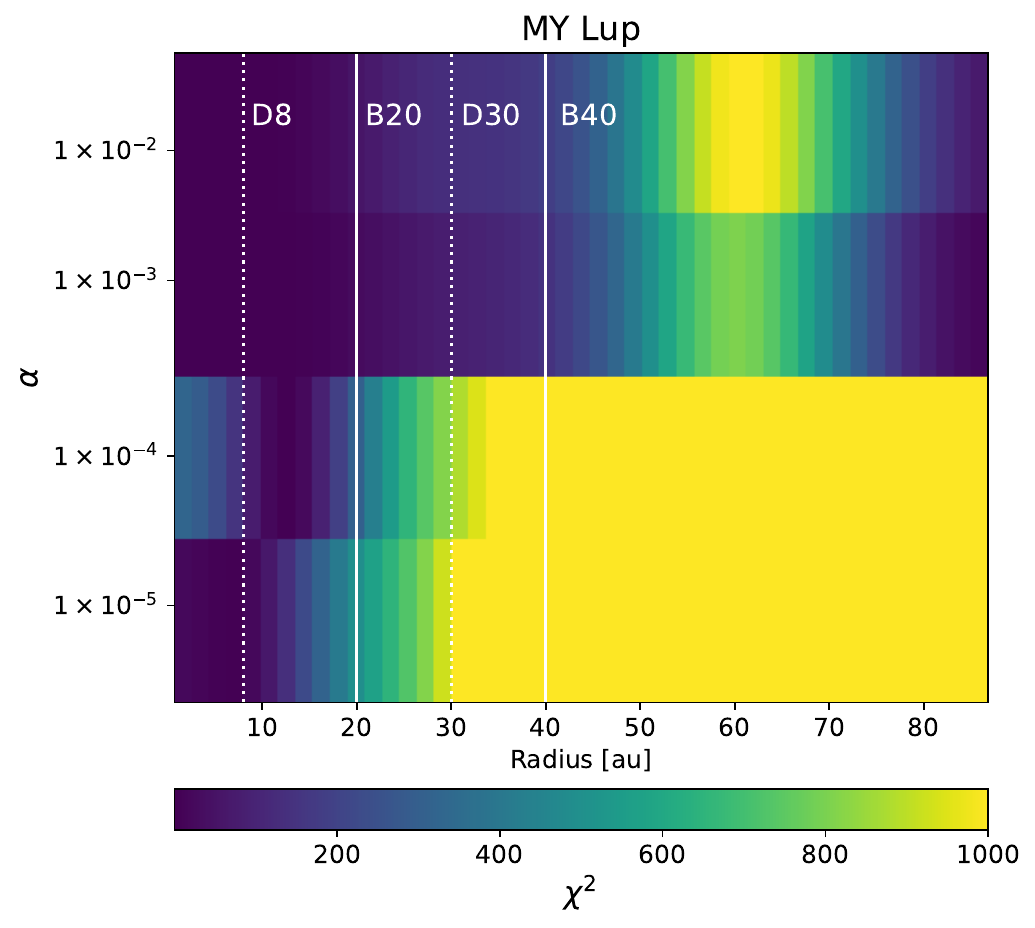}
\includegraphics[width=0.965\columnwidth]{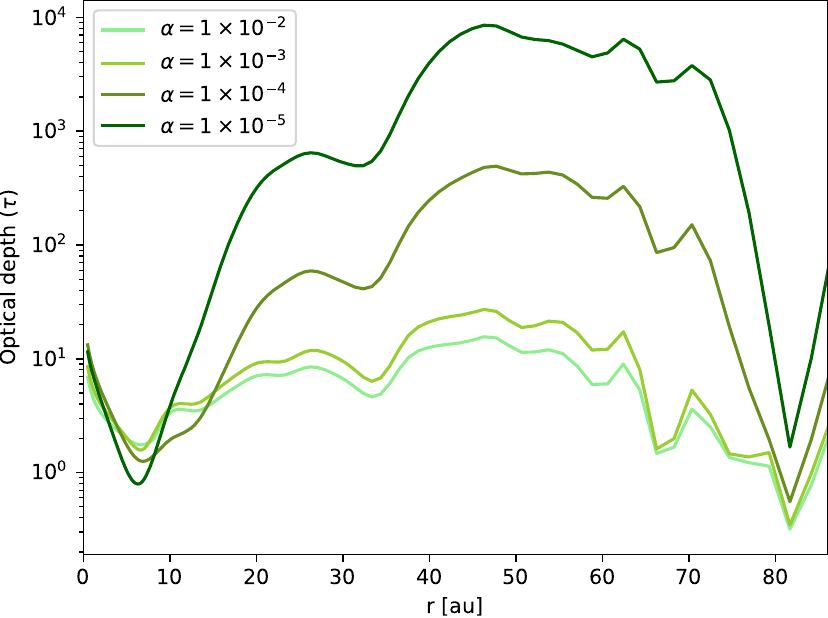}
\caption{Left panel shows $\chi^{2}$ between radial intensity profiles of the models and observations along the semi-minor axis for MY Lup. Models are computed for the astrosilicates prescription. Right panel shows vertical optical depth of the same models as a function of radius }
\label{fig:alpha-astrosil}
\end{figure*}

To exemplify the idea that different input dust opacities can lead to different vertical turbulence $\alpha$ results, we have calculated models of the ALMA observations for MY Lup, for a different prescription of dust opacity, which consists of 40\% graphite, 60\% astronomical silicates, and vacuum \citep{Draine2003a, Draine2003b} (shown in Fig.\,\ref{fig:opacities}, and hereafter referred to as "astrosilicates prescription"). This composition is motivated by the prescription considered in the recent article by \cite{Villenave2025}, 
which is similar, but not identical; since we could not reproduce exactly the same opacities used in \cite{Villenave2025}, using the grain compositions reported in that article. Fig.\,\ref{fig:profilesastrosil} shows the radial intensity profiles calculated for all models, following the same method described in Sect.\,\ref{sec:radiativetransfer}. This figure shows that two of the models can reproduce the shape of the radial intensity profile along the semi-major axis, and the two most settled models cannot reproduce the shape of the intensity profile. As the disc becomes optically thick for the most settled models, the procedure from the present paper does not allow us to find a good match for the dust surface density profile in these cases. We observe that the two most settled models have a very large optical depth compared to the less settled models in Fig.\,\ref{fig:alpha-astrosil}. Fig.\,\ref{fig:alpha-astrosil} shows that between the two less settled models, the model for alpha $\alpha = 1 \times10^{-3}$ is the best fit for the data of MY Lup. From these new models for MY Lup, we also show that different opacity prescriptions for the same fixed $\alpha$ will provide different vertical scale heights (Fig.\,\ref{fig:hdust1} for MY Lup), for the grain size at which the thermal emission is more efficient at the corresponding ALMA wavelength. The vertical height being smaller when the dust opacities peak at a larger grain size (for the same $\alpha$). In this example, the opacities peak at a larger grain size for the Ricci prescription.

\section{Spectral Energy Distribution}
\label{sec:discussionsed}
\begin{figure*}
\includegraphics[width=1\columnwidth]{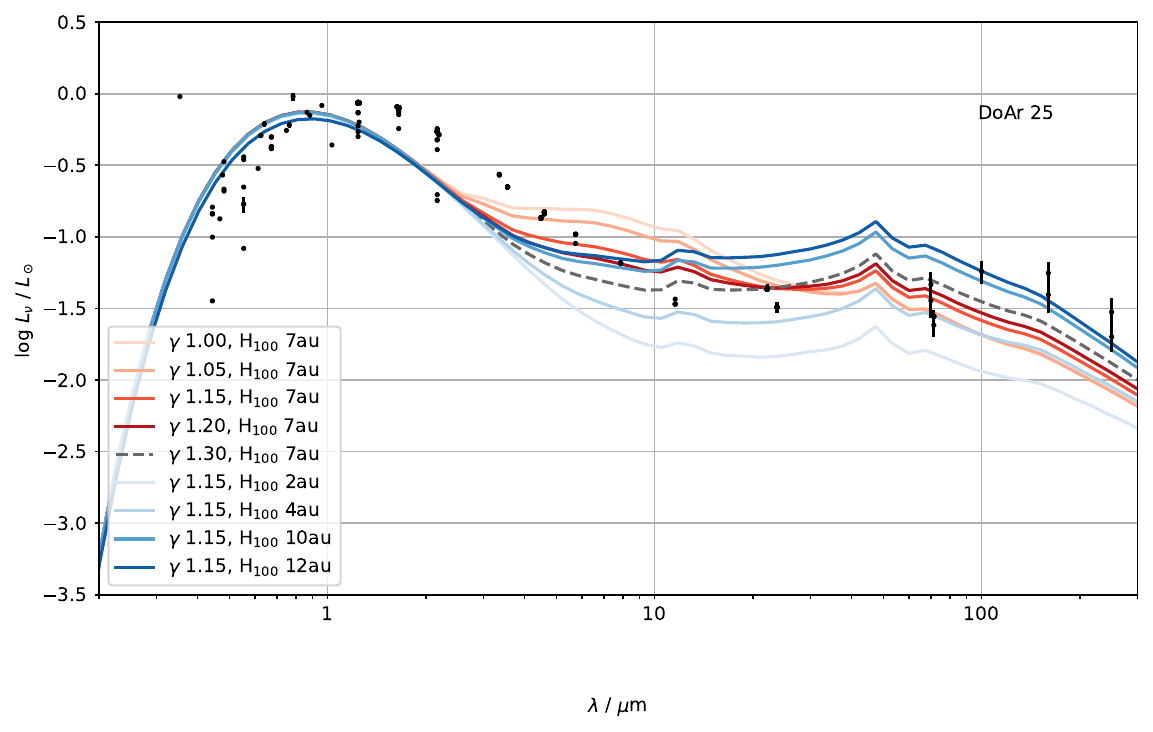}
\includegraphics[width=1\columnwidth]{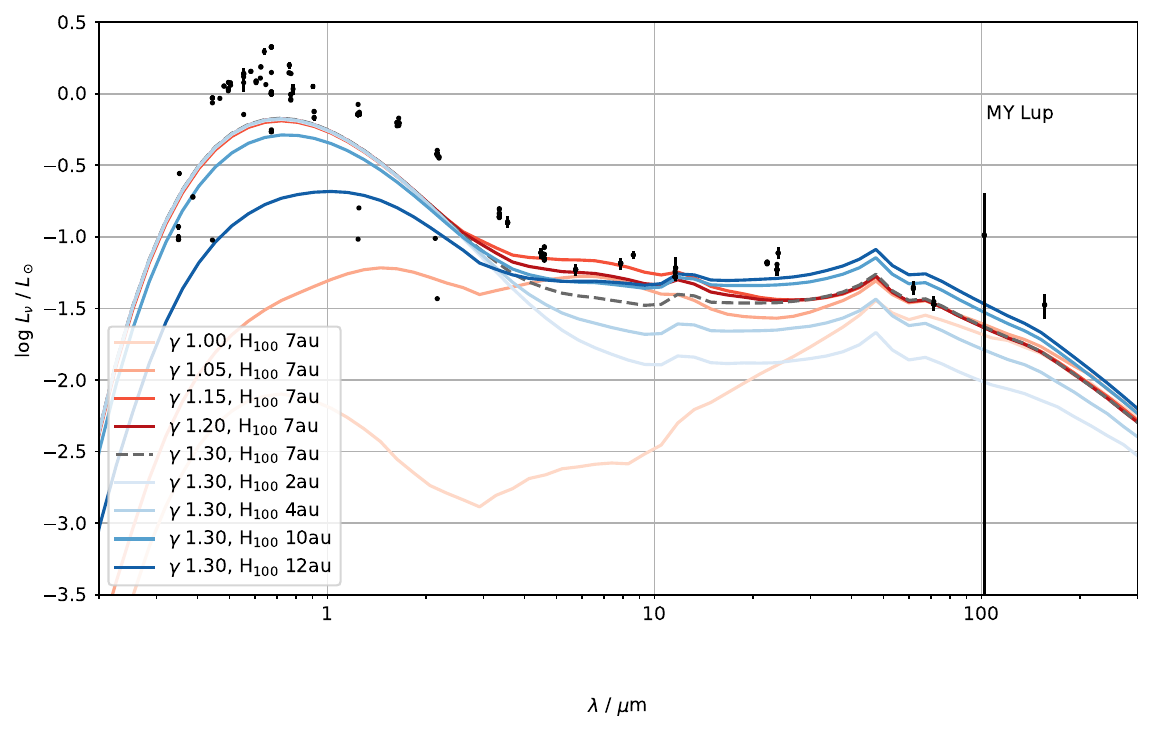}
\includegraphics[width=1\columnwidth]{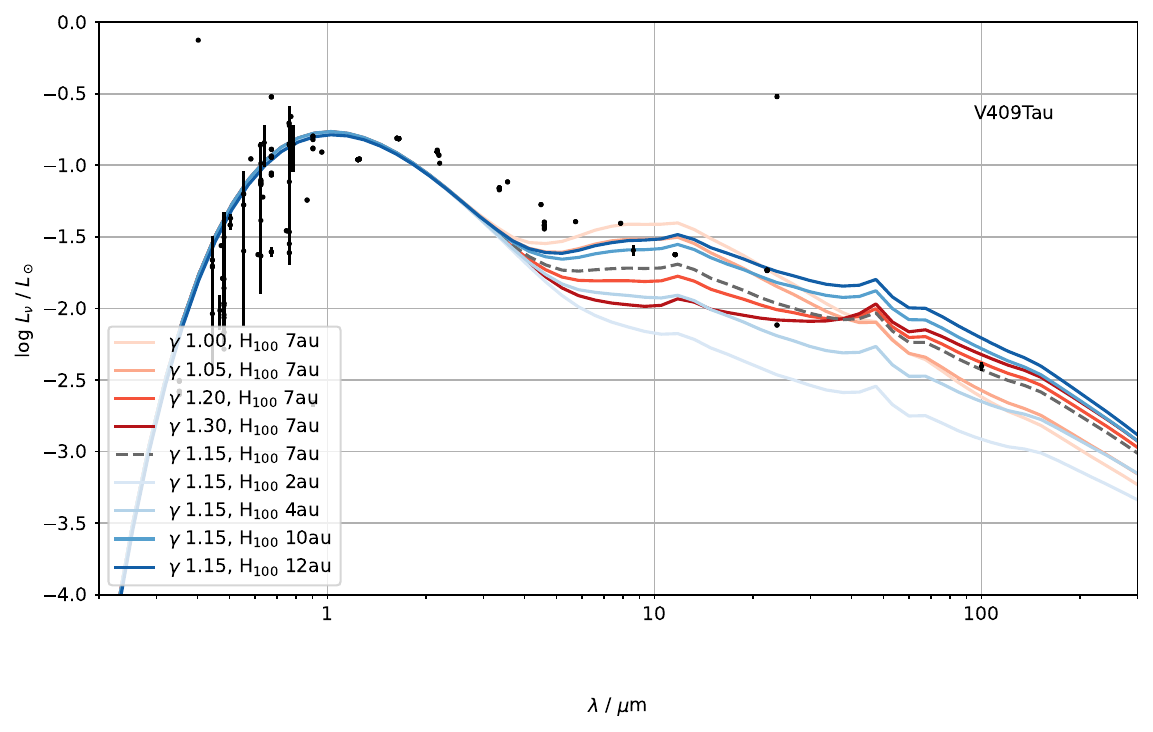}
\includegraphics[width=1\columnwidth]{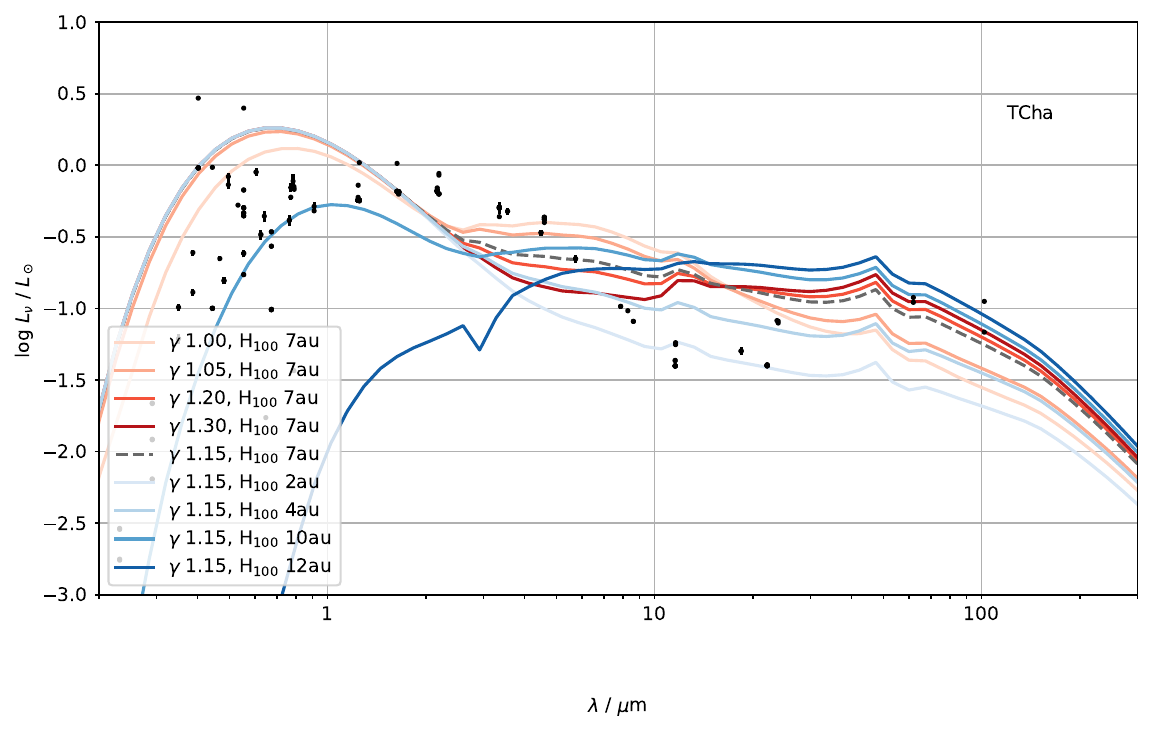}
\includegraphics[width=1.02\columnwidth]{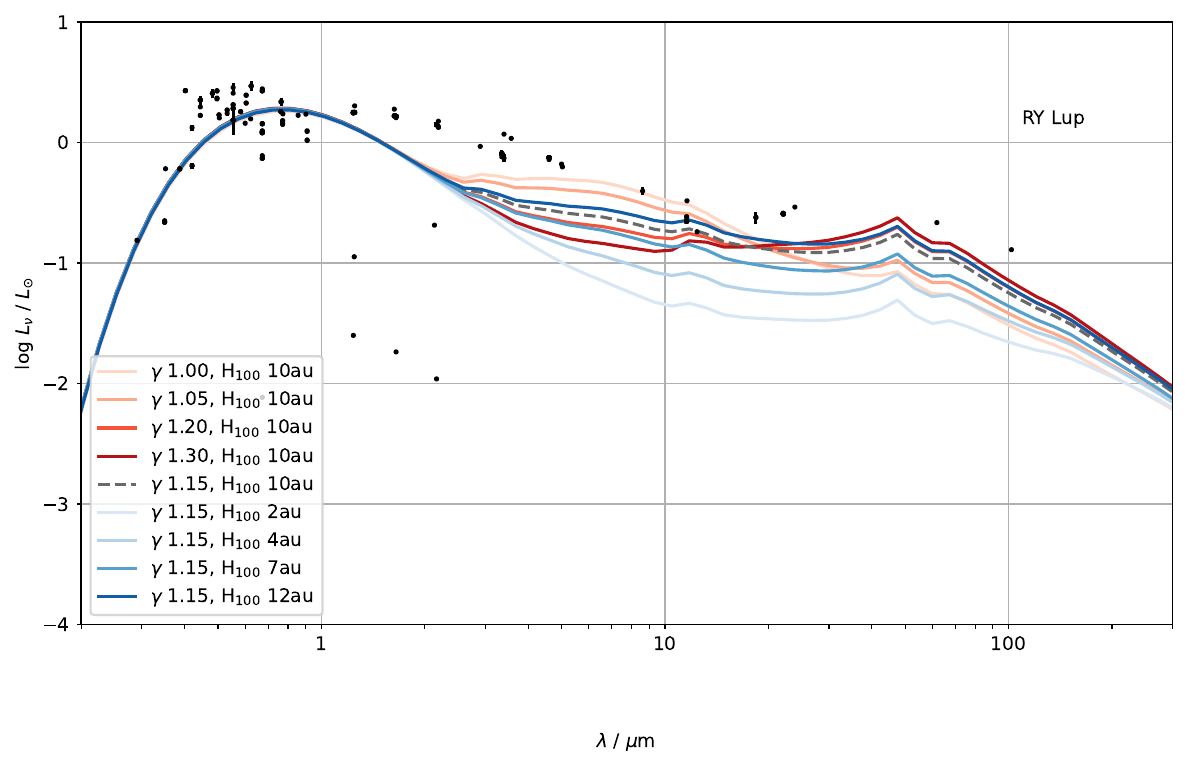}
\includegraphics[width=0.94\columnwidth]{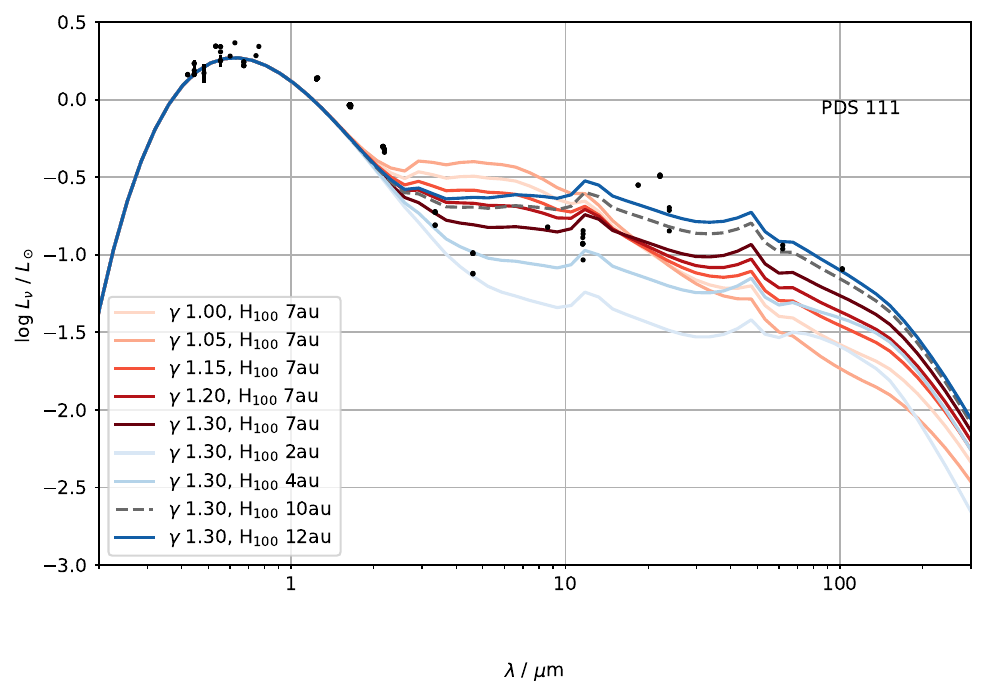}
\caption{Spectral energy distribution (SED) for all of the sources in our sample, modeled for different combinations of flaring index ($\gamma$), and gas scale height at 100 au (H$_{100}$). SEDs modeled for a fixed H$_{100}$ of 7 or 10 au are shown in red tones, and SEDs modeled for a fixed $\gamma$ are shown in blue tones. The model for the final combination of H$_{100}$ and $\gamma$ selected for ALMA data the modeling is shown with a grey dashed line. Photometric data is also shown in black dots, with the error bars. All photometric data was downloaded from the VizieR database of astronomical catalogues \citep{10.26093/cds/vizier,vizier2000}.}
\label{fig:seds}
\end{figure*}
To ensure that we select geometric parameters of flaring ($\gamma$) and scale height of the gas at 100 au (H$_{100}$) that are suitable for each disc in the ALMA data modeling, before fixing all parameters in the final simulations (Table\,\ref{tab:table3}), we performed several test simulations for different combinations of geometric parameters $\gamma$ and H$_{100}$ to reproduce the general shape of the SED. These tests are performed using the methodology presented in Sec.\,\ref{sec:radiativetransfer}; the 2D dust density distribution is calculated for 15 iterations in all cases, and the spectra are calculated using RADMC-3D after computing the dust temperature. For all sources, we first fixed H$_{100}$ to 7 or 10 au, and modeled the SED for a range of $\gamma$ parameters fixed to 1, 1.05, 1.15, 1.20, and 1.30. These results are shown in Fig.\,\ref{fig:seds} with red tones. From these tests performed for a fixed H$_{100}$, we select a parameter $\gamma$ that reproduces the general shape of each SED, and then perform additional tests for H$_{100}$. Here, we explore the parameter space of H$_{100}$ equal to 2, 4, 7, 10, and 12 au. The results of the latest tests mentioned are shown in Fig.\,\ref{fig:seds} with blue tones. After the second round of tests, we select an H$_{100}$ value, for the previous fixed $\gamma$ that reproduces the general shape of the SED for each source. The final geometric parameters selected for each disc are presented in Table\,\ref{tab:table3}. All photometric data that we used for these tests were downloaded from the VizieR database of astronomical catalogues \citep{10.26093/cds/vizier,vizier2000}.

\section{Testing other parameters for $\gamma$ and H$_{100}$ for MY Lup and T Cha}
\label{sec:geometrical-parameters}
\begin{figure*}
\includegraphics[width=0.73\columnwidth]{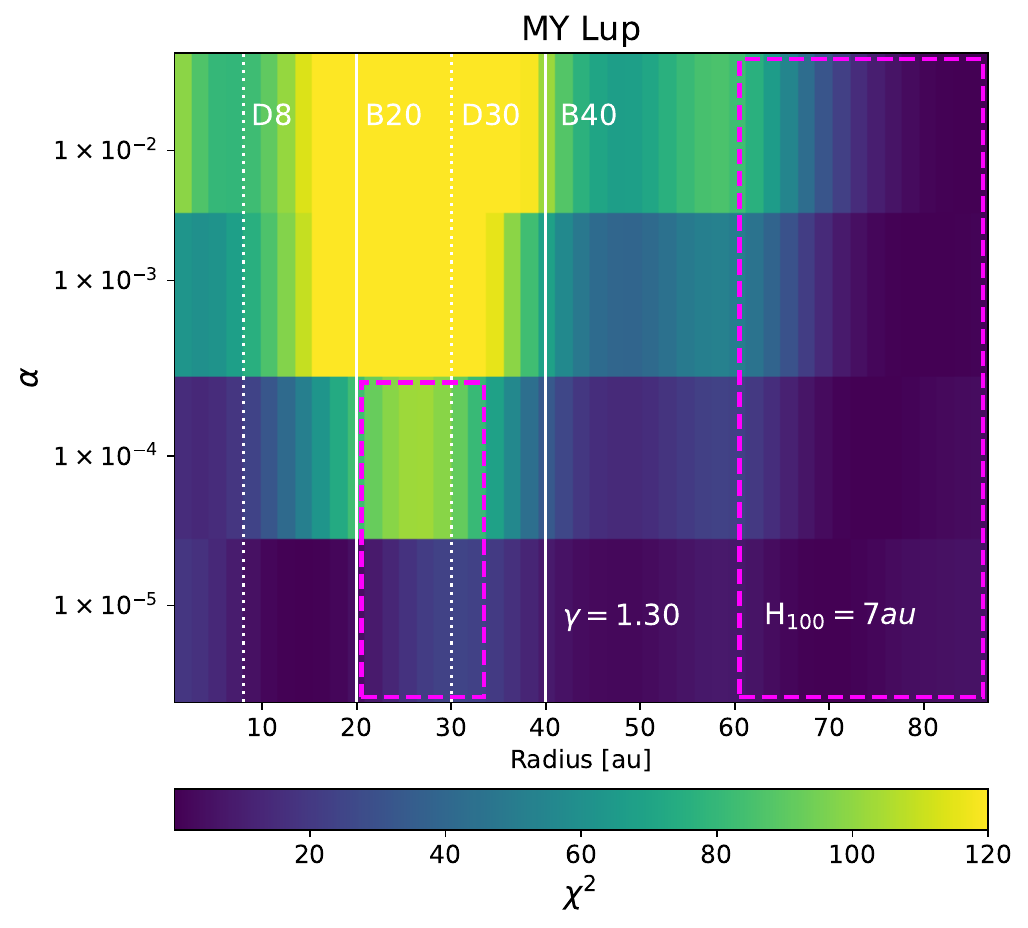}
\includegraphics[width=0.73\columnwidth]{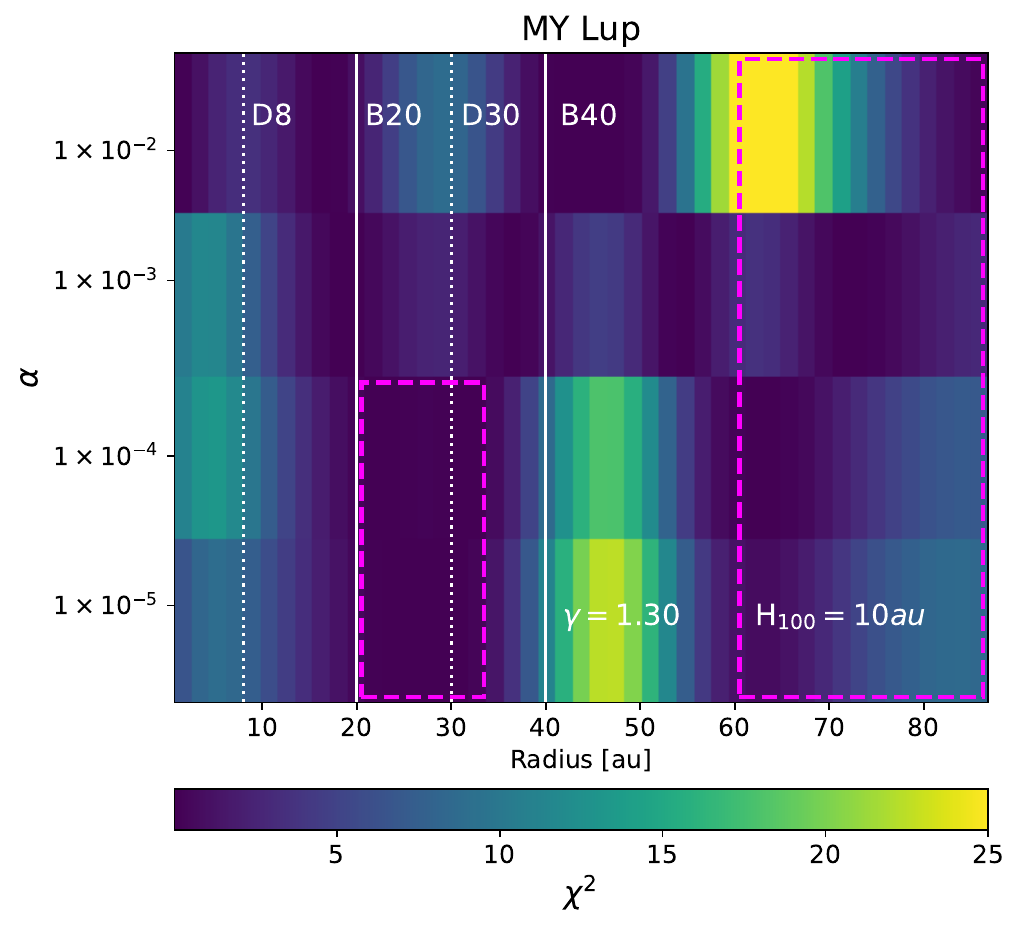}
\includegraphics[width=0.73\columnwidth]{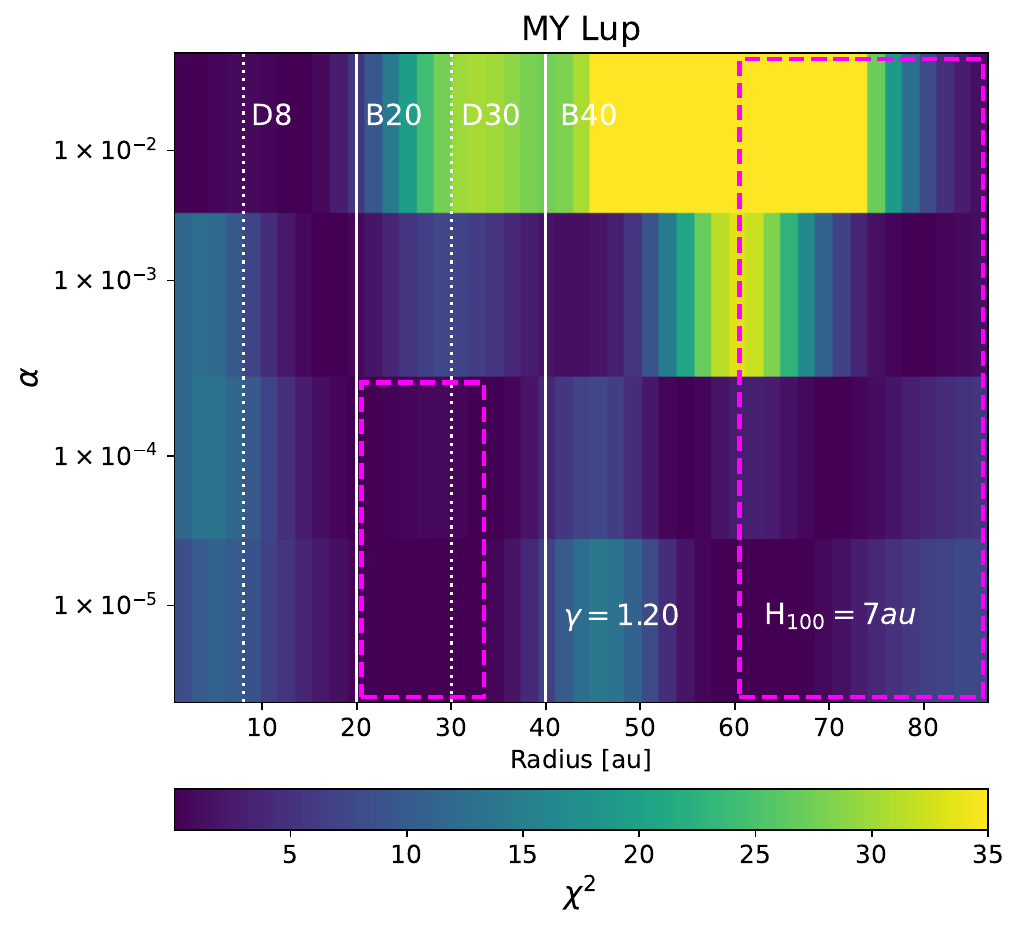}
\includegraphics[width=0.73\columnwidth]{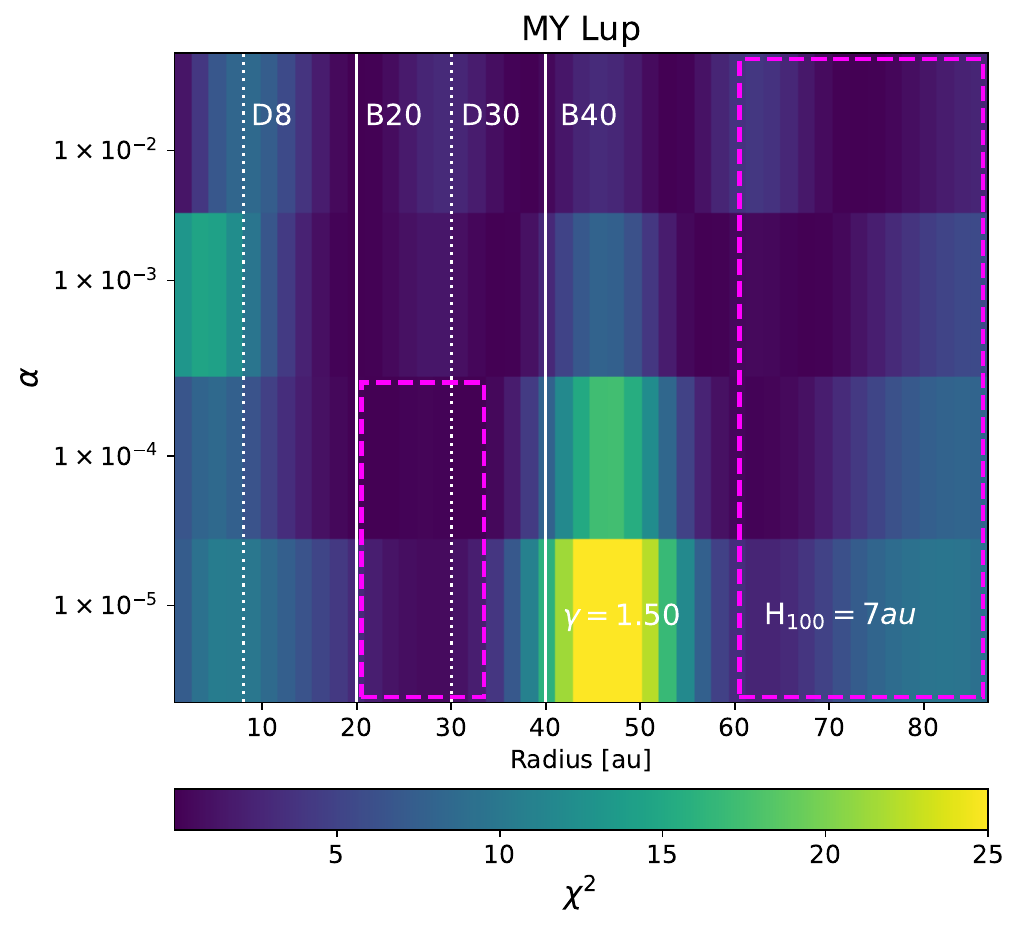}
\includegraphics[width=0.757\columnwidth]{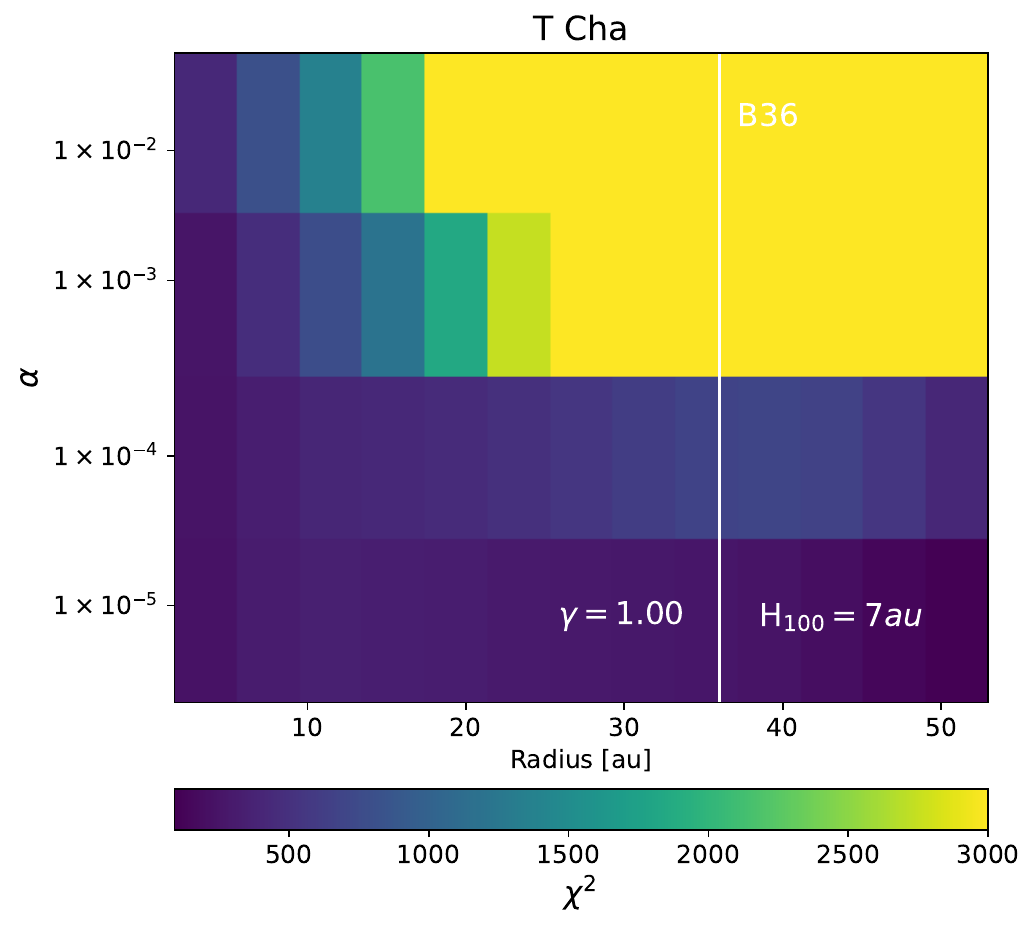}
\includegraphics[width=0.73\columnwidth]{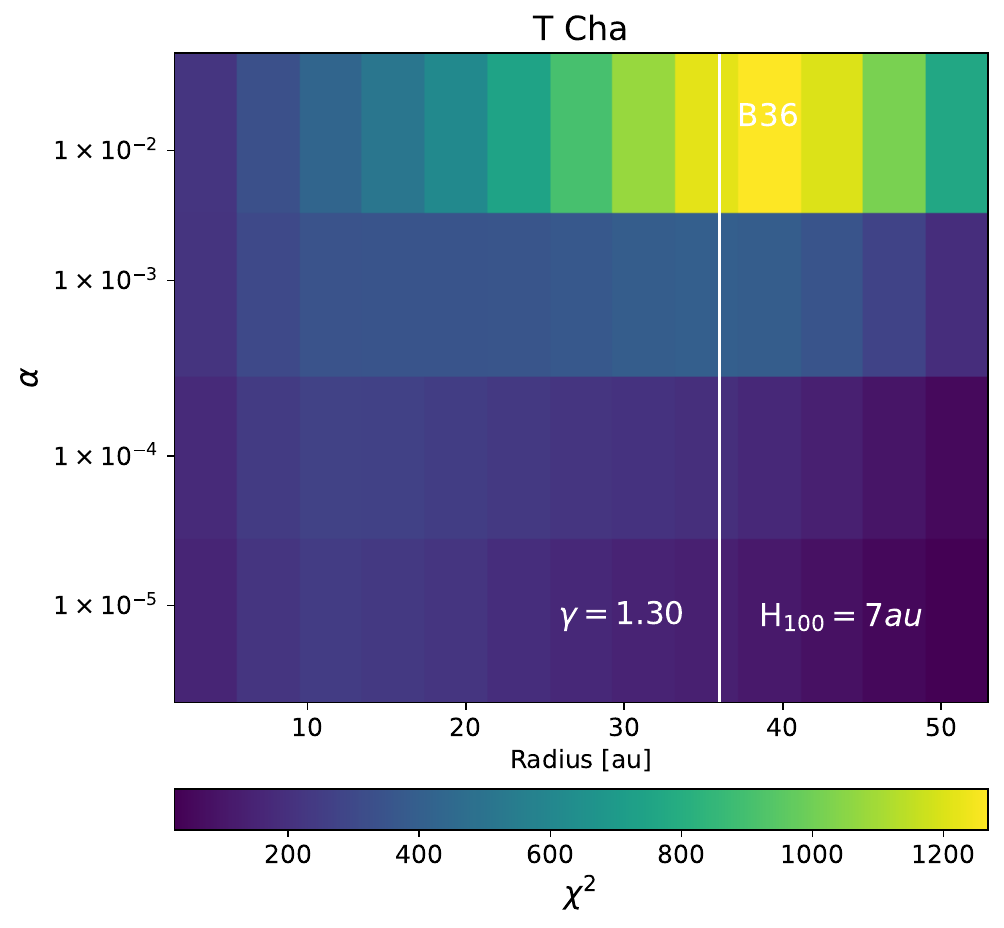}
\includegraphics[width=0.73\columnwidth]{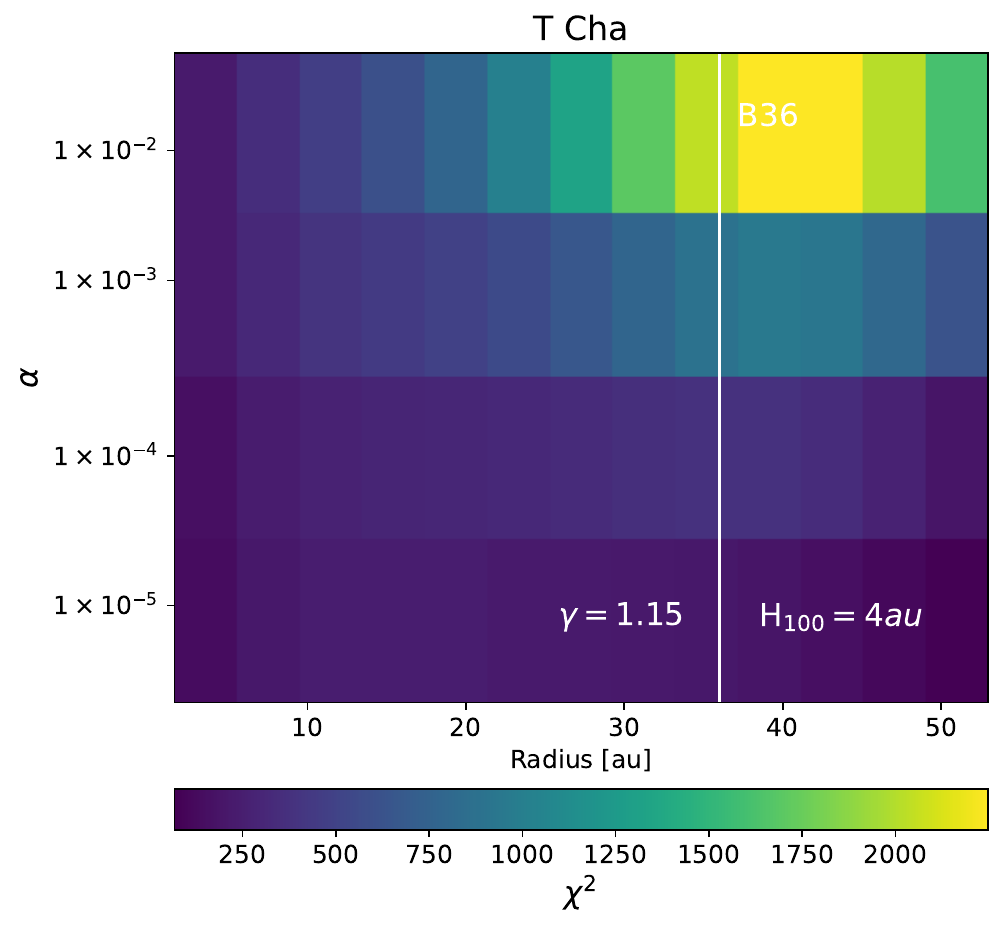}
\includegraphics[width=0.73\columnwidth]{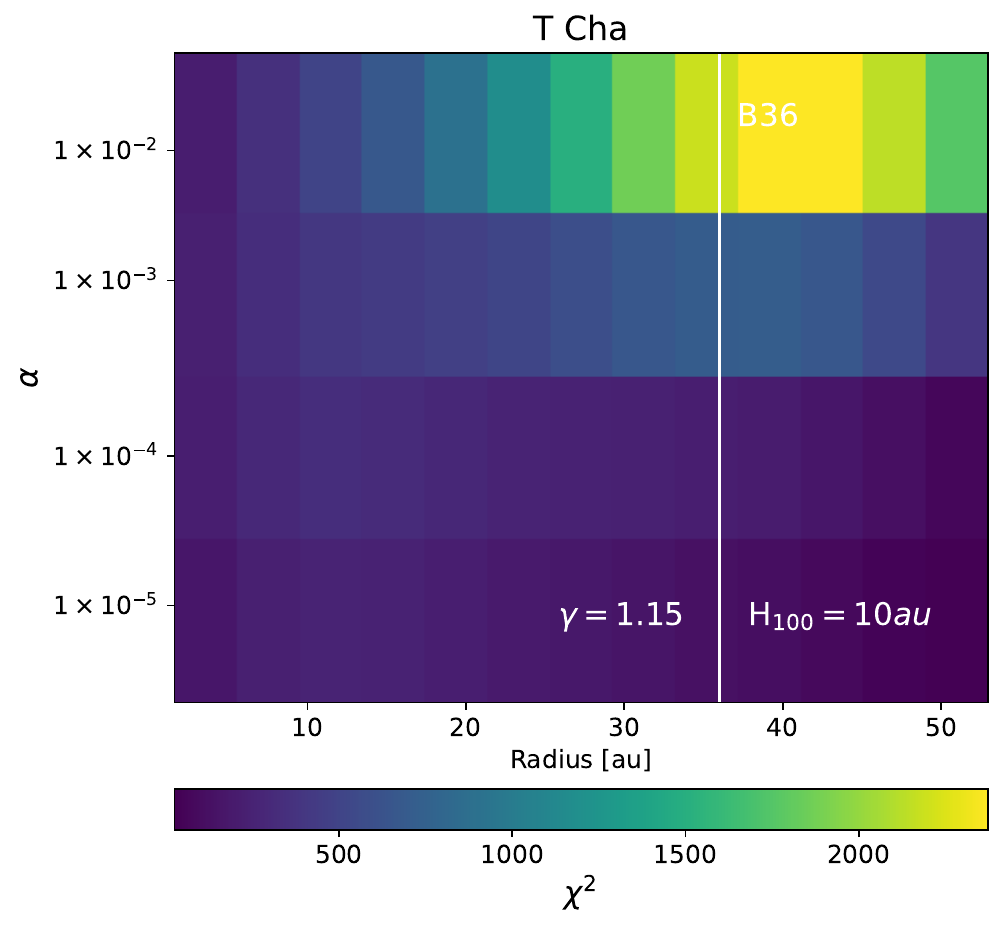}
\caption{All panels show $\chi^{2}$ between radial intensity profiles of the models and observations along the semi-minor axis for MY Lup and T Cha. Models are computed for the Ricci opacities prescription, and different geometric parameters $\gamma$, and H$_{100}$, which are indicated at the bottom right in each figure. Pink rectangles highlight the trends that are observed from all simulations for MY Lup.}
\label{fig:test-geometrical-params}
\end{figure*}

To investigate the potential influence of the geometric parameters $\gamma$ and H$_{100}$ on the final constraints made on the strength of dust settling and vertical turbulence, we tested several geometric parameters for MY Lup and T Cha. We selected MY Lup because the observations do not completely resolve the structures and the results of dust scale height and vertical turbulence remain tentative; and T Cha because our results are more robust in this case, since we resolve the structures and the observations have lower optical depth. The initial assumptions for MY Lup were $\gamma$ of 1.30 and H$_{100}$ of 7 au (Table\,\ref{tab:table3}), since these values fit the SED (Fig.\,\ref{sec:discussionsed}). We consider the initial model and assumptions as a reference point and vary $\gamma$ and H$_{100}$ to different values. We computed a model for a fixed $\gamma$ of 1.30 au for an H$_{100}$ of 10 au; and we also computed two models for a fixed H$_{100}$ of 7 au for a $\gamma$ of 1.20 and 1.30. We also confirmed that the SED of the models is similar to the observed SED for all the new combinations of parameters. These three new models are shown together with the model presented earlier in this paper for comparison in Fig.\,\ref{fig:test-geometrical-params}. From these tests, we conclude that geometry assumptions can influence the $\alpha$ constraints when substructures remain unresolved, but we highlight that some initial trends are maintained and observed in all the models for MY Lup. In particular, low $\alpha$ ($\lesssim 1\times10^{-4}$) is observed in all panels at $20\lesssim r\lesssim34$, and the increasing alpha is also observed in al panels at $r\gtrsim60$. We also performed similar tests for T Cha, considered the initial model and assumptions as a reference point, and also vary $\gamma$ and H$_{100}$ to different values. The results in \ref{fig:test-geometrical-params}, show that our constraints are more robust in this case because all panels show the dominance of low $\alpha$. This is possibly because it is easier to observe the gap-ring contrast in this source and because the emission is optically thinner at Band 3 from ALMA.


\bsp	
\label{lastpage}
\end{document}